\documentclass[showkeys, floatfix, noshowpacs,
	 preprintnumbers, twocolumn, amsmath, amssymb, aps, prl, 10pt, superscriptaddress
]{revtex4-2}
\pdfoutput=1

\newcommand{\sectioning}[1]{\textit{\textbf{#1}}}
\usepackage{color}
\usepackage{graphicx}
\usepackage{soul}

\begin{document}

\title{Non-linear Boson Sampling} 

\author{Nicol\`o Spagnolo}
\affiliation{Dipartimento di Fisica, Sapienza Universit\`{a} di Roma, Piazzale Aldo Moro 5, I-00185 Roma, Italy}
\author{Daniel J. Brod}
\affiliation{Instituto de F\'isica, Universidade Federal Fluminense, Niter\'oi, RJ, 24210-340, Brazil}
\author{Ernesto F. Galv\~ao}
\affiliation{International Iberian Nanotechnology Laboratory (INL), Ave. Mestre Jose Veiga, 4715-330, Braga, Portugal}
\affiliation{Instituto de Fisica, Universidade Federal Fluminense, Niter\'oi, RJ, 24210-340, Brazil}
\author{Fabio Sciarrino}
\affiliation{Dipartimento di Fisica, Sapienza Universit\`{a} di Roma, Piazzale Aldo Moro 5, I-00185 Roma, Italy}

\begin{abstract}
Boson Sampling is a task that is conjectured to be computationally hard for a classical computer, but which can be efficiently solved by linear-optical interferometers with Fock state inputs. Significant advances have been reported in the last few years, with demonstrations of small- and medium-scale devices, as well as implementations of variants such as Gaussian Boson Sampling. Besides the relevance of this class of computational models in the quest for  unambiguous experimental demonstrations of quantum advantage, recent results have also proposed first applications for hybrid quantum computing. Here, we introduce the adoption of non-linear photon-photon interactions in the Boson Sampling framework, and analyze the enhancement in complexity via an explicit linear-optical simulation scheme. By extending the computational expressivity of Boson Sampling, the introduction of non-linearities promises to disclose novel functionalities for this class of quantum devices. Hence, our results are expected to lead to new applications of near-term, restricted photonic quantum computers.
\end{abstract}

\maketitle

\sectioning{Introduction. --} Quantum technologies promise to provide speed-up in several fields, ranging from intrinsically secure long-distance quantum communication \cite{Yin17} to a novel generation of high-precision sensors \cite{Giovannetti11}, and enhanced computational and simulation capabilities \cite{NielsenBook}. Among the currently developed experimental platforms, in the last few years photonic technologies have recently experienced a technological boost in all fundamental components, namely photon sources, manipulation and detection \cite{Flamini18}. 

Recent studies have focused on identifying suitable dedicated classically-hard tasks, with the aim of leveraging the necessary technological resources and system size to reach the quantum advantage regime \cite{Arute19Supremacy,Zhong20}. Such regime corresponds to the scenario where a quantum device solves a given task faster than any classical counterpart. Within this context, a computational problem named Boson Sampling \cite{AA} has been defined as a promising approach. This problem, that consists in sampling from the output distribution of a system of $n$ non-interacting bosons undergoing linear evolution, is a classically-hard task (in $n$) while it can be naturally solved by a linear optical photonic system. Such sampling problem has also subsequently inspired other classes of sampling problems \cite{Boixo18,Harrow17} suitable to be solved with different quantum hardware \cite{Bernien17,Neill18,Arute19Supremacy}.

Starting from the original proposal \cite{AA}, several experimental implementations of Boson Sampling instances \cite{Broome13,Spring13,Tillmann13,Crespi13,Spagnolo14,Carolan14,Carolan15,Loredo17,He17,Wang17bs,Wang18,Wang19Det} and of recently proposed variants \cite{Bentivegna15,Zhong18,Wang18b,Paesani19,Zhong19} have been reported \cite{Brod19Review}. In particular, hybrid algorithms based on Gaussian Boson Sampling have been proposed for various tasks: quantum simulation \cite{Huh15,Huh17,Chin18,Paesani19,Banchi19}, optimization problems \cite{Arrazola18b}, point processes \cite{Jahangiri20point} graph theory \cite{Arrazola18,Bradler18,Bradler18b}, and quantum optical neural networks \cite{Steibrecher19NN}. Very recently, impressive experimental implementations of Gaussian Boson sampling have been reported \cite{Zhong20, Arrazola21}.  Besides the technological advances reported in the last few years, several studies have also focused on studying and improving classical simulation of Boson Sampling \cite{Neville17,Clifford18,Wu16,Gupt18,Lundow19,Clifford2020Faster}, and on defining the limits for simulability in the presence of imperfections, in particular losses and partial photon distinguishability \cite{Aaronson15,Arkhipov15,Leverrier15,Oszmaniec18,Garcia-Patron17,Renema17,Renema18,Qi19}. All these studies aimed at establishing a classical benchmarking framework for Boson Sampling, and currently place the threshold for quantum advantage in such a system to $n \sim 50$ photons in a network composed by $m \sim n^2 = 2500$ optical modes. Recent improvements in photon sources \cite{Michler2282,Ding_quantum_dot,Somaschi2016,Michler17,Huber18,Basso19} enabled first Boson Sampling experiments with a number of detected photons up to $n=14$ \cite{Wang19Det}. However, reaching the quantum advantage regime with a photonic platform solving the original formulation of the task \cite{AA} still requires a technological leap to enhance single-photon generation rates and indistinguishability, and to reduce losses in the current platforms for linear-optical networks.

In this Letter, we introduce the adoption of non-linear interactions at the few-photon level within the Boson Sampling framework as a route to increase the complexity and reduce the threshold for the quantum advantage regime. This possibility is encouraged by recent advances showing the first experimental demonstrations of non-linear photonic processes within solid state devices \cite{DeSantis17}. We will first describe the introduction of non-linearities within the otherwise linear evolution. Then, we will provide an upper bound on the complexity of the enhanced devices via a simulation scheme based on auxiliary, linear-optical gadgets. We will discuss both the asymptotic and finite cases, leveraging results from the well-established linear Boson Sampling framework \cite{AA}.

\begin{figure*}[ht!]
\centering
\includegraphics[width=0.99\textwidth]{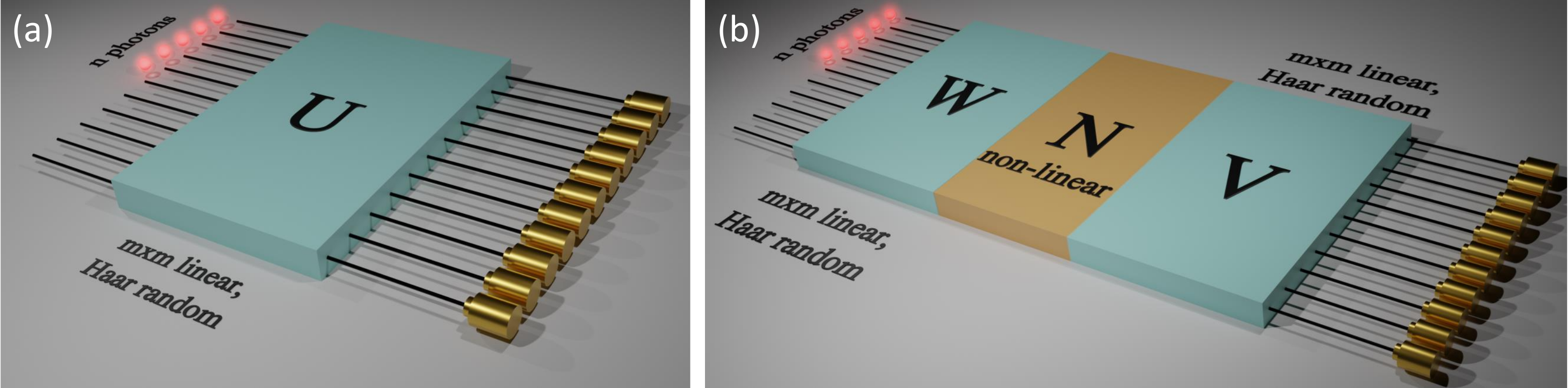}
\caption{(a) Scheme for a Boson Sampling apparatus according to the original formulation \cite{AA}, where $n$ input photons impinge in a linear network described by unitary matrx $U$. (b) Scheme for non-linear Boson Sampling which includes a non-linear evolution step $N$ between two Haar-random linear steps $W$ and $V$.}
\label{fig:scheme}
\end{figure*}

\sectioning{Boson Sampling. --} Boson Sampling \cite{AA} is a computational task which corresponds to sampling from the output distribution of $n$ indistinguishable, non-interacting photons after evolution through a $m$-mode linear network [see Fig. \ref{fig:scheme}(a)].
Given an interferometer described by a unitary matrix $U$, the transition amplitude from input $\vert S \rangle$ to output state $\vert T \rangle$ can be written as:
\begin{equation}
\label{eq:BS}
\mathcal{A}_{U}\big(\vert S \rangle \rightarrow \vert T \rangle\big) = \langle S \vert \varphi(U) \vert T \rangle = \frac{\mathrm{per}(U_{S,T})}{\sqrt{\prod_{i,j=1}^{m} s_{i}! t_{j}!}},
\end{equation}
where $\mathrm{per}(A) = \sum_{\sigma \in S_{n}} \prod_{i=1}^{n} a_{i,\sigma(i)}$ is the matrix permanent, $\{s_i\} (\{t_i\})$ are the occupation numbers of states $\vert S \rangle (\vert T \rangle)$, and $U_{S,T}$ is the $n \times n$ matrix obtained by selecting rows and columns of $U$ according to the $(s_{1}, \ldots, s_{m})$ and $(t_{1}, \ldots, t_{m})$ respectively. Calculation of permanents of matrices with complex entries is in the \#P-hard computational complexity class \cite{Valiant79}. In Eq.\ \eqref{eq:BS}, $\varphi(U)$ is the unitary transformation acting on the Hilbert space $\mathcal{H}_{m,n}$ of $n$ photons in $m$ modes, that corresponds to the linear evolution $U$ on the optical modes.  Due to the linearity of the evolution, $\varphi(U)$ is an homomorphism \cite{Aaronson11}. This means that, if a given evolution is the sequence of two linear networks $W$ and $V$, the overall evolution can be written in terms of permanents of submatrices of $U = V W$. 

In Ref.\ \cite{AA} it was shown that sampling (even approximately) from the output distribution of such a system is classically hard if (i) the input state $\vert S \rangle = \vert s_{1}, \ldots s_{m} \rangle$ has at most one photon per mode, (ii) $U$ is drawn randomly from the uniform Haar measure, and (iii) the number of modes $m$ and photons $n$ satisfy $m \gg n^{6}$.

\sectioning{Non-linear Boson Sampling. --} Let us now consider the scheme of Fig.\ \ref{fig:scheme}(b). An input state $\vert S \rangle = \vert s_{1}, \ldots s_{m} \rangle$ of $n$ indistinguishable photons undergoes a $m$-mode evolution divided in three steps. While steps 1 and 3 are linear evolutions $V$ and $W$ drawn from the Haar ensemble, the intermediate step 2 now consists of a non-linear evolution $N$. This $N$ transforms a state $\vert R \rangle = \vert r_{1}, \ldots r_{m} \rangle$ as $\vert R \rangle \stackrel{N}{\rightarrow}\sum_{R \in \Phi_{m,n}} \mathcal{N}_{r_{1} \ldots r_{m}}^{q_{1} \ldots q_{m}} \vert Q \rangle$, where $\vert Q \rangle = \vert q_{1}, \ldots q_{m} \rangle$ and $\Phi_{m,n}$ is the set of tuples corresponding to $n$ photons in $m$ modes. In this equation, function $\mathcal{N}_{r_{1} \ldots r_{m}}^{q_{1} \ldots q_{m}}$ represents the transition amplitude $\mathcal{A}_{N}(\vert R \rangle \rightarrow \vert Q \rangle)$ determined by the non-linear evolution. We assume  $\mathcal{N}_{r_{1} \ldots r_{m}}^{q_{1} \ldots q_{m}}$ has an efficient classical description, e.g., it is given by the composition of a small number of few-mode non-linear transformations, or by a Hamiltonian with a simple form in terms of the field operators.

Let us now write the overall transformation of input state $\vert S \rangle = \vert s_{1}, \ldots s_{m} \rangle$ according to the three-step evolution $W \rightarrow N \rightarrow V$, which includes linear transformations $W,V$ of the form given by Eq. (\ref{eq:BS}), and the non-linear $\mathcal{N}_{r_{1} \ldots r_{m}}^{q_{1} \ldots q_{m}}$: 
\begin{equation}
\label{eq:gen_nonlin}
\begin{aligned}
\mathcal{A}&_{W,N,V}\big( \vert S \rangle \rightarrow \vert T \rangle \big) = \\
&=\sum_{R \in \Phi_{m,n}}\sum_{Q \in \Phi_{m,n}} \frac{\mathrm{per}(W_{S,R}) \mathcal{N}_{r_{1} \ldots r_{m}}^{q_{1} \ldots q_{m}} \mathrm{per}(V_{Q,T})}{\sqrt{\prod_{i,j,k,l=1}^{m}s_{i}! r_{j}! q_{k}! t_{l}!}}.
\end{aligned}
\end{equation}

This amplitude is written as a Feynman path sum over all possible basis states just before and after the non-linear evolution step. If the permanent distribution was peaked, it might be possible to obtain a good approximation to Eq.\ \eqref{eq:gen_nonlin} by summing over only the dominant terms. Haar random matrices, however, display an anti-concentrated, relatively flat distribution \cite{AA}. In \cite{SuppMat} we provide numerical evidence for this, showing that to account for $(90\%, 95\%, 99\%)$ of the total probability mass function, we need to calculate the probabilities associated with respective fractions $\sim (0.5, 0.6, 0.8)$ of all possible outcomes; moreover, this behavior is nearly independent of $n$ and $m$.

Of course, there may be computational shortcuts to evaluating Eq.\ \eqref{eq:gen_nonlin}, other than the explicit sum over paths. For example, if we replace the non-linear term $N$ by a linear term, the amplitude can be evaluated as a single permanent. This motivates us to investigate different ways to assess the complexity of non-linear Boson Sampling.

\sectioning{Single-mode non-linear phase shift gate. --} Let us proceed by studying a specific example of non-linear evolution $N$ consisting of a single non-linear phase gate introduced in mode $x$. The unitary operator describing this gate can be written as $\hat{U}_{\mathrm{nlp}} = \exp(-\imath \hat{n}^2_{x} \phi)$. Its action on a generic $m$-mode state $\vert R \rangle$ leads to a function $\mathcal{N}$ of the form $\mathcal{N}_{r_{1} \ldots r_{m}}^{q_{1} \ldots q_{m}} = \exp(-\imath r_{x}^{2} \phi) \prod_{i=1}^{m} \delta_{r_{i},q_{i}}.$ Inserting this choice of non-linear evolution into the general expression (\ref{eq:gen_nonlin}) we obtain:
\begin{equation}
\label{eq:nonlinear_singlephase}
\begin{aligned}
\mathcal{A}^{\mathrm{nlp}}_{W,N,V} &\big( \vert S \rangle \rightarrow \vert T \rangle \big) = \sum_{R \in \Phi_{m,n}} \exp(-\imath r_{x}^{2} \phi) \times \\
& \times \frac{\mathrm{per}(W_{S,R}) \mathrm{per}(V_{R,T})}{\sqrt{\prod_{i=1}^{m}s_{i}! \big(\prod_{j=1}^{m}r_{j}!\big)^{2} \prod_{l=1}^{m}t_{l}!}},
\end{aligned}
\end{equation}
Eq.\ \eqref{eq:nonlinear_singlephase} can be rearranged in the following form (see \cite{SuppMat}):
\begin{equation}
\label{eq:nonlinear_singlephase_simpl}
\begin{aligned}
&\mathcal{A}^{\mathrm{nlp}}_{W,N,V} \big( \vert S \rangle \rightarrow \vert T \rangle \big) = 
\frac{\mathrm{per}(\bar{U}_{S,T})}{\sqrt{\prod_{i=1}^{m}s_{i}! \prod_{l=1}^{m}t_{l}!}} + \\
&+ \sum_{r_{x}>1} \frac{\mathrm{per}(W_{S,R}) [\exp(-\imath r_{x}^{2} \phi) - \exp(-\imath r_{x} \phi)] \mathrm{per}(V_{R,T})}{\sqrt{\prod_{i=1}^{m}s_{i}! \big(\prod_{j=1}^{m}r_{j}!\big)^{2} \prod_{l=1}^{m}t_{l}!}}.
\end{aligned}
\end{equation}
Here, $\bar{U}$ is a unitary transformation composed by the sequence $W$, $F$ and $V$, where $F$ replaces the non-linear phase in layer $N$ with a linear phase shift described by the operator $\exp(-\imath \hat{n}_{x} \phi)$. Equation \eqref{eq:nonlinear_singlephase_simpl} clearly shows that the departure from linear evolution is due only to bunching terms, corresponding to more than a single photon in mode $x$.

\sectioning{Bounding complexity via linear-optical simulation using auxiliary photons. --} An upper bound on the complexity of non-linear Boson Sampling can be obtained by devising a specific linear-optical simulation algorithm, that we discuss below for the case of a single-mode non-linear phase.

The simulation is based on the results by Scheel \textit{et al.} \cite{Scheel03}, describing how auxiliary photons and modes can be used, together with linear optics, to induce effective non-linear gates. In particular, given a single mode state in the photon number basis $\vert \psi_{\mathrm{in}} \rangle = \sum_{i} c_{i} \vert i \rangle$, it is possible to apply a polynomial of degree $k$ in the photon number operator $P_{k}(\hat{n})$ to $\vert \psi_{\mathrm{in}} \rangle$ by injecting the state in mode $1$ of a suitably chosen $(k+1)$-mode linear-optical gadget described by unitary $U_{\mathrm{eff}}$, where the auxiliary modes $j=2, \ldots, k+1$ are injected with a single photon state $\vert 1 \rangle_{j}$. The desired output state $\vert \psi_{\mathrm{out}} \rangle  = P_{k}(\hat{n}) \vert \psi_{\mathrm{in}} \rangle$ is obtained upon conditional detection of a single photon on each of the auxiliary modes. If the input state has a maximum number of $l$ photons $\vert \chi_{\mathrm{in}} \rangle = c_{0} \vert 0 \rangle + \ldots + c_{l} \vert l \rangle$, a polynomial of degree $l$ in $\hat{n}$ is sufficient to obtain the general evolution from $\vert \chi_{\mathrm{in}} \rangle$ to $\vert \chi_{\mathrm{out}} \rangle = c^{\prime}_{0} \vert 0 \rangle + \ldots + c^{\prime}_{l} \vert l \rangle$ with arbitrary coefficients $\{c^{\prime}_{0}, \ldots, c^{\prime}_{k}\}$ \cite{footnote_measInduced}. The success probability of the operation is equal to $\mathrm{Pr}_{\mathrm{succ}} = \vert \mathrm{per} (U_{\mathrm{eff}}^{0,1, \ldots, 1}) \vert^{2}$ \cite{Scheel03}, where $U_{\mathrm{eff}}^{0,1, \ldots, 1}$ is the $k \times k$ submatrix of $U_{\mathrm{eff}}$ obtained by removing row 1 and column 1 from the full matrix.

Finding the effective linear-optical simulation unitary $U_{\mathrm{eff}}$ has been done previously only for a few types of gates and small $k$ \cite{KLM01, Scheel03, Uskov09, Uskov10}, as the computational effort seems to scale exponentially with $k$. Nevertheless, even limited non-linear gate simulations can be quite versatile, as it is known that almost any non-linear gate can be combined with linear optics to generate arbitrary non-linear gates \cite{ZimborasO17} - for details, see \cite{SuppMat}.  

\begin{figure}[ht!]
\centering
\includegraphics[width = 0.49 \textwidth]{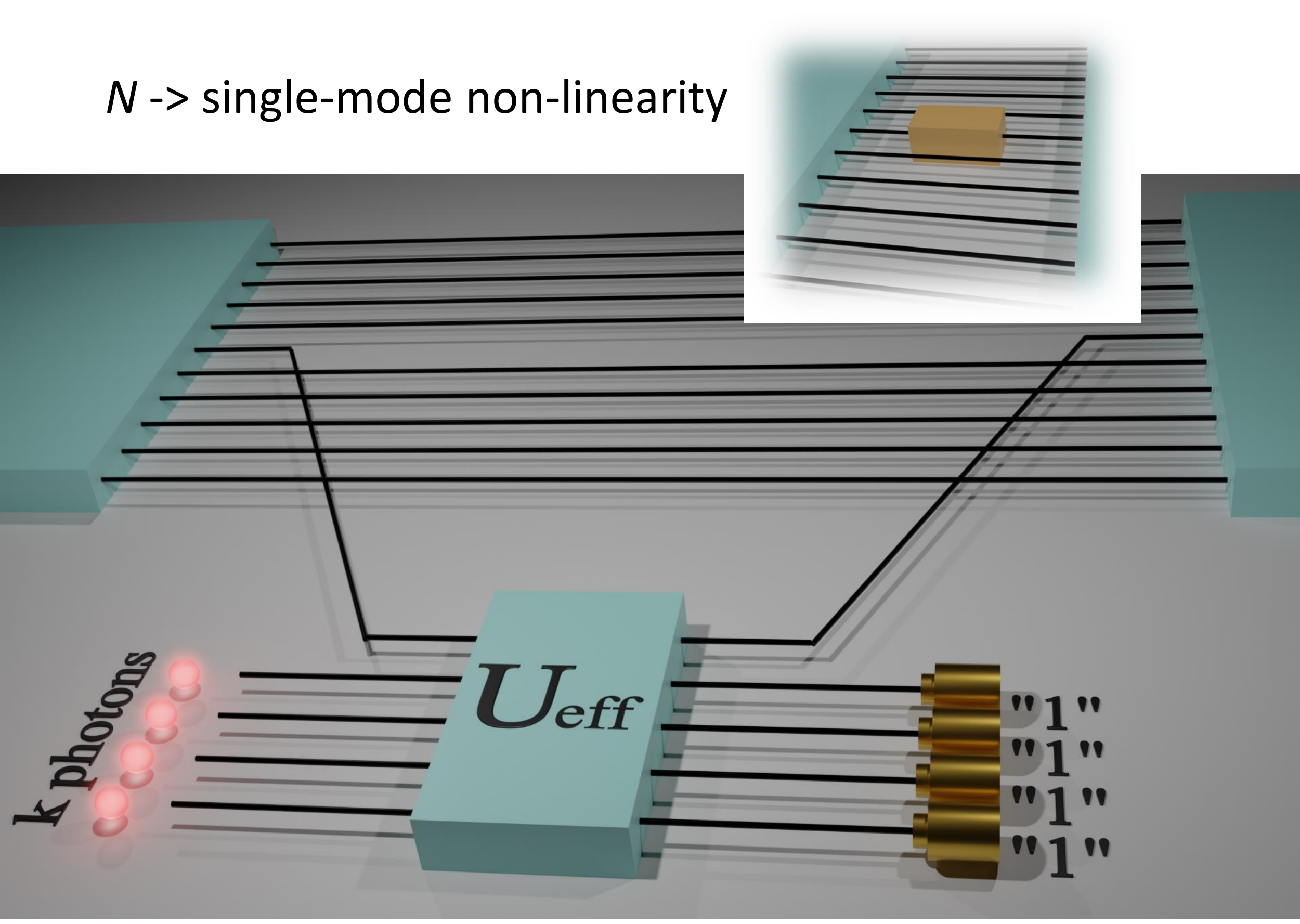}
\caption{Scheme for linear optics simulation of non-linear Boson Sampling with a single-mode non-linearity $N$ on mode $x$ between two Haar-random linear unitaries $W$ and $V$. Mode $x$ after linear transformation $W$ is injected in the first port of a $(k+1)$-mode linear optical gadget described by unitary $U_{\mathrm{eff}}$, whose other input ports are injected with $k$ single photon states ($k \leq n$). Detection of one photon in each of the $k$ auxiliary modes of the gadget heralds a successful simulation of the single-mode non-linearity in mode $x$ of the original interferometer (see inset).}
\label{fig:meas_ind_scheme}
\end{figure}

In Fig.\ \ref{fig:meas_ind_scheme} we describe the linear-optical, postselection-based gadget that can be used to simulate single-mode non-linear gates. We see that the $k$-mode linear optical gadget (with $k \leq n$) replaces the single-mode non-linear gate. In the gadget, mode $x$ and the $k$ single photons undergo the effective unitary $U_{\mathrm{eff}}$. This linear-optical simulation approximates the non-linear Boson Sampling evolution upon detection of $k$ photons at the auxiliary output modes.

Using the state-of-the-art weak classical simulation algorithm of Clifford and Clifford \cite{Clifford18}, we can simulate the enlarged ($n+k$, $m+k$) linear optical system, postselecting only those events where a single photon is measured in each of the auxiliary modes $(m+1, \ldots, m+k)$ (see \cite{SuppMat} for more details). This results in a classical simulation algorithm for the non-linear Boson Sampling experiment.

\begin{figure*}[ht!]
\centering
\includegraphics[width = 0.99 \textwidth]{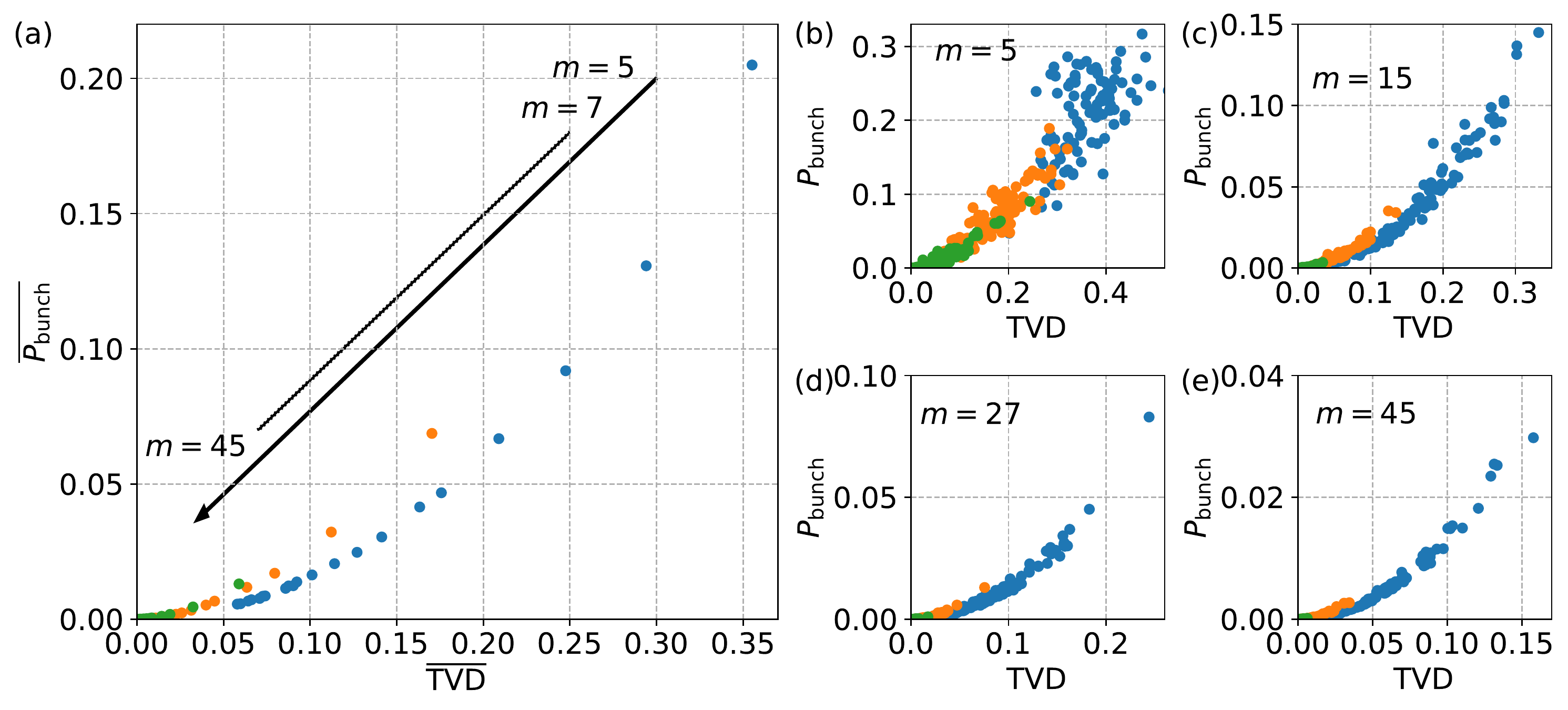}
\caption{Analysis of total variation distance (TVD) between the non-linear process and its linear-optical simulation, and bunching probability at the non-linear phase-shift site for $n=4$ photons. The TVD is calculated between the exact probability distribution [Eq. \eqref{eq:nonlinear_singlephase_simpl}], and the post-selected linear-optical simulation using $k<n$ auxiliary photons. In these plots, the exact distribution is calculated using the linear-optical simulation scheme with $n$ auxiliary photons. $P_{\mathrm{bunch}}$ is the probability of having more than $k$ photons at the non-linearity site. (a) Parametric plot showing a correlation between $\overline{\mathrm{TVD}}$ and $\overline{P_{\mathrm{bunch}}}$, averaged over $100$ different evolutions $W$ and $V$ for fixed $\phi = \pi/2$.  Each point corresponds to a different value of $m$, in the range $[5, 7, \ldots, 45]$. (b)-(e) Correlation between TVD and $P_{\mathrm{bunch}}$ for fixed number of modes $m$. Here, each point corresponds to different evolutions $W$ and $V$ for fixed $\phi = \pi/2$.
In all plots: blue points corresponds to $k=1$, orange points to $k=2$, and green points to $k=3$.}
\label{fig:meas_ind}
\end{figure*}

Let us now discuss some issues that arise when using this scheme to simulate non-linear Boson Sampling in either the asymptotic regime of large numbers of photons/modes, or in the finite setting.

\sectioning{Non-linearities in the asymptotic setting. --}
Assuming uniformly drawn, Haar-random interferometer unitaries, it has been shown that the appropriate scaling between the original number of modes $m$ and number of photons $n$ will result in asymptotic suppression of multi-photon collisions. More precisely: if $m=O(n^{j/(j-1)})$, then $(j+1)$-fold collisions are suppressed, when $n,m$ go to infinity \cite{Arkhipov12}. In particular, this will be true for the photon occupation numbers at the non-linear gates. So, by choosing $m=O(n^{k/(k+1)})$, at most $k$ photons will asymptotically be present at each non-linear gate, which means the linear-optical simulation (or classical simulation based on it) can be done with only $k$ auxiliary photons per non-linear gate. As we will soon show, such a simulation for small $k$, e.g. $k=2,3,4$ can be readily obtained. These simulations using $k=2$ are sufficient for an asymptotically perfect simulation for the usual Boson Sampling regime of $m=O(n^2)$. In other words, in this setting there is a precise correspondence between one single-mode non-linear phase gate and two extra auxiliary photons. More generally, the scaling of $m$ with $n$ dictates how many auxiliary photons are needed for asymptotically perfect simulation of a non-linear phase gate $N$.

\sectioning{Non-linearities in the finite setting. --}
The setting with finite $n,m$ is experimentally relevant, and in this case there will be no strict suppression of multi-photon collisions at the non-linear gates. Setting $k=n$ results in an exact classical simulation of non-linear Boson Sampling. When $k < n$ we will have only an approximate simulation. As an example, when $m \sim n^2$ numerical results suggest that a number of auxiliary photons equal to $k=2,3$ should provide a sufficiently accurate simulation given the large effective suppression of bunching at the output of Haar-random unitaries (see \cite{SuppMat}).

There are two main features that increase the simulation complexity. First, finding an effective unitary $U_{\mathrm{eff}}$ that uses $k$ photons for a linear-optical simulation seems to require the computation of permanents of $k \times k$ matrices \cite{Scheel03, SuppMat}, which results in a classical runtime that increases exponentially with $k$. The other cost incurred is the postselection overhead. From all the simulated events on the enlarged linear-optical set-up with $(n+k)$ photons, we only use events where the $k$ auxiliary photons were detected at the linear-optical simulation gadget. This will happen with a probability $\mathrm{Pr}_{\mathrm{succ}} = \vert \mathrm{per} (U_{\mathrm{eff}}^{0,1, \ldots, 1}) \vert^{2} = p$. There is some evidence that the maximum value of $p$ tends to decrease as $k$ increases \cite{Eisert2005}.

We have performed numerical simulations of non-linear Boson Sampling with a single non-linear phase in the finite setting, using the classical algorithm based on linear-optical simulation. The results are shown in Fig. \ref{fig:meas_ind} for ($n=3$, $m=5,9,16,27$) and ($n=4$, $m=16$). As expected, having $k=n$ results in exact sampling from the non-linear process, and in fact (once the appropriate gadget $U_{\mathrm{eff}}$ has been determined) is numerically found to be more computationally effective than directly using Eq. (\ref{eq:gen_nonlin}).  For fixed $n,k$, note that the simulation error decreases with increasing $m$, since bunching events become rarer. These results suggest that the crucial parameter for the simulation complexity is the scaling between $n$ and $m$ (see also \cite{SuppMat}). The regime when $m=O(n)$ is particularly interesting, as there is a trade-off between a faster classical simulation algorithm \cite{Clifford2020Faster}, and the increased complexity required to find the linear-optical gadget unitary $U_{\mathrm{eff}}$ for larger $k$.

If the number of auxiliary photons $k<n$, the simulation scheme based on linear-optical gadgets will be only approximate, due to non-linear dynamics of more than $k$ photons.  The key open point in this scenario is to quantify the simulation error incurred. In Fig. \ref{fig:meas_ind} we provide a numerical study on how the simulation error depends on $n, m, k$, as quantified by the total variation distance (TVD) between the exact non-linear evolution and its simulation using $k<n$ auxiliary photons. We observe a strong correlation between the TVD and the probability of bunching at the non-linearity. An open interesting research question is to obtain a quantitative description of this dependence between TVD and bunching, for instance by using bounds on bunching in the uniformly random, Haar ensemble of unitaries.

\sectioning{Discussion. --}
We have proposed the adoption of non-linear gates within the framework of Boson Sampling as a way to increase the computational complexity of the model. We have shown how to quantify this complexity using a linear-optical simulation with postselection, which itself can be simulated classically.  For large number $m$ of modes and $n$ of photons, suppressed bunching allows asymptotically perfect simulation at a cost of two extra photons per non-linear phase gate introduced, if we assume $m=O(n^2)$. For finite $m,n$ and single-mode non-linear phase gates, we identify the probability of bunching, governed by the scaling of $m$ as a function of $n$, as the key factor affecting the complexity of our proposed simulation scheme.

The non-linear Boson Sampling model we propose is inherently more expressive than linear Boson Sampling. In light of the recent developments regarding first application of Boson Sampling and its variants for hybrid quantum computational models, we expect that having access to increased functionalities enabled by non-linearities can be turned into useful advantage for tasks solvable with linear Boson Sampling, as well as propose altogether new tasks solvable by noisy, intermediate-scale quantum (NISQ) devices. In parallel, an important research direction regards development of more efficient simulation schemes for non-linear gates, which is directly relevant not only to the model we propose, but to general photonic quantum computation.

\begin{acknowledgments}
\sectioning{Acknowledgments. --} This work is supported by the ERC Advanced Grant QU-BOSS (QUantum advantage via non-linear BOSon Sampling, grant agreement no. 884676), by the European Union’s Horizon 2020 research and innovation program through the FET project PHOQUSING (“PHOtonic Quantum SamplING machine” - Grant Agreement No. 899544) and by FCT – Fundação para Ciência e Tecnologia, via project CEECINST/00062/2018. We acknowledge useful comments from Scott Aaronson.
\end{acknowledgments}


\begin{thebibliography}{72}%
\makeatletter
\providecommand \@ifxundefined [1]{%
 \@ifx{#1\undefined}
}%
\providecommand \@ifnum [1]{%
 \ifnum #1\expandafter \@firstoftwo
 \else \expandafter \@secondoftwo
 \fi
}%
\providecommand \@ifx [1]{%
 \ifx #1\expandafter \@firstoftwo
 \else \expandafter \@secondoftwo
 \fi
}%
\providecommand \natexlab [1]{#1}%
\providecommand \enquote  [1]{``#1''}%
\providecommand \bibnamefont  [1]{#1}%
\providecommand \bibfnamefont [1]{#1}%
\providecommand \citenamefont [1]{#1}%
\providecommand \href@noop [0]{\@secondoftwo}%
\providecommand \href [0]{\begingroup \@sanitize@url \@href}%
\providecommand \@href[1]{\@@startlink{#1}\@@href}%
\providecommand \@@href[1]{\endgroup#1\@@endlink}%
\providecommand \@sanitize@url [0]{\catcode `\\12\catcode `\$12\catcode
  `\&12\catcode `\#12\catcode `\^12\catcode `\_12\catcode `\%12\relax}%
\providecommand \@@startlink[1]{}%
\providecommand \@@endlink[0]{}%
\providecommand \url  [0]{\begingroup\@sanitize@url \@url }%
\providecommand \@url [1]{\endgroup\@href {#1}{\urlprefix }}%
\providecommand \urlprefix  [0]{URL }%
\providecommand \Eprint [0]{\href }%
\providecommand \doibase [0]{https://doi.org/}%
\providecommand \selectlanguage [0]{\@gobble}%
\providecommand \bibinfo  [0]{\@secondoftwo}%
\providecommand \bibfield  [0]{\@secondoftwo}%
\providecommand \translation [1]{[#1]}%
\providecommand \BibitemOpen [0]{}%
\providecommand \bibitemStop [0]{}%
\providecommand \bibitemNoStop [0]{.\EOS\space}%
\providecommand \EOS [0]{\spacefactor3000\relax}%
\providecommand \BibitemShut  [1]{\csname bibitem#1\endcsname}%
\let\auto@bib@innerbib\@empty
\bibitem [{\citenamefont {J.~Yin~and}\ \emph {et~al.}(2017)\citenamefont
  {J.~Yin~and}, \citenamefont {Li}, \citenamefont {Liao}, \citenamefont
  {L.~Zhang~and}, \citenamefont {Cai}, \citenamefont {Liu}, \citenamefont {Li},
  \citenamefont {Dai}, \citenamefont {Li}, \citenamefont {Lu}, \citenamefont
  {Gong}, \citenamefont {Xu}, \citenamefont {Li}, \citenamefont {Li},
  \citenamefont {Yin}, \citenamefont {Jiang}, \citenamefont {Li}, \citenamefont
  {Jia}, \citenamefont {Ren}, \citenamefont {He}, \citenamefont {Zhou},
  \citenamefont {Zhang}, \citenamefont {Wang}, \citenamefont {Chang},
  \citenamefont {Zhu}, \citenamefont {Liu}, \citenamefont {Chen}, \citenamefont
  {Lu}, \citenamefont {Shu}, \citenamefont {Peng}, \citenamefont {Wang},\ and\
  \citenamefont {Pan}}]{Yin17}%
  \BibitemOpen
  \bibfield  {author} {\bibinfo {author} {\bibfnamefont {Y.~C.}\ \bibnamefont
  {J.~Yin~and}}, \bibinfo {author} {\bibfnamefont {Y.-H.}\ \bibnamefont {Li}},
  \bibinfo {author} {\bibfnamefont {S.-K.}\ \bibnamefont {Liao}}, \bibinfo
  {author} {\bibfnamefont {J.-G.~R.}\ \bibnamefont {L.~Zhang~and}}, \bibinfo
  {author} {\bibfnamefont {W.-Q.}\ \bibnamefont {Cai}}, \bibinfo {author}
  {\bibfnamefont {W.-Y.}\ \bibnamefont {Liu}}, \bibinfo {author} {\bibfnamefont
  {B.}~\bibnamefont {Li}}, \bibinfo {author} {\bibfnamefont {H.}~\bibnamefont
  {Dai}}, \bibinfo {author} {\bibfnamefont {G.-B.}\ \bibnamefont {Li}},
  \bibinfo {author} {\bibfnamefont {Q.-M.}\ \bibnamefont {Lu}}, \bibinfo
  {author} {\bibfnamefont {Y.-H.}\ \bibnamefont {Gong}}, \bibinfo {author}
  {\bibfnamefont {Y.}~\bibnamefont {Xu}}, \bibinfo {author} {\bibfnamefont
  {S.-L.}\ \bibnamefont {Li}}, \bibinfo {author} {\bibfnamefont {F.-Z.}\
  \bibnamefont {Li}}, \bibinfo {author} {\bibfnamefont {Y.-Y.}\ \bibnamefont
  {Yin}}, \bibinfo {author} {\bibfnamefont {Z.-Q.}\ \bibnamefont {Jiang}},
  \bibinfo {author} {\bibfnamefont {M.}~\bibnamefont {Li}}, \bibinfo {author}
  {\bibfnamefont {J.-J.}\ \bibnamefont {Jia}}, \bibinfo {author} {\bibfnamefont
  {G.}~\bibnamefont {Ren}}, \bibinfo {author} {\bibfnamefont {D.}~\bibnamefont
  {He}}, \bibinfo {author} {\bibfnamefont {Y.-L.}\ \bibnamefont {Zhou}},
  \bibinfo {author} {\bibfnamefont {X.-X.}\ \bibnamefont {Zhang}}, \bibinfo
  {author} {\bibfnamefont {N.}~\bibnamefont {Wang}}, \bibinfo {author}
  {\bibfnamefont {X.}~\bibnamefont {Chang}}, \bibinfo {author} {\bibfnamefont
  {Z.-C.}\ \bibnamefont {Zhu}}, \bibinfo {author} {\bibfnamefont {N.-L.}\
  \bibnamefont {Liu}}, \bibinfo {author} {\bibfnamefont {Y.-A.}\ \bibnamefont
  {Chen}}, \bibinfo {author} {\bibfnamefont {C.-Y.}\ \bibnamefont {Lu}},
  \bibinfo {author} {\bibfnamefont {R.}~\bibnamefont {Shu}}, \bibinfo {author}
  {\bibfnamefont {C.-Z.}\ \bibnamefont {Peng}}, \bibinfo {author}
  {\bibfnamefont {J.-Y.}\ \bibnamefont {Wang}},\ and\ \bibinfo {author}
  {\bibfnamefont {J.-W.}\ \bibnamefont {Pan}},\ }\href
  {https://doi.org/10.1126/science.aan3211} {\bibfield  {journal} {\bibinfo
  {journal} {Science}\ }\textbf {\bibinfo {volume} {356}},\ \bibinfo {pages}
  {1140} (\bibinfo {year} {2017})}\BibitemShut {NoStop}%
\bibitem [{\citenamefont {Giovannetti}\ \emph {et~al.}(2011)\citenamefont
  {Giovannetti}, \citenamefont {Lloyd},\ and\ \citenamefont
  {Maccone}}]{Giovannetti11}%
  \BibitemOpen
  \bibfield  {author} {\bibinfo {author} {\bibfnamefont {V.}~\bibnamefont
  {Giovannetti}}, \bibinfo {author} {\bibfnamefont {S.}~\bibnamefont {Lloyd}},\
  and\ \bibinfo {author} {\bibfnamefont {L.}~\bibnamefont {Maccone}},\ }\href
  {https://doi.org/10.1038/nphoton.2011.35} {\bibfield  {journal} {\bibinfo
  {journal} {Nat. Photon.}\ }\textbf {\bibinfo {volume} {5}},\ \bibinfo {pages}
  {222} (\bibinfo {year} {2011})}\BibitemShut {NoStop}%
\bibitem [{\citenamefont {Nielsen}\ and\ \citenamefont
  {Chuang}(2010)}]{NielsenBook}%
  \BibitemOpen
  \bibfield  {author} {\bibinfo {author} {\bibfnamefont {M.}~\bibnamefont
  {Nielsen}}\ and\ \bibinfo {author} {\bibfnamefont {I.}~\bibnamefont
  {Chuang}},\ }\href@noop {} {\emph {\bibinfo {title} {Quantum Computation and
  Quantum Information}}}\ (\bibinfo  {publisher} {Cambridge University Press},\
  \bibinfo {year} {2010})\BibitemShut {NoStop}%
\bibitem [{\citenamefont {Flamini}\ \emph {et~al.}(2019)\citenamefont
  {Flamini}, \citenamefont {Spagnolo},\ and\ \citenamefont
  {Sciarrino}}]{Flamini18}%
  \BibitemOpen
  \bibfield  {author} {\bibinfo {author} {\bibfnamefont {F.}~\bibnamefont
  {Flamini}}, \bibinfo {author} {\bibfnamefont {N.}~\bibnamefont {Spagnolo}},\
  and\ \bibinfo {author} {\bibfnamefont {F.}~\bibnamefont {Sciarrino}},\ }\href
  {https://doi.org/10.1088/1361-6633/aad5b2} {\bibfield  {journal} {\bibinfo
  {journal} {Rep. Prog. Phys.}\ }\textbf {\bibinfo {volume} {82}},\ \bibinfo
  {pages} {016001} (\bibinfo {year} {2019})}\BibitemShut {NoStop}%
\bibitem [{\citenamefont {Arute}\ \emph {et~al.}(2019)\citenamefont {Arute},
  \citenamefont {Arya}, \citenamefont {Babbush}, \citenamefont {Bacon},
  \citenamefont {Bardin}, \citenamefont {Barends}, \citenamefont {Biswas},
  \citenamefont {Boixo}, \citenamefont {Brandao}, \citenamefont {Buell},
  \citenamefont {Burkett}, \citenamefont {Chen}, \citenamefont {Chen},
  \citenamefont {Chiaro}, \citenamefont {Collins}, \citenamefont {Courtney},
  \citenamefont {Dunsworth}, \citenamefont {Farhi}, \citenamefont {Foxen},
  \citenamefont {Fowler}, \citenamefont {Gidney}, \citenamefont {Giustina},
  \citenamefont {Graff}, \citenamefont {Guerin}, \citenamefont {Habegger},
  \citenamefont {Harrigan}, \citenamefont {Hartmann}, \citenamefont {Ho},
  \citenamefont {Hoffmann}, \citenamefont {Huang}, \citenamefont {Humble},
  \citenamefont {Isakov}, \citenamefont {Jeffrey}, \citenamefont {Jiang},
  \citenamefont {Kafri}, \citenamefont {Kechedzhi}, \citenamefont {Kelly},
  \citenamefont {Klimov}, \citenamefont {Knysh}, \citenamefont {Korotkov},
  \citenamefont {Kostritsa}, \citenamefont {Landhuis}, \citenamefont
  {Lindmark}, \citenamefont {Lucero}, \citenamefont {Lyakh}, \citenamefont
  {Mandr\'a}, \citenamefont {McClean}, \citenamefont {McEwen}, \citenamefont
  {Megrant}, \citenamefont {Mi}, \citenamefont {Michielsen}, \citenamefont
  {Mohseni}, \citenamefont {Mutus}, \citenamefont {Naaman}, \citenamefont
  {Neeley}, \citenamefont {Neill}, \citenamefont {Niu}, \citenamefont {Ostby},
  \citenamefont {Petukhov}, \citenamefont {Platt}, \citenamefont {Quintana},
  \citenamefont {Roushan}, \citenamefont {Rubin}, \citenamefont {Sank},
  \citenamefont {Satzinger}, \citenamefont {Smelyanskiy}, \citenamefont {Sung},
  \citenamefont {Trevithick}, \citenamefont {Vainsencher}, \citenamefont
  {Villalonga}, \citenamefont {White}, \citenamefont {Yao}, \citenamefont
  {Yeh}, \citenamefont {Zalcman}, \citenamefont {Neven},\ and\ \citenamefont
  {Martinis}}]{Arute19Supremacy}%
  \BibitemOpen
  \bibfield  {author} {\bibinfo {author} {\bibfnamefont {F.}~\bibnamefont
  {Arute}}, \bibinfo {author} {\bibfnamefont {K.}~\bibnamefont {Arya}},
  \bibinfo {author} {\bibfnamefont {R.}~\bibnamefont {Babbush}}, \bibinfo
  {author} {\bibfnamefont {D.}~\bibnamefont {Bacon}}, \bibinfo {author}
  {\bibfnamefont {J.~C.}\ \bibnamefont {Bardin}}, \bibinfo {author}
  {\bibfnamefont {R.}~\bibnamefont {Barends}}, \bibinfo {author} {\bibfnamefont
  {R.}~\bibnamefont {Biswas}}, \bibinfo {author} {\bibfnamefont
  {S.}~\bibnamefont {Boixo}}, \bibinfo {author} {\bibfnamefont {F.~G. S.~L.}\
  \bibnamefont {Brandao}}, \bibinfo {author} {\bibfnamefont {D.~A.}\
  \bibnamefont {Buell}}, \bibinfo {author} {\bibfnamefont {B.}~\bibnamefont
  {Burkett}}, \bibinfo {author} {\bibfnamefont {Y.}~\bibnamefont {Chen}},
  \bibinfo {author} {\bibfnamefont {Z.}~\bibnamefont {Chen}}, \bibinfo {author}
  {\bibfnamefont {B.}~\bibnamefont {Chiaro}}, \bibinfo {author} {\bibfnamefont
  {R.}~\bibnamefont {Collins}}, \bibinfo {author} {\bibfnamefont
  {W.}~\bibnamefont {Courtney}}, \bibinfo {author} {\bibfnamefont
  {A.}~\bibnamefont {Dunsworth}}, \bibinfo {author} {\bibfnamefont
  {E.}~\bibnamefont {Farhi}}, \bibinfo {author} {\bibfnamefont
  {B.}~\bibnamefont {Foxen}}, \bibinfo {author} {\bibfnamefont
  {A.}~\bibnamefont {Fowler}}, \bibinfo {author} {\bibfnamefont
  {C.}~\bibnamefont {Gidney}}, \bibinfo {author} {\bibfnamefont
  {M.}~\bibnamefont {Giustina}}, \bibinfo {author} {\bibfnamefont
  {R.}~\bibnamefont {Graff}}, \bibinfo {author} {\bibfnamefont
  {K.}~\bibnamefont {Guerin}}, \bibinfo {author} {\bibfnamefont
  {S.}~\bibnamefont {Habegger}}, \bibinfo {author} {\bibfnamefont {M.~P.}\
  \bibnamefont {Harrigan}}, \bibinfo {author} {\bibfnamefont {M.~J.}\
  \bibnamefont {Hartmann}}, \bibinfo {author} {\bibfnamefont {A.}~\bibnamefont
  {Ho}}, \bibinfo {author} {\bibfnamefont {M.}~\bibnamefont {Hoffmann}},
  \bibinfo {author} {\bibfnamefont {T.}~\bibnamefont {Huang}}, \bibinfo
  {author} {\bibfnamefont {T.~S.}\ \bibnamefont {Humble}}, \bibinfo {author}
  {\bibfnamefont {S.~V.}\ \bibnamefont {Isakov}}, \bibinfo {author}
  {\bibfnamefont {E.}~\bibnamefont {Jeffrey}}, \bibinfo {author} {\bibfnamefont
  {Z.}~\bibnamefont {Jiang}}, \bibinfo {author} {\bibfnamefont
  {D.}~\bibnamefont {Kafri}}, \bibinfo {author} {\bibfnamefont
  {K.}~\bibnamefont {Kechedzhi}}, \bibinfo {author} {\bibfnamefont
  {J.}~\bibnamefont {Kelly}}, \bibinfo {author} {\bibfnamefont {P.~V.}\
  \bibnamefont {Klimov}}, \bibinfo {author} {\bibfnamefont {S.}~\bibnamefont
  {Knysh}}, \bibinfo {author} {\bibfnamefont {A.}~\bibnamefont {Korotkov}},
  \bibinfo {author} {\bibfnamefont {F.}~\bibnamefont {Kostritsa}}, \bibinfo
  {author} {\bibfnamefont {D.}~\bibnamefont {Landhuis}}, \bibinfo {author}
  {\bibfnamefont {M.}~\bibnamefont {Lindmark}}, \bibinfo {author}
  {\bibfnamefont {E.}~\bibnamefont {Lucero}}, \bibinfo {author} {\bibfnamefont
  {D.}~\bibnamefont {Lyakh}}, \bibinfo {author} {\bibfnamefont
  {S.}~\bibnamefont {Mandr\'a}}, \bibinfo {author} {\bibfnamefont {J.~R.}\
  \bibnamefont {McClean}}, \bibinfo {author} {\bibfnamefont {M.}~\bibnamefont
  {McEwen}}, \bibinfo {author} {\bibfnamefont {A.}~\bibnamefont {Megrant}},
  \bibinfo {author} {\bibfnamefont {X.}~\bibnamefont {Mi}}, \bibinfo {author}
  {\bibfnamefont {K.}~\bibnamefont {Michielsen}}, \bibinfo {author}
  {\bibfnamefont {M.}~\bibnamefont {Mohseni}}, \bibinfo {author} {\bibfnamefont
  {J.}~\bibnamefont {Mutus}}, \bibinfo {author} {\bibfnamefont
  {O.}~\bibnamefont {Naaman}}, \bibinfo {author} {\bibfnamefont
  {M.}~\bibnamefont {Neeley}}, \bibinfo {author} {\bibfnamefont
  {C.}~\bibnamefont {Neill}}, \bibinfo {author} {\bibfnamefont {M.~Y.}\
  \bibnamefont {Niu}}, \bibinfo {author} {\bibfnamefont {E.}~\bibnamefont
  {Ostby}}, \bibinfo {author} {\bibfnamefont {A.}~\bibnamefont {Petukhov}},
  \bibinfo {author} {\bibfnamefont {J.~C.}\ \bibnamefont {Platt}}, \bibinfo
  {author} {\bibfnamefont {C.}~\bibnamefont {Quintana}}, \bibinfo {author}
  {\bibfnamefont {E.~G. R.~P.}\ \bibnamefont {Roushan}}, \bibinfo {author}
  {\bibfnamefont {N.~C.}\ \bibnamefont {Rubin}}, \bibinfo {author}
  {\bibfnamefont {D.}~\bibnamefont {Sank}}, \bibinfo {author} {\bibfnamefont
  {K.~J.}\ \bibnamefont {Satzinger}}, \bibinfo {author} {\bibfnamefont
  {V.}~\bibnamefont {Smelyanskiy}}, \bibinfo {author} {\bibfnamefont {K.~J.}\
  \bibnamefont {Sung}}, \bibinfo {author} {\bibfnamefont {M.~D.}\ \bibnamefont
  {Trevithick}}, \bibinfo {author} {\bibfnamefont {A.}~\bibnamefont
  {Vainsencher}}, \bibinfo {author} {\bibfnamefont {B.}~\bibnamefont
  {Villalonga}}, \bibinfo {author} {\bibfnamefont {T.}~\bibnamefont {White}},
  \bibinfo {author} {\bibfnamefont {Z.~J.}\ \bibnamefont {Yao}}, \bibinfo
  {author} {\bibfnamefont {P.}~\bibnamefont {Yeh}}, \bibinfo {author}
  {\bibfnamefont {A.}~\bibnamefont {Zalcman}}, \bibinfo {author} {\bibfnamefont
  {H.}~\bibnamefont {Neven}},\ and\ \bibinfo {author} {\bibfnamefont {J.~M.}\
  \bibnamefont {Martinis}},\ }\href {https://doi.org/10.1038/s41586-019-1666-5}
  {\bibfield  {journal} {\bibinfo  {journal} {Nature}\ }\textbf {\bibinfo
  {volume} {574}},\ \bibinfo {pages} {505} (\bibinfo {year}
  {2019})}\BibitemShut {NoStop}%
\bibitem [{\citenamefont {Zhong}\ \emph {et~al.}(2020)\citenamefont {Zhong},
  \citenamefont {Wang}, \citenamefont {Deng}, \citenamefont {Chen},
  \citenamefont {Peng}, \citenamefont {Luo}, \citenamefont {Qin}, \citenamefont
  {Wu}, \citenamefont {Ding}, \citenamefont {Hu}, \citenamefont {Hu},
  \citenamefont {Yang}, \citenamefont {Zhang}, \citenamefont {Li},
  \citenamefont {Yuxuan~Li}, \citenamefont {Gan}, \citenamefont {Yang},
  \citenamefont {You}, \citenamefont {Wang}, \citenamefont {Li}, \citenamefont
  {Liu}, \citenamefont {Lu},\ and\ \citenamefont {Pan}}]{Zhong20}%
  \BibitemOpen
  \bibfield  {author} {\bibinfo {author} {\bibfnamefont {H.-S.}\ \bibnamefont
  {Zhong}}, \bibinfo {author} {\bibfnamefont {H.}~\bibnamefont {Wang}},
  \bibinfo {author} {\bibfnamefont {Y.-H.}\ \bibnamefont {Deng}}, \bibinfo
  {author} {\bibfnamefont {M.-C.}\ \bibnamefont {Chen}}, \bibinfo {author}
  {\bibfnamefont {L.-C.}\ \bibnamefont {Peng}}, \bibinfo {author}
  {\bibfnamefont {Y.-H.}\ \bibnamefont {Luo}}, \bibinfo {author} {\bibfnamefont
  {J.}~\bibnamefont {Qin}}, \bibinfo {author} {\bibfnamefont {D.}~\bibnamefont
  {Wu}}, \bibinfo {author} {\bibfnamefont {X.}~\bibnamefont {Ding}}, \bibinfo
  {author} {\bibfnamefont {Y.}~\bibnamefont {Hu}}, \bibinfo {author}
  {\bibfnamefont {P.}~\bibnamefont {Hu}}, \bibinfo {author} {\bibfnamefont
  {X.-Y.}\ \bibnamefont {Yang}}, \bibinfo {author} {\bibfnamefont {W.-J.}\
  \bibnamefont {Zhang}}, \bibinfo {author} {\bibfnamefont {H.}~\bibnamefont
  {Li}}, \bibinfo {author} {\bibfnamefont {X.~J.}\ \bibnamefont {Yuxuan~Li}},
  \bibinfo {author} {\bibfnamefont {L.}~\bibnamefont {Gan}}, \bibinfo {author}
  {\bibfnamefont {G.}~\bibnamefont {Yang}}, \bibinfo {author} {\bibfnamefont
  {L.}~\bibnamefont {You}}, \bibinfo {author} {\bibfnamefont {Z.}~\bibnamefont
  {Wang}}, \bibinfo {author} {\bibfnamefont {L.}~\bibnamefont {Li}}, \bibinfo
  {author} {\bibfnamefont {N.-L.}\ \bibnamefont {Liu}}, \bibinfo {author}
  {\bibfnamefont {C.-Y.}\ \bibnamefont {Lu}},\ and\ \bibinfo {author}
  {\bibfnamefont {J.-W.}\ \bibnamefont {Pan}},\ }\href
  {https://doi.org/10.1126/science.abe8770} {\bibfield  {journal} {\bibinfo
  {journal} {Science}\ }\textbf {\bibinfo {volume} {370}},\ \bibinfo {pages}
  {1460} (\bibinfo {year} {2020})}\BibitemShut {NoStop}%
\bibitem [{\citenamefont {Aaronson}\ and\ \citenamefont {Arkhipov}(2011)}]{AA}%
  \BibitemOpen
  \bibfield  {author} {\bibinfo {author} {\bibfnamefont {S.}~\bibnamefont
  {Aaronson}}\ and\ \bibinfo {author} {\bibfnamefont {A.}~\bibnamefont
  {Arkhipov}},\ }in\ \href {https://doi.org/10.1145/1993636.1993682} {\emph
  {\bibinfo {booktitle} {Proceedings of the 43rd annual ACM symposium on Theory
  of Computing}}},\ \bibinfo {editor} {edited by\ \bibinfo {editor}
  {\bibfnamefont {A.}~\bibnamefont {Press}}}\ (\bibinfo {year} {2011})\ pp.\
  \bibinfo {pages} {333--342}\BibitemShut {NoStop}%
\bibitem [{\citenamefont {Boixo}\ \emph {et~al.}(2018)\citenamefont {Boixo},
  \citenamefont {Isakov}, \citenamefont {Smelyanskiy}, \citenamefont {Babbush},
  \citenamefont {Ding}, \citenamefont {Jiang}, \citenamefont {Bremner},
  \citenamefont {Martinis},\ and\ \citenamefont {Neven}}]{Boixo18}%
  \BibitemOpen
  \bibfield  {author} {\bibinfo {author} {\bibfnamefont {S.}~\bibnamefont
  {Boixo}}, \bibinfo {author} {\bibfnamefont {S.~V.}\ \bibnamefont {Isakov}},
  \bibinfo {author} {\bibfnamefont {V.~N.}\ \bibnamefont {Smelyanskiy}},
  \bibinfo {author} {\bibfnamefont {R.}~\bibnamefont {Babbush}}, \bibinfo
  {author} {\bibfnamefont {N.}~\bibnamefont {Ding}}, \bibinfo {author}
  {\bibfnamefont {Z.}~\bibnamefont {Jiang}}, \bibinfo {author} {\bibfnamefont
  {M.~J.}\ \bibnamefont {Bremner}}, \bibinfo {author} {\bibfnamefont {J.~M.}\
  \bibnamefont {Martinis}},\ and\ \bibinfo {author} {\bibfnamefont
  {H.}~\bibnamefont {Neven}},\ }\href
  {https://doi.org/10.1038/s41567-018-0124-x} {\bibfield  {journal} {\bibinfo
  {journal} {Nat. Phys.}\ }\textbf {\bibinfo {volume} {14}},\ \bibinfo {pages}
  {595} (\bibinfo {year} {2018})}\BibitemShut {NoStop}%
\bibitem [{\citenamefont {Harrow}\ and\ \citenamefont
  {Montanaro}(2017)}]{Harrow17}%
  \BibitemOpen
  \bibfield  {author} {\bibinfo {author} {\bibfnamefont {A.~W.}\ \bibnamefont
  {Harrow}}\ and\ \bibinfo {author} {\bibfnamefont {A.}~\bibnamefont
  {Montanaro}},\ }\href {https://doi.org/10.1038/nature23458} {\bibfield
  {journal} {\bibinfo  {journal} {Nature}\ }\textbf {\bibinfo {volume} {549}},\
  \bibinfo {pages} {203} (\bibinfo {year} {2017})}\BibitemShut {NoStop}%
\bibitem [{\citenamefont {Bernien}\ \emph {et~al.}(2017)\citenamefont
  {Bernien}, \citenamefont {Schwartz}, \citenamefont {Keesling}, \citenamefont
  {Levine}, \citenamefont {Omran}, \citenamefont {Pichler}, \citenamefont
  {Choi}, \citenamefont {Zibrov}, \citenamefont {Endres}, \citenamefont
  {Greiner}, \citenamefont {Vuletic},\ and\ \citenamefont {Lukin}}]{Bernien17}%
  \BibitemOpen
  \bibfield  {author} {\bibinfo {author} {\bibfnamefont {H.}~\bibnamefont
  {Bernien}}, \bibinfo {author} {\bibfnamefont {S.}~\bibnamefont {Schwartz}},
  \bibinfo {author} {\bibfnamefont {A.}~\bibnamefont {Keesling}}, \bibinfo
  {author} {\bibfnamefont {H.}~\bibnamefont {Levine}}, \bibinfo {author}
  {\bibfnamefont {A.}~\bibnamefont {Omran}}, \bibinfo {author} {\bibfnamefont
  {H.}~\bibnamefont {Pichler}}, \bibinfo {author} {\bibfnamefont
  {S.}~\bibnamefont {Choi}}, \bibinfo {author} {\bibfnamefont {A.~S.}\
  \bibnamefont {Zibrov}}, \bibinfo {author} {\bibfnamefont {M.}~\bibnamefont
  {Endres}}, \bibinfo {author} {\bibfnamefont {M.}~\bibnamefont {Greiner}},
  \bibinfo {author} {\bibfnamefont {V.}~\bibnamefont {Vuletic}},\ and\ \bibinfo
  {author} {\bibfnamefont {M.~D.}\ \bibnamefont {Lukin}},\ }\href
  {https://doi.org/10.1038/nature24622} {\bibfield  {journal} {\bibinfo
  {journal} {Nature}\ }\textbf {\bibinfo {volume} {551}},\ \bibinfo {pages}
  {579} (\bibinfo {year} {2017})}\BibitemShut {NoStop}%
\bibitem [{\citenamefont {C.~Neill~and}\ \emph {et~al.}(2018)\citenamefont
  {C.~Neill~and}, \citenamefont {Boixo}, \citenamefont {Isakov}, \citenamefont
  {Smelyanskiy}, \citenamefont {Megrant}, \citenamefont {Chiaro}, \citenamefont
  {Dunsworth}, \citenamefont {Arya}, \citenamefont {Barends}, \citenamefont
  {Burkett}, \citenamefont {Chen}, \citenamefont {Chen}, \citenamefont
  {Fowler}, \citenamefont {Foxen}, \citenamefont {Giustina}, \citenamefont
  {Graff}, \citenamefont {Jeffrey}, \citenamefont {Huang}, \citenamefont
  {Kelly}, \citenamefont {Klimov}, \citenamefont {Lucero}, \citenamefont
  {Mutus}, \citenamefont {Neeley}, \citenamefont {Quintana}, \citenamefont
  {Sank}, \citenamefont {Vainsencher}, \citenamefont {Wenner}, \citenamefont
  {White}, \citenamefont {Neven},\ and\ \citenamefont {Martinis}}]{Neill18}%
  \BibitemOpen
  \bibfield  {author} {\bibinfo {author} {\bibfnamefont {K.~K.}\ \bibnamefont
  {C.~Neill~and}, \bibfnamefont {P.~Roushan~and}}, \bibinfo {author}
  {\bibfnamefont {S.}~\bibnamefont {Boixo}}, \bibinfo {author} {\bibfnamefont
  {S.~V.}\ \bibnamefont {Isakov}}, \bibinfo {author} {\bibfnamefont
  {V.}~\bibnamefont {Smelyanskiy}}, \bibinfo {author} {\bibfnamefont
  {A.}~\bibnamefont {Megrant}}, \bibinfo {author} {\bibfnamefont
  {B.}~\bibnamefont {Chiaro}}, \bibinfo {author} {\bibfnamefont
  {A.}~\bibnamefont {Dunsworth}}, \bibinfo {author} {\bibfnamefont
  {K.}~\bibnamefont {Arya}}, \bibinfo {author} {\bibfnamefont {R.}~\bibnamefont
  {Barends}}, \bibinfo {author} {\bibfnamefont {B.}~\bibnamefont {Burkett}},
  \bibinfo {author} {\bibfnamefont {Y.}~\bibnamefont {Chen}}, \bibinfo {author}
  {\bibfnamefont {Z.}~\bibnamefont {Chen}}, \bibinfo {author} {\bibfnamefont
  {A.}~\bibnamefont {Fowler}}, \bibinfo {author} {\bibfnamefont
  {B.}~\bibnamefont {Foxen}}, \bibinfo {author} {\bibfnamefont
  {M.}~\bibnamefont {Giustina}}, \bibinfo {author} {\bibfnamefont
  {R.}~\bibnamefont {Graff}}, \bibinfo {author} {\bibfnamefont
  {E.}~\bibnamefont {Jeffrey}}, \bibinfo {author} {\bibfnamefont
  {T.}~\bibnamefont {Huang}}, \bibinfo {author} {\bibfnamefont
  {J.}~\bibnamefont {Kelly}}, \bibinfo {author} {\bibfnamefont
  {P.}~\bibnamefont {Klimov}}, \bibinfo {author} {\bibfnamefont
  {E.}~\bibnamefont {Lucero}}, \bibinfo {author} {\bibfnamefont
  {J.}~\bibnamefont {Mutus}}, \bibinfo {author} {\bibfnamefont
  {M.}~\bibnamefont {Neeley}}, \bibinfo {author} {\bibfnamefont
  {C.}~\bibnamefont {Quintana}}, \bibinfo {author} {\bibfnamefont
  {D.}~\bibnamefont {Sank}}, \bibinfo {author} {\bibfnamefont {A.}~\bibnamefont
  {Vainsencher}}, \bibinfo {author} {\bibfnamefont {J.}~\bibnamefont {Wenner}},
  \bibinfo {author} {\bibfnamefont {T.~C.}\ \bibnamefont {White}}, \bibinfo
  {author} {\bibfnamefont {H.}~\bibnamefont {Neven}},\ and\ \bibinfo {author}
  {\bibfnamefont {J.~M.}\ \bibnamefont {Martinis}},\ }\href
  {https://doi.org/10.1126/science.aao4309} {\bibfield  {journal} {\bibinfo
  {journal} {Science}\ }\textbf {\bibinfo {volume} {360}},\ \bibinfo {pages}
  {195} (\bibinfo {year} {2018})}\BibitemShut {NoStop}%
\bibitem [{\citenamefont {Broome}\ \emph {et~al.}(2013)\citenamefont {Broome},
  \citenamefont {Fedrizzi}, \citenamefont {Rahimi-Keshari}, \citenamefont
  {Dove}, \citenamefont {Aaronson}, \citenamefont {Ralph},\ and\ \citenamefont
  {White}}]{Broome13}%
  \BibitemOpen
  \bibfield  {author} {\bibinfo {author} {\bibfnamefont {M.~A.}\ \bibnamefont
  {Broome}}, \bibinfo {author} {\bibfnamefont {A.}~\bibnamefont {Fedrizzi}},
  \bibinfo {author} {\bibfnamefont {S.}~\bibnamefont {Rahimi-Keshari}},
  \bibinfo {author} {\bibfnamefont {J.}~\bibnamefont {Dove}}, \bibinfo {author}
  {\bibfnamefont {S.}~\bibnamefont {Aaronson}}, \bibinfo {author}
  {\bibfnamefont {T.~C.}\ \bibnamefont {Ralph}},\ and\ \bibinfo {author}
  {\bibfnamefont {A.~G.}\ \bibnamefont {White}},\ }\href
  {https://doi.org/10.1126/science.1231440} {\bibfield  {journal} {\bibinfo
  {journal} {Science}\ }\textbf {\bibinfo {volume} {339}},\ \bibinfo {pages}
  {794} (\bibinfo {year} {2013})}\BibitemShut {NoStop}%
\bibitem [{\citenamefont {Spring}\ \emph {et~al.}(2013)\citenamefont {Spring},
  \citenamefont {Metcalf}, \citenamefont {Humphreys}, \citenamefont
  {Kolthammer}, \citenamefont {Jin}, \citenamefont {Barbieri}, \citenamefont
  {Datta}, \citenamefont {Thomas-Peter}, \citenamefont {Langford},
  \citenamefont {Kundys}, \citenamefont {Gates}, \citenamefont {Smith},\ and\
  \citenamefont {Walmsley}}]{Spring13}%
  \BibitemOpen
  \bibfield  {author} {\bibinfo {author} {\bibfnamefont {J.~B.}\ \bibnamefont
  {Spring}}, \bibinfo {author} {\bibfnamefont {B.~J.}\ \bibnamefont {Metcalf}},
  \bibinfo {author} {\bibfnamefont {P.}~\bibnamefont {Humphreys}}, \bibinfo
  {author} {\bibfnamefont {W.~S.}\ \bibnamefont {Kolthammer}}, \bibinfo
  {author} {\bibfnamefont {X.~M.}\ \bibnamefont {Jin}}, \bibinfo {author}
  {\bibfnamefont {M.}~\bibnamefont {Barbieri}}, \bibinfo {author}
  {\bibfnamefont {A.}~\bibnamefont {Datta}}, \bibinfo {author} {\bibfnamefont
  {N.}~\bibnamefont {Thomas-Peter}}, \bibinfo {author} {\bibfnamefont {N.~K.}\
  \bibnamefont {Langford}}, \bibinfo {author} {\bibfnamefont {D.}~\bibnamefont
  {Kundys}}, \bibinfo {author} {\bibfnamefont {J.~C.}\ \bibnamefont {Gates}},
  \bibinfo {author} {\bibfnamefont {B.~J.}\ \bibnamefont {Smith}},\ and\
  \bibinfo {author} {\bibfnamefont {I.~A.}\ \bibnamefont {Walmsley}},\ }\href
  {https://doi.org/10.1126/science.1231692} {\bibfield  {journal} {\bibinfo
  {journal} {Science}\ }\textbf {\bibinfo {volume} {339}},\ \bibinfo {pages}
  {798} (\bibinfo {year} {2013})}\BibitemShut {NoStop}%
\bibitem [{\citenamefont {Tillmann}\ \emph {et~al.}(2013)\citenamefont
  {Tillmann}, \citenamefont {Dakic}, \citenamefont {Heilmann}, \citenamefont
  {Nolte}, \citenamefont {Szameit},\ and\ \citenamefont
  {Walther}}]{Tillmann13}%
  \BibitemOpen
  \bibfield  {author} {\bibinfo {author} {\bibfnamefont {M.}~\bibnamefont
  {Tillmann}}, \bibinfo {author} {\bibfnamefont {B.}~\bibnamefont {Dakic}},
  \bibinfo {author} {\bibfnamefont {R.}~\bibnamefont {Heilmann}}, \bibinfo
  {author} {\bibfnamefont {S.}~\bibnamefont {Nolte}}, \bibinfo {author}
  {\bibfnamefont {A.}~\bibnamefont {Szameit}},\ and\ \bibinfo {author}
  {\bibfnamefont {P.}~\bibnamefont {Walther}},\ }\href
  {https://doi.org/10.1038/nphoton.2013.102} {\bibfield  {journal} {\bibinfo
  {journal} {Nat. Photon.}\ }\textbf {\bibinfo {volume} {7}},\ \bibinfo {pages}
  {540} (\bibinfo {year} {2013})}\BibitemShut {NoStop}%
\bibitem [{\citenamefont {Crespi}\ \emph {et~al.}(2013)\citenamefont {Crespi},
  \citenamefont {Osellame}, \citenamefont {Ramponi}, \citenamefont {Brod},
  \citenamefont {Galv\~{a}o}, \citenamefont {Spagnolo}, \citenamefont
  {Vitelli}, \citenamefont {Maiorino}, \citenamefont {Mataloni},\ and\
  \citenamefont {Sciarrino}}]{Crespi13}%
  \BibitemOpen
  \bibfield  {author} {\bibinfo {author} {\bibfnamefont {A.}~\bibnamefont
  {Crespi}}, \bibinfo {author} {\bibfnamefont {R.}~\bibnamefont {Osellame}},
  \bibinfo {author} {\bibfnamefont {R.}~\bibnamefont {Ramponi}}, \bibinfo
  {author} {\bibfnamefont {D.~J.}\ \bibnamefont {Brod}}, \bibinfo {author}
  {\bibfnamefont {E.~F.}\ \bibnamefont {Galv\~{a}o}}, \bibinfo {author}
  {\bibfnamefont {N.}~\bibnamefont {Spagnolo}}, \bibinfo {author}
  {\bibfnamefont {C.}~\bibnamefont {Vitelli}}, \bibinfo {author} {\bibfnamefont
  {E.}~\bibnamefont {Maiorino}}, \bibinfo {author} {\bibfnamefont
  {P.}~\bibnamefont {Mataloni}},\ and\ \bibinfo {author} {\bibfnamefont
  {F.}~\bibnamefont {Sciarrino}},\ }\href
  {https://doi.org/10.1038/nphoton.2013.112} {\bibfield  {journal} {\bibinfo
  {journal} {Nat. Photon.}\ }\textbf {\bibinfo {volume} {7}},\ \bibinfo {pages}
  {545} (\bibinfo {year} {2013})}\BibitemShut {NoStop}%
\bibitem [{\citenamefont {Spagnolo}\ \emph {et~al.}(2014)\citenamefont
  {Spagnolo}, \citenamefont {Vitelli}, \citenamefont {Bentivegna},
  \citenamefont {Brod}, \citenamefont {Crespi}, \citenamefont {Flamini},
  \citenamefont {Giacomini}, \citenamefont {Milani}, \citenamefont {Ramponi},
  \citenamefont {Mataloni}, \citenamefont {Osellame}, \citenamefont
  {Galv\~{a}o},\ and\ \citenamefont {Sciarrino}}]{Spagnolo14}%
  \BibitemOpen
  \bibfield  {author} {\bibinfo {author} {\bibfnamefont {N.}~\bibnamefont
  {Spagnolo}}, \bibinfo {author} {\bibfnamefont {C.}~\bibnamefont {Vitelli}},
  \bibinfo {author} {\bibfnamefont {M.}~\bibnamefont {Bentivegna}}, \bibinfo
  {author} {\bibfnamefont {D.~J.}\ \bibnamefont {Brod}}, \bibinfo {author}
  {\bibfnamefont {A.}~\bibnamefont {Crespi}}, \bibinfo {author} {\bibfnamefont
  {F.}~\bibnamefont {Flamini}}, \bibinfo {author} {\bibfnamefont
  {S.}~\bibnamefont {Giacomini}}, \bibinfo {author} {\bibfnamefont
  {G.}~\bibnamefont {Milani}}, \bibinfo {author} {\bibfnamefont
  {R.}~\bibnamefont {Ramponi}}, \bibinfo {author} {\bibfnamefont
  {P.}~\bibnamefont {Mataloni}}, \bibinfo {author} {\bibfnamefont
  {R.}~\bibnamefont {Osellame}}, \bibinfo {author} {\bibfnamefont {E.~F.}\
  \bibnamefont {Galv\~{a}o}},\ and\ \bibinfo {author} {\bibfnamefont
  {F.}~\bibnamefont {Sciarrino}},\ }\href
  {https://doi.org/10.1038/nphoton.2014.135} {\bibfield  {journal} {\bibinfo
  {journal} {Nat. Photon.}\ }\textbf {\bibinfo {volume} {8}},\ \bibinfo {pages}
  {615} (\bibinfo {year} {2014})}\BibitemShut {NoStop}%
\bibitem [{\citenamefont {Carolan}\ \emph {et~al.}(2014)\citenamefont
  {Carolan}, \citenamefont {Meinecke}, \citenamefont {Shadbolt}, \citenamefont
  {Russell}, \citenamefont {Ismail}, \citenamefont {Worhoff}, \citenamefont
  {Rudolph}, \citenamefont {Thompson}, \citenamefont {O'Brien}, \citenamefont
  {Matthews},\ and\ \citenamefont {Laing}}]{Carolan14}%
  \BibitemOpen
  \bibfield  {author} {\bibinfo {author} {\bibfnamefont {J.}~\bibnamefont
  {Carolan}}, \bibinfo {author} {\bibfnamefont {J.~D.~A.}\ \bibnamefont
  {Meinecke}}, \bibinfo {author} {\bibfnamefont {P.~J.}\ \bibnamefont
  {Shadbolt}}, \bibinfo {author} {\bibfnamefont {N.~J.}\ \bibnamefont
  {Russell}}, \bibinfo {author} {\bibfnamefont {N.}~\bibnamefont {Ismail}},
  \bibinfo {author} {\bibfnamefont {K.}~\bibnamefont {Worhoff}}, \bibinfo
  {author} {\bibfnamefont {T.}~\bibnamefont {Rudolph}}, \bibinfo {author}
  {\bibfnamefont {M.~G.}\ \bibnamefont {Thompson}}, \bibinfo {author}
  {\bibfnamefont {J.~L.}\ \bibnamefont {O'Brien}}, \bibinfo {author}
  {\bibfnamefont {J.~C.~F.}\ \bibnamefont {Matthews}},\ and\ \bibinfo {author}
  {\bibfnamefont {A.}~\bibnamefont {Laing}},\ }\href
  {https://doi.org/10.1038/nphoton.2014.152} {\bibfield  {journal} {\bibinfo
  {journal} {Nat. Photon.}\ }\textbf {\bibinfo {volume} {8}},\ \bibinfo {pages}
  {621} (\bibinfo {year} {2014})}\BibitemShut {NoStop}%
\bibitem [{\citenamefont {Carolan}\ \emph {et~al.}(2015)\citenamefont
  {Carolan}, \citenamefont {Harrold}, \citenamefont {Sparrow}, \citenamefont
  {Martin-Lopez}, \citenamefont {Russell}, \citenamefont {Silverstone},
  \citenamefont {Shadbolt}, \citenamefont {Matsuda}, \citenamefont {Oguma},
  \citenamefont {Itoh}, \citenamefont {Marshall}, \citenamefont {Thompson},
  \citenamefont {Matthews}, \citenamefont {Hashimoto}, \citenamefont
  {O'Brien},\ and\ \citenamefont {Laing}}]{Carolan15}%
  \BibitemOpen
  \bibfield  {author} {\bibinfo {author} {\bibfnamefont {J.}~\bibnamefont
  {Carolan}}, \bibinfo {author} {\bibfnamefont {C.}~\bibnamefont {Harrold}},
  \bibinfo {author} {\bibfnamefont {C.}~\bibnamefont {Sparrow}}, \bibinfo
  {author} {\bibfnamefont {E.}~\bibnamefont {Martin-Lopez}}, \bibinfo {author}
  {\bibfnamefont {N.~J.}\ \bibnamefont {Russell}}, \bibinfo {author}
  {\bibfnamefont {J.~W.}\ \bibnamefont {Silverstone}}, \bibinfo {author}
  {\bibfnamefont {P.~J.}\ \bibnamefont {Shadbolt}}, \bibinfo {author}
  {\bibfnamefont {N.}~\bibnamefont {Matsuda}}, \bibinfo {author} {\bibfnamefont
  {M.}~\bibnamefont {Oguma}}, \bibinfo {author} {\bibfnamefont
  {M.}~\bibnamefont {Itoh}}, \bibinfo {author} {\bibfnamefont {G.~D.}\
  \bibnamefont {Marshall}}, \bibinfo {author} {\bibfnamefont {M.~G.}\
  \bibnamefont {Thompson}}, \bibinfo {author} {\bibfnamefont {J.~C.~F.}\
  \bibnamefont {Matthews}}, \bibinfo {author} {\bibfnamefont {T.}~\bibnamefont
  {Hashimoto}}, \bibinfo {author} {\bibfnamefont {J.~L.}\ \bibnamefont
  {O'Brien}},\ and\ \bibinfo {author} {\bibfnamefont {A.}~\bibnamefont
  {Laing}},\ }\href {https://doi.org/10.1126/science.aab3642} {\bibfield
  {journal} {\bibinfo  {journal} {Science}\ }\textbf {\bibinfo {volume}
  {349}},\ \bibinfo {pages} {711} (\bibinfo {year} {2015})}\BibitemShut
  {NoStop}%
\bibitem [{\citenamefont {Loredo}\ \emph {et~al.}(2017)\citenamefont {Loredo},
  \citenamefont {Broome}, \citenamefont {Hilaire}, \citenamefont {Gazzano},
  \citenamefont {Sagnes}, \citenamefont {Lemaitre}, \citenamefont {Almeida},
  \citenamefont {Senellart},\ and\ \citenamefont {White}}]{Loredo17}%
  \BibitemOpen
  \bibfield  {author} {\bibinfo {author} {\bibfnamefont {J.~C.}\ \bibnamefont
  {Loredo}}, \bibinfo {author} {\bibfnamefont {M.~A.}\ \bibnamefont {Broome}},
  \bibinfo {author} {\bibfnamefont {P.}~\bibnamefont {Hilaire}}, \bibinfo
  {author} {\bibfnamefont {O.}~\bibnamefont {Gazzano}}, \bibinfo {author}
  {\bibfnamefont {I.}~\bibnamefont {Sagnes}}, \bibinfo {author} {\bibfnamefont
  {A.}~\bibnamefont {Lemaitre}}, \bibinfo {author} {\bibfnamefont {M.~P.}\
  \bibnamefont {Almeida}}, \bibinfo {author} {\bibfnamefont {P.}~\bibnamefont
  {Senellart}},\ and\ \bibinfo {author} {\bibfnamefont {A.~G.}\ \bibnamefont
  {White}},\ }\href {https://doi.org/10.1103/PhysRevLett.118.130503} {\bibfield
   {journal} {\bibinfo  {journal} {Phys. Rev. Lett.}\ }\textbf {\bibinfo
  {volume} {118}},\ \bibinfo {pages} {130503} (\bibinfo {year}
  {2017})}\BibitemShut {NoStop}%
\bibitem [{\citenamefont {He}\ \emph {et~al.}(2017)\citenamefont {He},
  \citenamefont {Ding}, \citenamefont {Su}, \citenamefont {Huang},
  \citenamefont {Qin}, \citenamefont {Wang}, \citenamefont {Unsleber},
  \citenamefont {Chen}, \citenamefont {Wang}, \citenamefont {He}, \citenamefont
  {Wang}, \citenamefont {Zhang}, \citenamefont {Chen}, \citenamefont
  {Schneider}, \citenamefont {Kamp}, \citenamefont {You}, \citenamefont {Wang},
  \citenamefont {H\"{o}fling}, \citenamefont {Lu},\ and\ \citenamefont
  {Pan}}]{He17}%
  \BibitemOpen
  \bibfield  {author} {\bibinfo {author} {\bibfnamefont {Y.}~\bibnamefont
  {He}}, \bibinfo {author} {\bibfnamefont {X.}~\bibnamefont {Ding}}, \bibinfo
  {author} {\bibfnamefont {Z.-E.}\ \bibnamefont {Su}}, \bibinfo {author}
  {\bibfnamefont {H.-L.}\ \bibnamefont {Huang}}, \bibinfo {author}
  {\bibfnamefont {J.}~\bibnamefont {Qin}}, \bibinfo {author} {\bibfnamefont
  {C.}~\bibnamefont {Wang}}, \bibinfo {author} {\bibfnamefont {S.}~\bibnamefont
  {Unsleber}}, \bibinfo {author} {\bibfnamefont {C.}~\bibnamefont {Chen}},
  \bibinfo {author} {\bibfnamefont {H.}~\bibnamefont {Wang}}, \bibinfo {author}
  {\bibfnamefont {Y.-M.}\ \bibnamefont {He}}, \bibinfo {author} {\bibfnamefont
  {X.-L.}\ \bibnamefont {Wang}}, \bibinfo {author} {\bibfnamefont {W.-J.}\
  \bibnamefont {Zhang}}, \bibinfo {author} {\bibfnamefont {S.-J.}\ \bibnamefont
  {Chen}}, \bibinfo {author} {\bibfnamefont {C.}~\bibnamefont {Schneider}},
  \bibinfo {author} {\bibfnamefont {M.}~\bibnamefont {Kamp}}, \bibinfo {author}
  {\bibfnamefont {L.-X.}\ \bibnamefont {You}}, \bibinfo {author} {\bibfnamefont
  {Z.}~\bibnamefont {Wang}}, \bibinfo {author} {\bibfnamefont {S.}~\bibnamefont
  {H\"{o}fling}}, \bibinfo {author} {\bibfnamefont {C.-Y.}\ \bibnamefont
  {Lu}},\ and\ \bibinfo {author} {\bibfnamefont {J.-W.}\ \bibnamefont {Pan}},\
  }\href {https://doi.org/10.1103/PhysRevLett.118.190501} {\bibfield  {journal}
  {\bibinfo  {journal} {Phys. Rev. Lett.}\ }\textbf {\bibinfo {volume} {118}},\
  \bibinfo {pages} {190501} (\bibinfo {year} {2017})}\BibitemShut {NoStop}%
\bibitem [{\citenamefont {Wang}\ \emph {et~al.}(2017)\citenamefont {Wang},
  \citenamefont {He}, \citenamefont {Li}, \citenamefont {Su}, \citenamefont
  {Li}, \citenamefont {Huang}, \citenamefont {Ding}, \citenamefont {Chen},
  \citenamefont {Liu}, \citenamefont {Qin}, \citenamefont {Li}, \citenamefont
  {He}, \citenamefont {Schneider}, \citenamefont {Kamp}, \citenamefont {Peng},
  \citenamefont {Hoefling}, \citenamefont {Lu},\ and\ \citenamefont
  {Pan}}]{Wang17bs}%
  \BibitemOpen
  \bibfield  {author} {\bibinfo {author} {\bibfnamefont {H.}~\bibnamefont
  {Wang}}, \bibinfo {author} {\bibfnamefont {Y.}~\bibnamefont {He}}, \bibinfo
  {author} {\bibfnamefont {Y.-H.}\ \bibnamefont {Li}}, \bibinfo {author}
  {\bibfnamefont {Z.-E.}\ \bibnamefont {Su}}, \bibinfo {author} {\bibfnamefont
  {B.}~\bibnamefont {Li}}, \bibinfo {author} {\bibfnamefont {H.-L.}\
  \bibnamefont {Huang}}, \bibinfo {author} {\bibfnamefont {X.}~\bibnamefont
  {Ding}}, \bibinfo {author} {\bibfnamefont {M.-C.}\ \bibnamefont {Chen}},
  \bibinfo {author} {\bibfnamefont {C.}~\bibnamefont {Liu}}, \bibinfo {author}
  {\bibfnamefont {J.}~\bibnamefont {Qin}}, \bibinfo {author} {\bibfnamefont
  {J.-P.}\ \bibnamefont {Li}}, \bibinfo {author} {\bibfnamefont {Y.-M.}\
  \bibnamefont {He}}, \bibinfo {author} {\bibfnamefont {C.}~\bibnamefont
  {Schneider}}, \bibinfo {author} {\bibfnamefont {M.}~\bibnamefont {Kamp}},
  \bibinfo {author} {\bibfnamefont {C.-Z.}\ \bibnamefont {Peng}}, \bibinfo
  {author} {\bibfnamefont {S.}~\bibnamefont {Hoefling}}, \bibinfo {author}
  {\bibfnamefont {C.-Y.}\ \bibnamefont {Lu}},\ and\ \bibinfo {author}
  {\bibfnamefont {J.-W.}\ \bibnamefont {Pan}},\ }\href
  {https://doi.org/10.1038/nphoton.2017.63} {\bibfield  {journal} {\bibinfo
  {journal} {Nat. Photon.}\ }\textbf {\bibinfo {volume} {11}},\ \bibinfo
  {pages} {361} (\bibinfo {year} {2017})}\BibitemShut {NoStop}%
\bibitem [{\citenamefont {Wang}\ \emph
  {et~al.}(2018{\natexlab{a}})\citenamefont {Wang}, \citenamefont {Li},
  \citenamefont {Jiang}, \citenamefont {He}, \citenamefont {Li}, \citenamefont
  {Ding}, \citenamefont {Chen}, \citenamefont {Qin}, \citenamefont {Peng},
  \citenamefont {Schneider}, \citenamefont {Kamp}, \citenamefont {Zhang},
  \citenamefont {Li}, \citenamefont {You}, \citenamefont {Wang}, \citenamefont
  {Dowling}, \citenamefont {H\"{o}fling}, \citenamefont {Lu},\ and\
  \citenamefont {Pan}}]{Wang18}%
  \BibitemOpen
  \bibfield  {author} {\bibinfo {author} {\bibfnamefont {H.}~\bibnamefont
  {Wang}}, \bibinfo {author} {\bibfnamefont {W.}~\bibnamefont {Li}}, \bibinfo
  {author} {\bibfnamefont {X.}~\bibnamefont {Jiang}}, \bibinfo {author}
  {\bibfnamefont {Y.-M.}\ \bibnamefont {He}}, \bibinfo {author} {\bibfnamefont
  {Y.-H.}\ \bibnamefont {Li}}, \bibinfo {author} {\bibfnamefont
  {X.}~\bibnamefont {Ding}}, \bibinfo {author} {\bibfnamefont {M.-C.}\
  \bibnamefont {Chen}}, \bibinfo {author} {\bibfnamefont {J.}~\bibnamefont
  {Qin}}, \bibinfo {author} {\bibfnamefont {C.-Z.}\ \bibnamefont {Peng}},
  \bibinfo {author} {\bibfnamefont {C.}~\bibnamefont {Schneider}}, \bibinfo
  {author} {\bibfnamefont {M.}~\bibnamefont {Kamp}}, \bibinfo {author}
  {\bibfnamefont {W.-J.}\ \bibnamefont {Zhang}}, \bibinfo {author}
  {\bibfnamefont {H.}~\bibnamefont {Li}}, \bibinfo {author} {\bibfnamefont
  {L.-X.}\ \bibnamefont {You}}, \bibinfo {author} {\bibfnamefont
  {Z.}~\bibnamefont {Wang}}, \bibinfo {author} {\bibfnamefont {J.~P.}\
  \bibnamefont {Dowling}}, \bibinfo {author} {\bibfnamefont {S.}~\bibnamefont
  {H\"{o}fling}}, \bibinfo {author} {\bibfnamefont {C.-Y.}\ \bibnamefont
  {Lu}},\ and\ \bibinfo {author} {\bibfnamefont {J.-W.}\ \bibnamefont {Pan}},\
  }\href {https://doi.org/10.1103/PhysRevLett.120.230502} {\bibfield  {journal}
  {\bibinfo  {journal} {Phys. Rev. Lett.}\ }\textbf {\bibinfo {volume} {120}},\
  \bibinfo {pages} {230502} (\bibinfo {year} {2018}{\natexlab{a}})}\BibitemShut
  {NoStop}%
\bibitem [{\citenamefont {Wang}\ \emph {et~al.}(2019)\citenamefont {Wang},
  \citenamefont {Qin}, \citenamefont {Ding}, \citenamefont {Chen},
  \citenamefont {Chen}, \citenamefont {You}, \citenamefont {He}, \citenamefont
  {Jiang}, \citenamefont {You}, \citenamefont {Wang}, \citenamefont
  {Schneider}, \citenamefont {Renema}, \citenamefont {H\"{o}fling},
  \citenamefont {Lu},\ and\ \citenamefont {Pan}}]{Wang19Det}%
  \BibitemOpen
  \bibfield  {author} {\bibinfo {author} {\bibfnamefont {H.}~\bibnamefont
  {Wang}}, \bibinfo {author} {\bibfnamefont {J.}~\bibnamefont {Qin}}, \bibinfo
  {author} {\bibfnamefont {X.}~\bibnamefont {Ding}}, \bibinfo {author}
  {\bibfnamefont {M.-C.}\ \bibnamefont {Chen}}, \bibinfo {author}
  {\bibfnamefont {S.}~\bibnamefont {Chen}}, \bibinfo {author} {\bibfnamefont
  {X.}~\bibnamefont {You}}, \bibinfo {author} {\bibfnamefont {Y.-M.}\
  \bibnamefont {He}}, \bibinfo {author} {\bibfnamefont {X.}~\bibnamefont
  {Jiang}}, \bibinfo {author} {\bibfnamefont {L.}~\bibnamefont {You}}, \bibinfo
  {author} {\bibfnamefont {Z.}~\bibnamefont {Wang}}, \bibinfo {author}
  {\bibfnamefont {C.}~\bibnamefont {Schneider}}, \bibinfo {author}
  {\bibfnamefont {J.~J.}\ \bibnamefont {Renema}}, \bibinfo {author}
  {\bibfnamefont {S.}~\bibnamefont {H\"{o}fling}}, \bibinfo {author}
  {\bibfnamefont {C.-Y.}\ \bibnamefont {Lu}},\ and\ \bibinfo {author}
  {\bibfnamefont {J.-W.}\ \bibnamefont {Pan}},\ }\href
  {https://doi.org/10.1103/PhysRevLett.123.250503} {\bibfield  {journal}
  {\bibinfo  {journal} {Phys. Rev. Lett.}\ }\textbf {\bibinfo {volume} {123}},\
  \bibinfo {pages} {250503} (\bibinfo {year} {2019})}\BibitemShut {NoStop}%
\bibitem [{\citenamefont {Bentivegna}\ \emph {et~al.}(2015)\citenamefont
  {Bentivegna}, \citenamefont {Spagnolo}, \citenamefont {Vitelli},
  \citenamefont {Flamini}, \citenamefont {Viggianiello}, \citenamefont
  {Latmiral}, \citenamefont {Mataloni}, \citenamefont {Brod}, \citenamefont
  {Galv\~{a}o}, \citenamefont {Crespi}, \citenamefont {Ramponi}, \citenamefont
  {Osellame},\ and\ \citenamefont {Sciarrino}}]{Bentivegna15}%
  \BibitemOpen
  \bibfield  {author} {\bibinfo {author} {\bibfnamefont {M.}~\bibnamefont
  {Bentivegna}}, \bibinfo {author} {\bibfnamefont {N.}~\bibnamefont
  {Spagnolo}}, \bibinfo {author} {\bibfnamefont {C.}~\bibnamefont {Vitelli}},
  \bibinfo {author} {\bibfnamefont {F.}~\bibnamefont {Flamini}}, \bibinfo
  {author} {\bibfnamefont {N.}~\bibnamefont {Viggianiello}}, \bibinfo {author}
  {\bibfnamefont {L.}~\bibnamefont {Latmiral}}, \bibinfo {author}
  {\bibfnamefont {P.}~\bibnamefont {Mataloni}}, \bibinfo {author}
  {\bibfnamefont {D.~J.}\ \bibnamefont {Brod}}, \bibinfo {author}
  {\bibfnamefont {E.~F.}\ \bibnamefont {Galv\~{a}o}}, \bibinfo {author}
  {\bibfnamefont {A.}~\bibnamefont {Crespi}}, \bibinfo {author} {\bibfnamefont
  {R.}~\bibnamefont {Ramponi}}, \bibinfo {author} {\bibfnamefont
  {R.}~\bibnamefont {Osellame}},\ and\ \bibinfo {author} {\bibfnamefont
  {F.}~\bibnamefont {Sciarrino}},\ }\href
  {https://doi.org/10.1126/sciadv.1400255} {\bibfield  {journal} {\bibinfo
  {journal} {Sci. Adv.}\ }\textbf {\bibinfo {volume} {1}},\ \bibinfo {pages}
  {e1400255} (\bibinfo {year} {2015})}\BibitemShut {NoStop}%
\bibitem [{\citenamefont {Zhong}\ \emph {et~al.}(2018)\citenamefont {Zhong},
  \citenamefont {Li}, \citenamefont {Li}, \citenamefont {Peng}, \citenamefont
  {Su}, \citenamefont {Hu}, \citenamefont {He}, \citenamefont {Ding},
  \citenamefont {Zhang}, \citenamefont {Li}, \citenamefont {Zhang},
  \citenamefont {Wang}, \citenamefont {You}, \citenamefont {Wang},
  \citenamefont {Jiang}, \citenamefont {Li}, \citenamefont {Chen},
  \citenamefont {Liu}, \citenamefont {Lu},\ and\ \citenamefont
  {Pan}}]{Zhong18}%
  \BibitemOpen
  \bibfield  {author} {\bibinfo {author} {\bibfnamefont {H.-S.}\ \bibnamefont
  {Zhong}}, \bibinfo {author} {\bibfnamefont {Y.}~\bibnamefont {Li}}, \bibinfo
  {author} {\bibfnamefont {W.}~\bibnamefont {Li}}, \bibinfo {author}
  {\bibfnamefont {L.-C.}\ \bibnamefont {Peng}}, \bibinfo {author}
  {\bibfnamefont {Z.-E.}\ \bibnamefont {Su}}, \bibinfo {author} {\bibfnamefont
  {Y.}~\bibnamefont {Hu}}, \bibinfo {author} {\bibfnamefont {Y.-M.}\
  \bibnamefont {He}}, \bibinfo {author} {\bibfnamefont {X.}~\bibnamefont
  {Ding}}, \bibinfo {author} {\bibfnamefont {W.-J.}\ \bibnamefont {Zhang}},
  \bibinfo {author} {\bibfnamefont {H.}~\bibnamefont {Li}}, \bibinfo {author}
  {\bibfnamefont {L.}~\bibnamefont {Zhang}}, \bibinfo {author} {\bibfnamefont
  {Z.}~\bibnamefont {Wang}}, \bibinfo {author} {\bibfnamefont {L.-X.}\
  \bibnamefont {You}}, \bibinfo {author} {\bibfnamefont {X.-L.}\ \bibnamefont
  {Wang}}, \bibinfo {author} {\bibfnamefont {X.}~\bibnamefont {Jiang}},
  \bibinfo {author} {\bibfnamefont {L.}~\bibnamefont {Li}}, \bibinfo {author}
  {\bibfnamefont {Y.-A.}\ \bibnamefont {Chen}}, \bibinfo {author}
  {\bibfnamefont {N.-L.}\ \bibnamefont {Liu}}, \bibinfo {author} {\bibfnamefont
  {C.-Y.}\ \bibnamefont {Lu}},\ and\ \bibinfo {author} {\bibfnamefont {J.-W.}\
  \bibnamefont {Pan}},\ }\href {https://doi.org/10.1103/PhysRevLett.121.250505}
  {\bibfield  {journal} {\bibinfo  {journal} {Phys. Rev. Lett.}\ }\textbf
  {\bibinfo {volume} {121}},\ \bibinfo {pages} {250505} (\bibinfo {year}
  {2018})}\BibitemShut {NoStop}%
\bibitem [{\citenamefont {Wang}\ \emph
  {et~al.}(2018{\natexlab{b}})\citenamefont {Wang}, \citenamefont {Jing},
  \citenamefont {Sun}, \citenamefont {Yang}, \citenamefont {Yu}, \citenamefont
  {Tamma}, \citenamefont {Bao},\ and\ \citenamefont {Pan}}]{Wang18b}%
  \BibitemOpen
  \bibfield  {author} {\bibinfo {author} {\bibfnamefont {X.-J.}\ \bibnamefont
  {Wang}}, \bibinfo {author} {\bibfnamefont {B.}~\bibnamefont {Jing}}, \bibinfo
  {author} {\bibfnamefont {P.-F.}\ \bibnamefont {Sun}}, \bibinfo {author}
  {\bibfnamefont {C.-W.}\ \bibnamefont {Yang}}, \bibinfo {author}
  {\bibfnamefont {Y.}~\bibnamefont {Yu}}, \bibinfo {author} {\bibfnamefont
  {V.}~\bibnamefont {Tamma}}, \bibinfo {author} {\bibfnamefont {X.-H.}\
  \bibnamefont {Bao}},\ and\ \bibinfo {author} {\bibfnamefont {J.-W.}\
  \bibnamefont {Pan}},\ }\href {https://doi.org/10.1103/PhysRevLett.121.080501}
  {\bibfield  {journal} {\bibinfo  {journal} {Phys. Rev. Lett.}\ }\textbf
  {\bibinfo {volume} {121}},\ \bibinfo {pages} {080501} (\bibinfo {year}
  {2018}{\natexlab{b}})}\BibitemShut {NoStop}%
\bibitem [{\citenamefont {Paesani}\ \emph {et~al.}(2019)\citenamefont
  {Paesani}, \citenamefont {Ding}, \citenamefont {Santagati}, \citenamefont
  {Chakhmakhchyan}, \citenamefont {Vigliar}, \citenamefont {Rottwitt},
  \citenamefont {Oxenl\"{o}we}, \citenamefont {Wang}, \citenamefont
  {Thompson},\ and\ \citenamefont {Laing}}]{Paesani19}%
  \BibitemOpen
  \bibfield  {author} {\bibinfo {author} {\bibfnamefont {S.}~\bibnamefont
  {Paesani}}, \bibinfo {author} {\bibfnamefont {Y.}~\bibnamefont {Ding}},
  \bibinfo {author} {\bibfnamefont {R.}~\bibnamefont {Santagati}}, \bibinfo
  {author} {\bibfnamefont {L.}~\bibnamefont {Chakhmakhchyan}}, \bibinfo
  {author} {\bibfnamefont {C.}~\bibnamefont {Vigliar}}, \bibinfo {author}
  {\bibfnamefont {K.}~\bibnamefont {Rottwitt}}, \bibinfo {author}
  {\bibfnamefont {L.~K.}\ \bibnamefont {Oxenl\"{o}we}}, \bibinfo {author}
  {\bibfnamefont {J.}~\bibnamefont {Wang}}, \bibinfo {author} {\bibfnamefont
  {M.~G.}\ \bibnamefont {Thompson}},\ and\ \bibinfo {author} {\bibfnamefont
  {A.}~\bibnamefont {Laing}},\ }\href
  {https://doi.org/doi.org/10.1038/s41567-019-0567-8} {\bibfield  {journal}
  {\bibinfo  {journal} {Nat. Phys.}\ }\textbf {\bibinfo {volume} {15}},\
  \bibinfo {pages} {925} (\bibinfo {year} {2019})}\BibitemShut {NoStop}%
\bibitem [{\citenamefont {Zhong}\ \emph {et~al.}(2019)\citenamefont {Zhong},
  \citenamefont {Peng}, \citenamefont {Li}, \citenamefont {Hu}, \citenamefont
  {Li}, \citenamefont {Qin}, \citenamefont {Wu}, \citenamefont {Zhang},
  \citenamefont {Li}, \citenamefont {Zhang}, \citenamefont {Wang},
  \citenamefont {You}, \citenamefont {Jiang}, \citenamefont {Li}, \citenamefont
  {Liu}, \citenamefont {P.Dowling}, \citenamefont {Lu},\ and\ \citenamefont
  {Pan}}]{Zhong19}%
  \BibitemOpen
  \bibfield  {author} {\bibinfo {author} {\bibfnamefont {H.-S.}\ \bibnamefont
  {Zhong}}, \bibinfo {author} {\bibfnamefont {L.-C.}\ \bibnamefont {Peng}},
  \bibinfo {author} {\bibfnamefont {Y.}~\bibnamefont {Li}}, \bibinfo {author}
  {\bibfnamefont {Y.}~\bibnamefont {Hu}}, \bibinfo {author} {\bibfnamefont
  {W.}~\bibnamefont {Li}}, \bibinfo {author} {\bibfnamefont {J.}~\bibnamefont
  {Qin}}, \bibinfo {author} {\bibfnamefont {D.}~\bibnamefont {Wu}}, \bibinfo
  {author} {\bibfnamefont {W.}~\bibnamefont {Zhang}}, \bibinfo {author}
  {\bibfnamefont {H.}~\bibnamefont {Li}}, \bibinfo {author} {\bibfnamefont
  {L.}~\bibnamefont {Zhang}}, \bibinfo {author} {\bibfnamefont
  {Z.}~\bibnamefont {Wang}}, \bibinfo {author} {\bibfnamefont {L.}~\bibnamefont
  {You}}, \bibinfo {author} {\bibfnamefont {X.}~\bibnamefont {Jiang}}, \bibinfo
  {author} {\bibfnamefont {L.}~\bibnamefont {Li}}, \bibinfo {author}
  {\bibfnamefont {N.-L.}\ \bibnamefont {Liu}}, \bibinfo {author} {\bibfnamefont
  {J.}~\bibnamefont {P.Dowling}}, \bibinfo {author} {\bibfnamefont {C.-Y.}\
  \bibnamefont {Lu}},\ and\ \bibinfo {author} {\bibfnamefont {J.-W.}\
  \bibnamefont {Pan}},\ }\href {https://doi.org/10.1016/j.scib.2019.04.007}
  {\bibfield  {journal} {\bibinfo  {journal} {Sci. Bull.}\ }\textbf {\bibinfo
  {volume} {64}},\ \bibinfo {pages} {511} (\bibinfo {year} {2019})}\BibitemShut
  {NoStop}%
\bibitem [{\citenamefont {Brod}\ \emph {et~al.}(2019)\citenamefont {Brod},
  \citenamefont {ao}, \citenamefont {Crespi}, \citenamefont {Osellame},
  \citenamefont {Spagnolo},\ and\ \citenamefont {Sciarrino}}]{Brod19Review}%
  \BibitemOpen
  \bibfield  {author} {\bibinfo {author} {\bibfnamefont {D.~J.}\ \bibnamefont
  {Brod}}, \bibinfo {author} {\bibfnamefont {E.~F.~G.}\ \bibnamefont {ao}},
  \bibinfo {author} {\bibfnamefont {A.}~\bibnamefont {Crespi}}, \bibinfo
  {author} {\bibfnamefont {R.}~\bibnamefont {Osellame}}, \bibinfo {author}
  {\bibfnamefont {N.}~\bibnamefont {Spagnolo}},\ and\ \bibinfo {author}
  {\bibfnamefont {F.}~\bibnamefont {Sciarrino}},\ }\href
  {https://doi.org/10.1117/1.AP.1.3.034001} {\bibfield  {journal} {\bibinfo
  {journal} {Adv. Phot.}\ }\textbf {\bibinfo {volume} {1}},\ \bibinfo {pages}
  {04001} (\bibinfo {year} {2019})}\BibitemShut {NoStop}%
\bibitem [{\citenamefont {Huh}\ \emph {et~al.}(2015)\citenamefont {Huh},
  \citenamefont {Guerreschi}, \citenamefont {Peropadre}, \citenamefont
  {McClean},\ and\ \citenamefont {Aspuru-Guzik}}]{Huh15}%
  \BibitemOpen
  \bibfield  {author} {\bibinfo {author} {\bibfnamefont {J.}~\bibnamefont
  {Huh}}, \bibinfo {author} {\bibfnamefont {G.~G.}\ \bibnamefont {Guerreschi}},
  \bibinfo {author} {\bibfnamefont {B.}~\bibnamefont {Peropadre}}, \bibinfo
  {author} {\bibfnamefont {J.~R.}\ \bibnamefont {McClean}},\ and\ \bibinfo
  {author} {\bibfnamefont {A.}~\bibnamefont {Aspuru-Guzik}},\ }\href
  {https://doi.org/10.1038/nphoton.2015.153} {\bibfield  {journal} {\bibinfo
  {journal} {Nat. Photon.}\ }\textbf {\bibinfo {volume} {9}},\ \bibinfo {pages}
  {615} (\bibinfo {year} {2015})}\BibitemShut {NoStop}%
\bibitem [{\citenamefont {Huh}\ and\ \citenamefont {Yung}(2017)}]{Huh17}%
  \BibitemOpen
  \bibfield  {author} {\bibinfo {author} {\bibfnamefont {J.}~\bibnamefont
  {Huh}}\ and\ \bibinfo {author} {\bibfnamefont {M.-H.}\ \bibnamefont {Yung}},\
  }\href {https://doi.org/10.1038/s41598-017-07770-z} {\bibfield  {journal}
  {\bibinfo  {journal} {Sci. Rep.}\ }\textbf {\bibinfo {volume} {7}},\ \bibinfo
  {pages} {7462} (\bibinfo {year} {2017})}\BibitemShut {NoStop}%
\bibitem [{\citenamefont {Chin}\ and\ \citenamefont {Huh}(2018)}]{Chin18}%
  \BibitemOpen
  \bibfield  {author} {\bibinfo {author} {\bibfnamefont {S.}~\bibnamefont
  {Chin}}\ and\ \bibinfo {author} {\bibfnamefont {J.}~\bibnamefont {Huh}},\
  }\href {https://doi.org/10.1088/1742-6596/1071/1/012009} {\bibfield
  {journal} {\bibinfo  {journal} {Journal of Physics: Conf. Series}\ }\textbf
  {\bibinfo {volume} {1071}},\ \bibinfo {pages} {012009} (\bibinfo {year}
  {2018})}\BibitemShut {NoStop}%
\bibitem [{\citenamefont {Banchi}\ \emph {et~al.}(2020)\citenamefont {Banchi},
  \citenamefont {Fingerhuth}, \citenamefont {Babej}, \citenamefont {Ing},\ and\
  \citenamefont {Arrazola}}]{Banchi19}%
  \BibitemOpen
  \bibfield  {author} {\bibinfo {author} {\bibfnamefont {L.}~\bibnamefont
  {Banchi}}, \bibinfo {author} {\bibfnamefont {M.}~\bibnamefont {Fingerhuth}},
  \bibinfo {author} {\bibfnamefont {T.}~\bibnamefont {Babej}}, \bibinfo
  {author} {\bibfnamefont {C.}~\bibnamefont {Ing}},\ and\ \bibinfo {author}
  {\bibfnamefont {J.~M.}\ \bibnamefont {Arrazola}},\ }\href
  {https://doi.org/10.1126/sciadv.aax1950} {\bibfield  {journal} {\bibinfo
  {journal} {Sci. Adv.}\ }\textbf {\bibinfo {volume} {6}},\ \bibinfo {pages}
  {eeax1950} (\bibinfo {year} {2020})}\BibitemShut {NoStop}%
\bibitem [{\citenamefont {Arrazola}\ \emph {et~al.}(2018)\citenamefont
  {Arrazola}, \citenamefont {Bromley},\ and\ \citenamefont
  {Rebentrost}}]{Arrazola18b}%
  \BibitemOpen
  \bibfield  {author} {\bibinfo {author} {\bibfnamefont {J.~M.}\ \bibnamefont
  {Arrazola}}, \bibinfo {author} {\bibfnamefont {R.~R.}\ \bibnamefont
  {Bromley}},\ and\ \bibinfo {author} {\bibfnamefont {P.}~\bibnamefont
  {Rebentrost}},\ }\href {https://doi.org/10.1103/PhysRevA.98.012322}
  {\bibfield  {journal} {\bibinfo  {journal} {Phys. Rev. A}\ }\textbf {\bibinfo
  {volume} {98}},\ \bibinfo {pages} {012322} (\bibinfo {year}
  {2018})}\BibitemShut {NoStop}%
\bibitem [{\citenamefont {Jahangiri}\ \emph {et~al.}(2020)\citenamefont
  {Jahangiri}, \citenamefont {Arrazola}, \citenamefont {Quesada},\ and\
  \citenamefont {Killoran}}]{Jahangiri20point}%
  \BibitemOpen
  \bibfield  {author} {\bibinfo {author} {\bibfnamefont {S.}~\bibnamefont
  {Jahangiri}}, \bibinfo {author} {\bibfnamefont {J.~M.}\ \bibnamefont
  {Arrazola}}, \bibinfo {author} {\bibfnamefont {N.}~\bibnamefont {Quesada}},\
  and\ \bibinfo {author} {\bibfnamefont {N.}~\bibnamefont {Killoran}},\ }\href
  {https://doi.org/10.1103/PhysRevE.101.022134} {\bibfield  {journal} {\bibinfo
   {journal} {Phys. Rev. E}\ }\textbf {\bibinfo {volume} {101}},\ \bibinfo
  {pages} {022134} (\bibinfo {year} {2020})}\BibitemShut {NoStop}%
\bibitem [{\citenamefont {Arrazola}\ and\ \citenamefont
  {Bromley}(2018)}]{Arrazola18}%
  \BibitemOpen
  \bibfield  {author} {\bibinfo {author} {\bibfnamefont {J.~M.}\ \bibnamefont
  {Arrazola}}\ and\ \bibinfo {author} {\bibfnamefont {T.~R.}\ \bibnamefont
  {Bromley}},\ }\href {https://doi.org/10.1103/PhysRevLett.121.030503}
  {\bibfield  {journal} {\bibinfo  {journal} {Phys. Rev. Lett.}\ }\textbf
  {\bibinfo {volume} {121}},\ \bibinfo {pages} {030503} (\bibinfo {year}
  {2018})}\BibitemShut {NoStop}%
\bibitem [{\citenamefont {Bradler}\ \emph {et~al.}(2018)\citenamefont
  {Bradler}, \citenamefont {Dallaire-Demers}, \citenamefont {Rebentrost},
  \citenamefont {Su},\ and\ \citenamefont {Weedbrook}}]{Bradler18}%
  \BibitemOpen
  \bibfield  {author} {\bibinfo {author} {\bibfnamefont {K.}~\bibnamefont
  {Bradler}}, \bibinfo {author} {\bibfnamefont {P.-L.}\ \bibnamefont
  {Dallaire-Demers}}, \bibinfo {author} {\bibfnamefont {P.}~\bibnamefont
  {Rebentrost}}, \bibinfo {author} {\bibfnamefont {D.}~\bibnamefont {Su}},\
  and\ \bibinfo {author} {\bibfnamefont {C.}~\bibnamefont {Weedbrook}},\ }\href
  {https://doi.org/10.1103/PhysRevA.98.032310} {\bibfield  {journal} {\bibinfo
  {journal} {Phys. Rev. A}\ }\textbf {\bibinfo {volume} {98}},\ \bibinfo
  {pages} {032310} (\bibinfo {year} {2018})}\BibitemShut {NoStop}%
\bibitem [{\citenamefont {Bradler}\ \emph {et~al.}()\citenamefont {Bradler},
  \citenamefont {Friedland}, \citenamefont {Izaac}, \citenamefont {Killoran},\
  and\ \citenamefont {Su}}]{Bradler18b}%
  \BibitemOpen
  \bibfield  {author} {\bibinfo {author} {\bibfnamefont {K.}~\bibnamefont
  {Bradler}}, \bibinfo {author} {\bibfnamefont {S.}~\bibnamefont {Friedland}},
  \bibinfo {author} {\bibfnamefont {J.}~\bibnamefont {Izaac}}, \bibinfo
  {author} {\bibfnamefont {N.}~\bibnamefont {Killoran}},\ and\ \bibinfo
  {author} {\bibfnamefont {D.}~\bibnamefont {Su}},\ }\bibinfo {note} {preprint
  at arXiv:1810.10644}\BibitemShut {NoStop}%
\bibitem [{\citenamefont {Steinbrecher}\ \emph {et~al.}(2019)\citenamefont
  {Steinbrecher}, \citenamefont {Olson}, \citenamefont {Englund},\ and\
  \citenamefont {Carolan}}]{Steibrecher19NN}%
  \BibitemOpen
  \bibfield  {author} {\bibinfo {author} {\bibfnamefont {G.~R.}\ \bibnamefont
  {Steinbrecher}}, \bibinfo {author} {\bibfnamefont {J.~P.}\ \bibnamefont
  {Olson}}, \bibinfo {author} {\bibfnamefont {D.}~\bibnamefont {Englund}},\
  and\ \bibinfo {author} {\bibfnamefont {J.}~\bibnamefont {Carolan}},\ }\href
  {https://doi.org/10.1038/s41534-019-0174-7} {\bibfield  {journal} {\bibinfo
  {journal} {npj Quantum Inf.}\ }\textbf {\bibinfo {volume} {5}},\ \bibinfo
  {pages} {60} (\bibinfo {year} {2019})}\BibitemShut {NoStop}%
\bibitem [{\citenamefont {Arrazola}\ \emph {et~al.}(2021)\citenamefont
  {Arrazola}, \citenamefont {Bergholm}, \citenamefont {Bradler}, \citenamefont
  {Bromley}, \citenamefont {Collins}, \citenamefont {Dhand}, \citenamefont
  {Fumagalli}, \citenamefont {Gerrits}, \citenamefont {Goussev}, \citenamefont
  {Helt}, \citenamefont {Hundal}, \citenamefont {Isacsson}, \citenamefont
  {Israel}, \citenamefont {Izaac}, \citenamefont {Jahangiri}, \citenamefont
  {Janik}, \citenamefont {Killoran}, \citenamefont {Kumar}, \citenamefont
  {Lavoie}, \citenamefont {Lita}, \citenamefont {Mahler}, \citenamefont
  {Menotti}, \citenamefont {Morrison}, \citenamefont {Nam}, \citenamefont
  {Neuhaus}, \citenamefont {Qi}, \citenamefont {Quesada}, \citenamefont
  {Repingon}, \citenamefont {Sabapathy}, \citenamefont {Schuld}, \citenamefont
  {Su}, \citenamefont {Swinarton}, \citenamefont {Szava}, \citenamefont {Tan},
  \citenamefont {Tan}, \citenamefont {Vaidya}, \citenamefont {Vernon},
  \citenamefont {Zabaneh},\ and\ \citenamefont {Zhang}}]{Arrazola21}%
  \BibitemOpen
  \bibfield  {author} {\bibinfo {author} {\bibfnamefont {J.}~\bibnamefont
  {Arrazola}}, \bibinfo {author} {\bibfnamefont {V.}~\bibnamefont {Bergholm}},
  \bibinfo {author} {\bibfnamefont {K.}~\bibnamefont {Bradler}}, \bibinfo
  {author} {\bibfnamefont {T.}~\bibnamefont {Bromley}}, \bibinfo {author}
  {\bibfnamefont {M.}~\bibnamefont {Collins}}, \bibinfo {author} {\bibfnamefont
  {I.}~\bibnamefont {Dhand}}, \bibinfo {author} {\bibfnamefont
  {A.}~\bibnamefont {Fumagalli}}, \bibinfo {author} {\bibfnamefont
  {T.}~\bibnamefont {Gerrits}}, \bibinfo {author} {\bibfnamefont
  {A.}~\bibnamefont {Goussev}}, \bibinfo {author} {\bibfnamefont
  {L.}~\bibnamefont {Helt}}, \bibinfo {author} {\bibfnamefont {J.}~\bibnamefont
  {Hundal}}, \bibinfo {author} {\bibfnamefont {T.}~\bibnamefont {Isacsson}},
  \bibinfo {author} {\bibfnamefont {R.}~\bibnamefont {Israel}}, \bibinfo
  {author} {\bibfnamefont {J.}~\bibnamefont {Izaac}}, \bibinfo {author}
  {\bibfnamefont {S.}~\bibnamefont {Jahangiri}}, \bibinfo {author}
  {\bibfnamefont {R.}~\bibnamefont {Janik}}, \bibinfo {author} {\bibfnamefont
  {N.}~\bibnamefont {Killoran}}, \bibinfo {author} {\bibfnamefont {S.~P.}\
  \bibnamefont {Kumar}}, \bibinfo {author} {\bibfnamefont {J.}~\bibnamefont
  {Lavoie}}, \bibinfo {author} {\bibfnamefont {A.~E.}\ \bibnamefont {Lita}},
  \bibinfo {author} {\bibfnamefont {D.~H.}\ \bibnamefont {Mahler}}, \bibinfo
  {author} {\bibfnamefont {M.}~\bibnamefont {Menotti}}, \bibinfo {author}
  {\bibfnamefont {B.}~\bibnamefont {Morrison}}, \bibinfo {author}
  {\bibfnamefont {S.~W.}\ \bibnamefont {Nam}}, \bibinfo {author} {\bibfnamefont
  {L.}~\bibnamefont {Neuhaus}}, \bibinfo {author} {\bibfnamefont {H.~Y.}\
  \bibnamefont {Qi}}, \bibinfo {author} {\bibfnamefont {N.}~\bibnamefont
  {Quesada}}, \bibinfo {author} {\bibfnamefont {A.}~\bibnamefont {Repingon}},
  \bibinfo {author} {\bibfnamefont {K.~K.}\ \bibnamefont {Sabapathy}}, \bibinfo
  {author} {\bibfnamefont {M.}~\bibnamefont {Schuld}}, \bibinfo {author}
  {\bibfnamefont {D.}~\bibnamefont {Su}}, \bibinfo {author} {\bibfnamefont
  {J.}~\bibnamefont {Swinarton}}, \bibinfo {author} {\bibfnamefont
  {A.}~\bibnamefont {Szava}}, \bibinfo {author} {\bibfnamefont
  {K.}~\bibnamefont {Tan}}, \bibinfo {author} {\bibfnamefont {P.}~\bibnamefont
  {Tan}}, \bibinfo {author} {\bibfnamefont {V.~D.}\ \bibnamefont {Vaidya}},
  \bibinfo {author} {\bibfnamefont {Z.}~\bibnamefont {Vernon}}, \bibinfo
  {author} {\bibfnamefont {Z.}~\bibnamefont {Zabaneh}},\ and\ \bibinfo {author}
  {\bibfnamefont {Y.}~\bibnamefont {Zhang}},\ }\href
  {https://doi.org/10.1038/s41586-021-03202-1} {\bibfield  {journal} {\bibinfo
  {journal} {Nature}\ }\textbf {\bibinfo {volume} {591}},\ \bibinfo {pages}
  {54} (\bibinfo {year} {2021})}\BibitemShut {NoStop}%
\bibitem [{\citenamefont {Neville}\ \emph {et~al.}(2017)\citenamefont
  {Neville}, \citenamefont {Sparrow}, \citenamefont {Clifford}, \citenamefont
  {Johnston}, \citenamefont {Birchall}, \citenamefont {Montanaro},\ and\
  \citenamefont {Laing}}]{Neville17}%
  \BibitemOpen
  \bibfield  {author} {\bibinfo {author} {\bibfnamefont {A.}~\bibnamefont
  {Neville}}, \bibinfo {author} {\bibfnamefont {C.}~\bibnamefont {Sparrow}},
  \bibinfo {author} {\bibfnamefont {R.}~\bibnamefont {Clifford}}, \bibinfo
  {author} {\bibfnamefont {E.}~\bibnamefont {Johnston}}, \bibinfo {author}
  {\bibfnamefont {P.~M.}\ \bibnamefont {Birchall}}, \bibinfo {author}
  {\bibfnamefont {A.}~\bibnamefont {Montanaro}},\ and\ \bibinfo {author}
  {\bibfnamefont {A.}~\bibnamefont {Laing}},\ }\href
  {https://doi.org/10.1038/nphys4270} {\bibfield  {journal} {\bibinfo
  {journal} {Nat. Phys.}\ }\textbf {\bibinfo {volume} {13}},\ \bibinfo {pages}
  {1153} (\bibinfo {year} {2017})}\BibitemShut {NoStop}%
\bibitem [{\citenamefont {Clifford}\ and\ \citenamefont
  {Clifford}(2018)}]{Clifford18}%
  \BibitemOpen
  \bibfield  {author} {\bibinfo {author} {\bibfnamefont {P.}~\bibnamefont
  {Clifford}}\ and\ \bibinfo {author} {\bibfnamefont {R.}~\bibnamefont
  {Clifford}},\ }in\ \href {https://doi.org/10.1137/1.9781611975031.10} {\emph
  {\bibinfo {booktitle} {SODA '18: Proc. 29\textsuperscript{th} ACM-SIAM Symp.
  on Discrete Algorithms}}}\ (\bibinfo {year} {2018})\ pp.\ \bibinfo {pages}
  {146--155}\BibitemShut {NoStop}%
\bibitem [{\citenamefont {Wu}\ \emph {et~al.}(2018)\citenamefont {Wu},
  \citenamefont {Liu}, \citenamefont {Zhang}, \citenamefont {Jin},
  \citenamefont {Wang}, \citenamefont {Wang},\ and\ \citenamefont
  {Yang}}]{Wu16}%
  \BibitemOpen
  \bibfield  {author} {\bibinfo {author} {\bibfnamefont {J.}~\bibnamefont
  {Wu}}, \bibinfo {author} {\bibfnamefont {Y.}~\bibnamefont {Liu}}, \bibinfo
  {author} {\bibfnamefont {B.}~\bibnamefont {Zhang}}, \bibinfo {author}
  {\bibfnamefont {X.}~\bibnamefont {Jin}}, \bibinfo {author} {\bibfnamefont
  {Y.}~\bibnamefont {Wang}}, \bibinfo {author} {\bibfnamefont {H.}~\bibnamefont
  {Wang}},\ and\ \bibinfo {author} {\bibfnamefont {X.}~\bibnamefont {Yang}},\
  }\href {https://doi.org/10.1093/nsr/nwy079} {\bibfield  {journal} {\bibinfo
  {journal} {Nat. Sci. Rev.}\ }\textbf {\bibinfo {volume} {5}},\ \bibinfo
  {pages} {715} (\bibinfo {year} {2018})}\BibitemShut {NoStop}%
\bibitem [{\citenamefont {Gupt}\ \emph {et~al.}(2020)\citenamefont {Gupt},
  \citenamefont {Arrazola}, \citenamefont {Quesada},\ and\ \citenamefont
  {Bromley}}]{Gupt18}%
  \BibitemOpen
  \bibfield  {author} {\bibinfo {author} {\bibfnamefont {B.}~\bibnamefont
  {Gupt}}, \bibinfo {author} {\bibfnamefont {J.~M.}\ \bibnamefont {Arrazola}},
  \bibinfo {author} {\bibfnamefont {N.}~\bibnamefont {Quesada}},\ and\ \bibinfo
  {author} {\bibfnamefont {T.~R.}\ \bibnamefont {Bromley}},\ }\href
  {https://doi.org/10.1007/s11128-020-02713-6} {\bibfield  {journal} {\bibinfo
  {journal} {Quantum Info. Proc.}\ }\textbf {\bibinfo {volume} {19}},\ \bibinfo
  {pages} {249} (\bibinfo {year} {2020})}\BibitemShut {NoStop}%
\bibitem [{\citenamefont {Lundow}\ and\ \citenamefont
  {Markstr\"{o}m}()}]{Lundow19}%
  \BibitemOpen
  \bibfield  {author} {\bibinfo {author} {\bibfnamefont {P.~H.}\ \bibnamefont
  {Lundow}}\ and\ \bibinfo {author} {\bibfnamefont {K.}~\bibnamefont
  {Markstr\"{o}m}},\ }\bibinfo {note} {preprint at
  arXiv:1904.06229}\BibitemShut {NoStop}%
\bibitem [{\citenamefont {Clifford}\ and\ \citenamefont
  {Clifford}()}]{Clifford2020Faster}%
  \BibitemOpen
  \bibfield  {author} {\bibinfo {author} {\bibfnamefont {P.}~\bibnamefont
  {Clifford}}\ and\ \bibinfo {author} {\bibfnamefont {R.}~\bibnamefont
  {Clifford}},\ }\bibinfo {note} {preprint at arXiv:2005.04214}\BibitemShut
  {NoStop}%
\bibitem [{\citenamefont {Aaronson}\ and\ \citenamefont
  {Brod}(2015)}]{Aaronson15}%
  \BibitemOpen
  \bibfield  {author} {\bibinfo {author} {\bibfnamefont {S.}~\bibnamefont
  {Aaronson}}\ and\ \bibinfo {author} {\bibfnamefont {D.~J.}\ \bibnamefont
  {Brod}},\ }\href {https://doi.org/10.1103/PhysRevA.93.012335} {\bibfield
  {journal} {\bibinfo  {journal} {Phys. Rev. A}\ }\textbf {\bibinfo {volume}
  {93}},\ \bibinfo {pages} {012335} (\bibinfo {year} {2015})}\BibitemShut
  {NoStop}%
\bibitem [{\citenamefont {Arkhipov}(2015)}]{Arkhipov15}%
  \BibitemOpen
  \bibfield  {author} {\bibinfo {author} {\bibfnamefont {A.}~\bibnamefont
  {Arkhipov}},\ }\href {https://doi.org/10.1103/PhysRevA.92.062326} {\bibfield
  {journal} {\bibinfo  {journal} {Phys. Rev. A}\ }\textbf {\bibinfo {volume}
  {92}},\ \bibinfo {pages} {062326} (\bibinfo {year} {2015})}\BibitemShut
  {NoStop}%
\bibitem [{\citenamefont {Leverrier}\ and\ \citenamefont
  {Garcia-Patron}(2015)}]{Leverrier15}%
  \BibitemOpen
  \bibfield  {author} {\bibinfo {author} {\bibfnamefont {A.}~\bibnamefont
  {Leverrier}}\ and\ \bibinfo {author} {\bibfnamefont {R.}~\bibnamefont
  {Garcia-Patron}},\ }\href {https://dl.acm.org/citation.cfm?id=2871409}
  {\bibfield  {journal} {\bibinfo  {journal} {Quantum Inf. Comput.}\ }\textbf
  {\bibinfo {volume} {15}},\ \bibinfo {pages} {0489} (\bibinfo {year}
  {2015})}\BibitemShut {NoStop}%
\bibitem [{\citenamefont {Oszmaniec}\ and\ \citenamefont
  {Brod}(2018)}]{Oszmaniec18}%
  \BibitemOpen
  \bibfield  {author} {\bibinfo {author} {\bibfnamefont {M.}~\bibnamefont
  {Oszmaniec}}\ and\ \bibinfo {author} {\bibfnamefont {D.~J.}\ \bibnamefont
  {Brod}},\ }\href {https://doi.org/10.1088/1367-2630/aadfa8} {\bibfield
  {journal} {\bibinfo  {journal} {New Journal of Physics}\ }\textbf {\bibinfo
  {volume} {20}},\ \bibinfo {pages} {092002} (\bibinfo {year}
  {2018})}\BibitemShut {NoStop}%
\bibitem [{\citenamefont {Garc\'{\i}a-Patr\'{o}n}\ \emph
  {et~al.}(2019)\citenamefont {Garc\'{\i}a-Patr\'{o}n}, \citenamefont
  {Renema},\ and\ \citenamefont {Shchesnovich}}]{Garcia-Patron17}%
  \BibitemOpen
  \bibfield  {author} {\bibinfo {author} {\bibfnamefont {R.}~\bibnamefont
  {Garc\'{\i}a-Patr\'{o}n}}, \bibinfo {author} {\bibfnamefont {J.~J.}\
  \bibnamefont {Renema}},\ and\ \bibinfo {author} {\bibfnamefont
  {V.}~\bibnamefont {Shchesnovich}},\ }\href
  {https://doi.org/10.22331/q-2019-08-05-169} {\bibfield  {journal} {\bibinfo
  {journal} {Quantum}\ }\textbf {\bibinfo {volume} {3}},\ \bibinfo {pages}
  {169} (\bibinfo {year} {2019})}\BibitemShut {NoStop}%
\bibitem [{\citenamefont {Renema}\ \emph {et~al.}(2018)\citenamefont {Renema},
  \citenamefont {Menssen}, \citenamefont {Clements}, \citenamefont {Triginer},
  \citenamefont {Kolthammer},\ and\ \citenamefont {Walmsley}}]{Renema17}%
  \BibitemOpen
  \bibfield  {author} {\bibinfo {author} {\bibfnamefont {J.~J.}\ \bibnamefont
  {Renema}}, \bibinfo {author} {\bibfnamefont {A.}~\bibnamefont {Menssen}},
  \bibinfo {author} {\bibfnamefont {W.~R.}\ \bibnamefont {Clements}}, \bibinfo
  {author} {\bibfnamefont {G.}~\bibnamefont {Triginer}}, \bibinfo {author}
  {\bibfnamefont {W.~S.}\ \bibnamefont {Kolthammer}},\ and\ \bibinfo {author}
  {\bibfnamefont {I.~A.}\ \bibnamefont {Walmsley}},\ }\href
  {https://doi.org/10.1103/PhysRevLett.120.220502} {\bibfield  {journal}
  {\bibinfo  {journal} {Phys. Rev. Lett.}\ }\textbf {\bibinfo {volume} {120}},\
  \bibinfo {pages} {220502} (\bibinfo {year} {2018})}\BibitemShut {NoStop}%
\bibitem [{\citenamefont {Renema}\ \emph {et~al.}()\citenamefont {Renema},
  \citenamefont {Shchesnovich},\ and\ \citenamefont
  {Garc\'{\i}a-Patr\'{o}n}}]{Renema18}%
  \BibitemOpen
  \bibfield  {author} {\bibinfo {author} {\bibfnamefont {J.~J.}\ \bibnamefont
  {Renema}}, \bibinfo {author} {\bibfnamefont {V.}~\bibnamefont
  {Shchesnovich}},\ and\ \bibinfo {author} {\bibfnamefont {R.}~\bibnamefont
  {Garc\'{\i}a-Patr\'{o}n}},\ }\bibinfo {note} {preprint at
  arXiv:1809.01953}\BibitemShut {NoStop}%
\bibitem [{\citenamefont {Qi}\ \emph {et~al.}(2020)\citenamefont {Qi},
  \citenamefont {Brod}, \citenamefont {Quesada},\ and\ \citenamefont
  {Garcia-Patron}}]{Qi19}%
  \BibitemOpen
  \bibfield  {author} {\bibinfo {author} {\bibfnamefont {H.}~\bibnamefont
  {Qi}}, \bibinfo {author} {\bibfnamefont {D.~J.}\ \bibnamefont {Brod}},
  \bibinfo {author} {\bibfnamefont {N.}~\bibnamefont {Quesada}},\ and\ \bibinfo
  {author} {\bibfnamefont {R.}~\bibnamefont {Garcia-Patron}},\ }\href
  {https://doi.org/10.1103/PhysRevLett.124.100502} {\bibfield  {journal}
  {\bibinfo  {journal} {Phys. Rev. Lett.}\ }\textbf {\bibinfo {volume} {124}},\
  \bibinfo {pages} {100502} (\bibinfo {year} {2020})}\BibitemShut {NoStop}%
\bibitem [{\citenamefont {Michler}\ \emph {et~al.}(2000)\citenamefont
  {Michler}, \citenamefont {Kiraz}, \citenamefont {Becher}, \citenamefont
  {Schoenfeld}, \citenamefont {Petroff}, \citenamefont {Zhang}, \citenamefont
  {Hu},\ and\ \citenamefont {Imamoglu}}]{Michler2282}%
  \BibitemOpen
  \bibfield  {author} {\bibinfo {author} {\bibfnamefont {P.}~\bibnamefont
  {Michler}}, \bibinfo {author} {\bibfnamefont {A.}~\bibnamefont {Kiraz}},
  \bibinfo {author} {\bibfnamefont {C.}~\bibnamefont {Becher}}, \bibinfo
  {author} {\bibfnamefont {W.~V.}\ \bibnamefont {Schoenfeld}}, \bibinfo
  {author} {\bibfnamefont {P.~M.}\ \bibnamefont {Petroff}}, \bibinfo {author}
  {\bibfnamefont {L.}~\bibnamefont {Zhang}}, \bibinfo {author} {\bibfnamefont
  {E.}~\bibnamefont {Hu}},\ and\ \bibinfo {author} {\bibfnamefont
  {A.}~\bibnamefont {Imamoglu}},\ }\href
  {https://doi.org/10.1126/science.290.5500.2282} {\bibfield  {journal}
  {\bibinfo  {journal} {Science}\ }\textbf {\bibinfo {volume} {290}},\ \bibinfo
  {pages} {2282} (\bibinfo {year} {2000})}\BibitemShut {NoStop}%
\bibitem [{\citenamefont {Ding}\ \emph {et~al.}(2016)\citenamefont {Ding},
  \citenamefont {He}, \citenamefont {Duan}, \citenamefont {Gregersen},
  \citenamefont {Chen}, \citenamefont {Unsleber}, \citenamefont {Maier},
  \citenamefont {Schneider}, \citenamefont {Kamp}, \citenamefont {H\"ofling},
  \citenamefont {Lu},\ and\ \citenamefont {Pan}}]{Ding_quantum_dot}%
  \BibitemOpen
  \bibfield  {author} {\bibinfo {author} {\bibfnamefont {X.}~\bibnamefont
  {Ding}}, \bibinfo {author} {\bibfnamefont {Y.}~\bibnamefont {He}}, \bibinfo
  {author} {\bibfnamefont {Z.-C.}\ \bibnamefont {Duan}}, \bibinfo {author}
  {\bibfnamefont {N.}~\bibnamefont {Gregersen}}, \bibinfo {author}
  {\bibfnamefont {M.-C.}\ \bibnamefont {Chen}}, \bibinfo {author}
  {\bibfnamefont {S.}~\bibnamefont {Unsleber}}, \bibinfo {author}
  {\bibfnamefont {S.}~\bibnamefont {Maier}}, \bibinfo {author} {\bibfnamefont
  {C.}~\bibnamefont {Schneider}}, \bibinfo {author} {\bibfnamefont
  {M.}~\bibnamefont {Kamp}}, \bibinfo {author} {\bibfnamefont {S.}~\bibnamefont
  {H\"ofling}}, \bibinfo {author} {\bibfnamefont {C.-Y.}\ \bibnamefont {Lu}},\
  and\ \bibinfo {author} {\bibfnamefont {J.-W.}\ \bibnamefont {Pan}},\ }\href
  {https://doi.org/10.1103/PhysRevLett.116.020401} {\bibfield  {journal}
  {\bibinfo  {journal} {Phys. Rev. Lett.}\ }\textbf {\bibinfo {volume} {116}},\
  \bibinfo {pages} {020401} (\bibinfo {year} {2016})}\BibitemShut {NoStop}%
\bibitem [{\citenamefont {Somaschi}\ \emph {et~al.}(2016)\citenamefont
  {Somaschi}, \citenamefont {Giesz}, \citenamefont {De~Santis}, \citenamefont
  {Loredo}, \citenamefont {Hornecker}, \citenamefont {Portalupi}, \citenamefont
  {Grange}, \citenamefont {Anton}, \citenamefont {Demory}, \citenamefont
  {Gomez}, \citenamefont {Sagnes}, \citenamefont {Lanzillotti-Kimura},
  \citenamefont {Lemaitre}, \citenamefont {Auffeves}, \citenamefont {White},
  \citenamefont {Lanco},\ and\ \citenamefont {Senellart}}]{Somaschi2016}%
  \BibitemOpen
  \bibfield  {author} {\bibinfo {author} {\bibfnamefont {N.}~\bibnamefont
  {Somaschi}}, \bibinfo {author} {\bibfnamefont {V.}~\bibnamefont {Giesz}},
  \bibinfo {author} {\bibfnamefont {L.}~\bibnamefont {De~Santis}}, \bibinfo
  {author} {\bibfnamefont {M.~P.}\ \bibnamefont {Loredo}, \bibfnamefont
  {J.~C.~Almeida}}, \bibinfo {author} {\bibfnamefont {G.}~\bibnamefont
  {Hornecker}}, \bibinfo {author} {\bibfnamefont {S.~L.}\ \bibnamefont
  {Portalupi}}, \bibinfo {author} {\bibfnamefont {T.}~\bibnamefont {Grange}},
  \bibinfo {author} {\bibfnamefont {C.}~\bibnamefont {Anton}}, \bibinfo
  {author} {\bibfnamefont {J.}~\bibnamefont {Demory}}, \bibinfo {author}
  {\bibfnamefont {C.}~\bibnamefont {Gomez}}, \bibinfo {author} {\bibfnamefont
  {I.}~\bibnamefont {Sagnes}}, \bibinfo {author} {\bibfnamefont {N.~D.}\
  \bibnamefont {Lanzillotti-Kimura}}, \bibinfo {author} {\bibfnamefont
  {A.}~\bibnamefont {Lemaitre}}, \bibinfo {author} {\bibfnamefont
  {A.}~\bibnamefont {Auffeves}}, \bibinfo {author} {\bibfnamefont {A.~G.}\
  \bibnamefont {White}}, \bibinfo {author} {\bibfnamefont {L.}~\bibnamefont
  {Lanco}},\ and\ \bibinfo {author} {\bibfnamefont {P.}~\bibnamefont
  {Senellart}},\ }\href {https://doi.org/10.1038/nphoton.2016.23} {\bibfield
  {journal} {\bibinfo  {journal} {Nat. Photon.}\ }\textbf {\bibinfo {volume}
  {10}},\ \bibinfo {pages} {340} (\bibinfo {year} {2016})}\BibitemShut
  {NoStop}%
\bibitem [{\citenamefont {Michler}(2017)}]{Michler17}%
  \BibitemOpen
  \bibinfo {editor} {\bibfnamefont {P.}~\bibnamefont {Michler}},\ ed.,\
  \href@noop {} {\emph {\bibinfo {title} {Quantum Dots for Quantum Information
  Technologies}}},\ Nano-Optics and Nanophotonics\ (\bibinfo  {publisher}
  {Springer, Cham},\ \bibinfo {year} {2017})\BibitemShut {NoStop}%
\bibitem [{\citenamefont {Huber}\ \emph {et~al.}(2018)\citenamefont {Huber},
  \citenamefont {Reindl}, \citenamefont {Aberl}, \citenamefont {Rastelli},\
  and\ \citenamefont {Trotta}}]{Huber18}%
  \BibitemOpen
  \bibfield  {author} {\bibinfo {author} {\bibfnamefont {D.}~\bibnamefont
  {Huber}}, \bibinfo {author} {\bibfnamefont {M.}~\bibnamefont {Reindl}},
  \bibinfo {author} {\bibfnamefont {J.}~\bibnamefont {Aberl}}, \bibinfo
  {author} {\bibfnamefont {A.}~\bibnamefont {Rastelli}},\ and\ \bibinfo
  {author} {\bibfnamefont {R.}~\bibnamefont {Trotta}},\ }\href
  {https://doi.org/10.1088/2040-8986/aac4c4} {\bibfield  {journal} {\bibinfo
  {journal} {J. Opt.}\ }\textbf {\bibinfo {volume} {20}},\ \bibinfo {pages}
  {073002} (\bibinfo {year} {2018})}\BibitemShut {NoStop}%
\bibitem [{\citenamefont {{Basso Basset}}\ \emph {et~al.}(2019)\citenamefont
  {{Basso Basset}}, \citenamefont {Rota}, \citenamefont {Schimpf},
  \citenamefont {Tedeschi}, \citenamefont {Zeuner}, \citenamefont {da~Silva},
  \citenamefont {Reindl}, \citenamefont {Zwiller}, \citenamefont {J\"{o}ns},
  \citenamefont {Rastelli},\ and\ \citenamefont {Trotta}}]{Basso19}%
  \BibitemOpen
  \bibfield  {author} {\bibinfo {author} {\bibfnamefont {F.}~\bibnamefont
  {{Basso Basset}}}, \bibinfo {author} {\bibfnamefont {M.~B.}\ \bibnamefont
  {Rota}}, \bibinfo {author} {\bibfnamefont {C.}~\bibnamefont {Schimpf}},
  \bibinfo {author} {\bibfnamefont {D.}~\bibnamefont {Tedeschi}}, \bibinfo
  {author} {\bibfnamefont {K.~D.}\ \bibnamefont {Zeuner}}, \bibinfo {author}
  {\bibfnamefont {S.~F.~C.}\ \bibnamefont {da~Silva}}, \bibinfo {author}
  {\bibfnamefont {M.}~\bibnamefont {Reindl}}, \bibinfo {author} {\bibfnamefont
  {V.}~\bibnamefont {Zwiller}}, \bibinfo {author} {\bibfnamefont {K.~D.}\
  \bibnamefont {J\"{o}ns}}, \bibinfo {author} {\bibfnamefont {A.}~\bibnamefont
  {Rastelli}},\ and\ \bibinfo {author} {\bibfnamefont {R.}~\bibnamefont
  {Trotta}},\ }\href {https://doi.org/10.1103/PhysRevLett.123.160501}
  {\bibfield  {journal} {\bibinfo  {journal} {Phys. Rev. Lett.}\ }\textbf
  {\bibinfo {volume} {123}},\ \bibinfo {pages} {160501} (\bibinfo {year}
  {2019})}\BibitemShut {NoStop}%
\bibitem [{\citenamefont {Santis}\ \emph {et~al.}(2017)\citenamefont {Santis},
  \citenamefont {Anton}, \citenamefont {Reznychenko}, \citenamefont {Somaschi},
  \citenamefont {Coppola}, \citenamefont {Senellart}, \citenamefont {Gomez},
  \citenamefont {Lemaitre}, \citenamefont {Sagnes}, \citenamefont {White},
  \citenamefont {Lanco}, \citenamefont {Auffeves},\ and\ \citenamefont
  {Senellart}}]{DeSantis17}%
  \BibitemOpen
  \bibfield  {author} {\bibinfo {author} {\bibfnamefont {L.~D.}\ \bibnamefont
  {Santis}}, \bibinfo {author} {\bibfnamefont {C.}~\bibnamefont {Anton}},
  \bibinfo {author} {\bibfnamefont {B.}~\bibnamefont {Reznychenko}}, \bibinfo
  {author} {\bibfnamefont {N.}~\bibnamefont {Somaschi}}, \bibinfo {author}
  {\bibfnamefont {G.}~\bibnamefont {Coppola}}, \bibinfo {author} {\bibfnamefont
  {J.}~\bibnamefont {Senellart}}, \bibinfo {author} {\bibfnamefont
  {C.}~\bibnamefont {Gomez}}, \bibinfo {author} {\bibfnamefont
  {A.}~\bibnamefont {Lemaitre}}, \bibinfo {author} {\bibfnamefont
  {I.}~\bibnamefont {Sagnes}}, \bibinfo {author} {\bibfnamefont {A.~G.}\
  \bibnamefont {White}}, \bibinfo {author} {\bibfnamefont {L.}~\bibnamefont
  {Lanco}}, \bibinfo {author} {\bibfnamefont {A.}~\bibnamefont {Auffeves}},\
  and\ \bibinfo {author} {\bibfnamefont {P.}~\bibnamefont {Senellart}},\ }\href
  {https://doi.org/10.1038/nnano.2017.85} {\bibfield  {journal} {\bibinfo
  {journal} {Nat. Nanotechnol.}\ }\textbf {\bibinfo {volume} {12}},\ \bibinfo
  {pages} {663} (\bibinfo {year} {2017})}\BibitemShut {NoStop}%
\bibitem [{\citenamefont {Valiant}(1979)}]{Valiant79}%
  \BibitemOpen
  \bibfield  {author} {\bibinfo {author} {\bibfnamefont {L.}~\bibnamefont
  {Valiant}},\ }\href {https://doi.org/10.1016/0304-3975(79)90044-6} {\bibfield
   {journal} {\bibinfo  {journal} {Theor. Comput. Sci.}\ }\textbf {\bibinfo
  {volume} {8}},\ \bibinfo {pages} {189} (\bibinfo {year} {1979})}\BibitemShut
  {NoStop}%
\bibitem [{\citenamefont {Aaronson}(2011)}]{Aaronson11}%
  \BibitemOpen
  \bibfield  {author} {\bibinfo {author} {\bibfnamefont {S.}~\bibnamefont
  {Aaronson}},\ }\href {https://doi.org/10.1098/rspa.2011.0232} {\bibfield
  {journal} {\bibinfo  {journal} {Proc. Roy. Soc. A}\ }\textbf {\bibinfo
  {volume} {467}},\ \bibinfo {pages} {3393} (\bibinfo {year}
  {2011})}\BibitemShut {NoStop}%
\bibitem [{Sup()}]{SuppMat}%
  \BibitemOpen
  \href@noop {} {}\bibinfo {note} {See Supplemental Material for more details
  on the theoretical derivations and on the performed numerical
  simulations.}\BibitemShut {Stop}%
\bibitem [{\citenamefont {Scheel}\ \emph {et~al.}(2003)\citenamefont {Scheel},
  \citenamefont {Nemoto}, \citenamefont {Munro},\ and\ \citenamefont
  {Knight}}]{Scheel03}%
  \BibitemOpen
  \bibfield  {author} {\bibinfo {author} {\bibfnamefont {S.}~\bibnamefont
  {Scheel}}, \bibinfo {author} {\bibfnamefont {K.}~\bibnamefont {Nemoto}},
  \bibinfo {author} {\bibfnamefont {W.~J.}\ \bibnamefont {Munro}},\ and\
  \bibinfo {author} {\bibfnamefont {P.~L.}\ \bibnamefont {Knight}},\ }\href
  {https://doi.org/10.1103/PhysRevA.68.032310} {\bibfield  {journal} {\bibinfo
  {journal} {Phys. Rev. A}\ }\textbf {\bibinfo {volume} {68}},\ \bibinfo
  {pages} {032310} (\bibinfo {year} {2003})}\BibitemShut {NoStop}%
\bibitem [{foo()}]{footnote_measInduced}%
  \BibitemOpen
  \href@noop {} {}\bibinfo {note} {The effective evolution induced by this
  method must be some degree-$l$ polynomial in $\hat{n}$, though it remains an
  open question whether \emph{any} degree-$l$ polynomial can be implemented
  using only $l$ auxiliary photons. This can be shown to hold for $l=2$.
  Suppose we have some input $\vert \chi_{\mathrm{in}} \rangle = c_0 \vert 0
  \rangle + c_1 \vert 1 \rangle + c_2 \vert 2 \rangle$. All $c_i$ must be
  phases. The phase $c_0$ can be fixed by an overall global phase, and the
  phase $c_1$ can be fixed by applying a (linear) phase shifter. As shown in
  \cite{SuppMat}, there is a gadget that implements any chosen value of $c_2$.
  Combining these facts, it follows that any nonlinear phase acting on at most
  2-photon states can be simulated by using 2 auxiliary photons.}\BibitemShut
  {Stop}%
\bibitem [{\citenamefont {Knill}\ \emph {et~al.}(2001)\citenamefont {Knill},
  \citenamefont {Laflamme},\ and\ \citenamefont {Milburn}}]{KLM01}%
  \BibitemOpen
  \bibfield  {author} {\bibinfo {author} {\bibfnamefont {E.}~\bibnamefont
  {Knill}}, \bibinfo {author} {\bibfnamefont {R.}~\bibnamefont {Laflamme}},\
  and\ \bibinfo {author} {\bibfnamefont {G.~J.}\ \bibnamefont {Milburn}},\
  }\href@noop {} {\bibfield  {journal} {\bibinfo  {journal} {Nature}\ }\textbf
  {\bibinfo {volume} {409}},\ \bibinfo {pages} {46} (\bibinfo {year}
  {2001})}\BibitemShut {NoStop}%
\bibitem [{\citenamefont {Uskov}\ \emph {et~al.}(2009)\citenamefont {Uskov},
  \citenamefont {Kaplan}, \citenamefont {Smith}, \citenamefont {Huver},\ and\
  \citenamefont {Dowling}}]{Uskov09}%
  \BibitemOpen
  \bibfield  {author} {\bibinfo {author} {\bibfnamefont {D.~B.}\ \bibnamefont
  {Uskov}}, \bibinfo {author} {\bibfnamefont {L.}~\bibnamefont {Kaplan}},
  \bibinfo {author} {\bibfnamefont {A.~M.}\ \bibnamefont {Smith}}, \bibinfo
  {author} {\bibfnamefont {S.~D.}\ \bibnamefont {Huver}},\ and\ \bibinfo
  {author} {\bibfnamefont {J.~P.}\ \bibnamefont {Dowling}},\ }\href
  {https://doi.org/10.1103/PhysRevA.79.042326} {\bibfield  {journal} {\bibinfo
  {journal} {Phys. Rev. A}\ }\textbf {\bibinfo {volume} {79}},\ \bibinfo
  {pages} {042326} (\bibinfo {year} {2009})}\BibitemShut {NoStop}%
\bibitem [{\citenamefont {Uskov}\ \emph {et~al.}(2010)\citenamefont {Uskov},
  \citenamefont {Smith},\ and\ \citenamefont {Kaplan}}]{Uskov10}%
  \BibitemOpen
  \bibfield  {author} {\bibinfo {author} {\bibfnamefont {D.~B.}\ \bibnamefont
  {Uskov}}, \bibinfo {author} {\bibfnamefont {A.~M.}\ \bibnamefont {Smith}},\
  and\ \bibinfo {author} {\bibfnamefont {L.}~\bibnamefont {Kaplan}},\ }\href
  {https://doi.org/10.1103/PhysRevA.81.012303} {\bibfield  {journal} {\bibinfo
  {journal} {Phys. Rev. A}\ }\textbf {\bibinfo {volume} {81}},\ \bibinfo
  {pages} {012303} (\bibinfo {year} {2010})}\BibitemShut {NoStop}%
\bibitem [{\citenamefont {Oszmaniec}\ and\ \citenamefont
  {Zimbor\'as}(2017)}]{ZimborasO17}%
  \BibitemOpen
  \bibfield  {author} {\bibinfo {author} {\bibfnamefont {M.}~\bibnamefont
  {Oszmaniec}}\ and\ \bibinfo {author} {\bibfnamefont {Z.}~\bibnamefont
  {Zimbor\'as}},\ }\href {https://doi.org/10.1103/PhysRevLett.119.220502}
  {\bibfield  {journal} {\bibinfo  {journal} {Phys. Rev. Lett.}\ }\textbf
  {\bibinfo {volume} {119}},\ \bibinfo {pages} {220502} (\bibinfo {year}
  {2017})}\BibitemShut {NoStop}%
\bibitem [{\citenamefont {Arkhipov}\ and\ \citenamefont
  {Kuperberg}(2012)}]{Arkhipov12}%
  \BibitemOpen
  \bibfield  {author} {\bibinfo {author} {\bibfnamefont {A.}~\bibnamefont
  {Arkhipov}}\ and\ \bibinfo {author} {\bibfnamefont {G.}~\bibnamefont
  {Kuperberg}},\ }\href@noop {} {\bibfield  {journal} {\bibinfo  {journal}
  {Geometry and Topology Monographs}\ }\textbf {\bibinfo {volume} {18}},\
  \bibinfo {pages} {1} (\bibinfo {year} {2012})}\BibitemShut {NoStop}%
\bibitem [{\citenamefont {Eisert}(2005)}]{Eisert2005}%
  \BibitemOpen
  \bibfield  {author} {\bibinfo {author} {\bibfnamefont {J.}~\bibnamefont
  {Eisert}},\ }\href {https://doi.org/10.1103/PhysRevLett.95.040502} {\bibfield
   {journal} {\bibinfo  {journal} {Phys. Rev. Lett.}\ }\textbf {\bibinfo
  {volume} {95}},\ \bibinfo {pages} {040502} (\bibinfo {year}
  {2005})}\BibitemShut {NoStop}%
\end{thebibliography}
\end{document}


\title{Supplemental Materials --- Non-linear Boson Sampling}

\author{Nicol\`o Spagnolo}
\affiliation{Dipartimento di Fisica, Sapienza Universit\`{a} di Roma, Piazzale Aldo Moro 5, I-00185 Roma, Italy}
\author{Daniel J. Brod}
\affiliation{Instituto de F\'isica, Universidade Federal Fluminense, Niter\'oi, RJ, 24210-340, Brazil}
\author{Ernesto F. Galv\~ao}
\affiliation{International Iberian Nanotechnology Laboratory (INL), Ave. Mestre Jose Veiga, 4715-330, Braga, Portugal}
\affiliation{Instituto de F\'isica, Universidade Federal Fluminense, Niter\'oi, RJ, 24210-340, Brazil}
\author{Fabio Sciarrino}
\affiliation{Dipartimento di Fisica, Sapienza Universit\`{a} di Roma, Piazzale Aldo Moro 5, I-00185 Roma, Italy}

\maketitle

\section{Transition amplitudes for non-linear Boson Sampling}

Here we derive the transition amplitude for non-linear Boson Sampling, corresponding to Eq.\ (2) reported in the main text. For the case of linear dynamics, the transformation is represented by a unitary matrix $U$ which describes the evolution of creation operators according to $a_{i}^{\dag} \rightarrow \sum_{j} U_{j,i} a_{j}^{\dag}$. 
A system of $n$ indistinguishable photons evolving via a linear transformation are characterized by input-output transition amplitudes of the form:
\begin{equation}
\label{eq:trans_ampl_linear}
\mathcal{A}_{U}\big(\vert S \rangle \rightarrow \vert T \rangle\big) = \langle S \vert \varphi(U) \vert T \rangle = \frac{\mathrm{per}(U_{S,T})}{\sqrt{\prod_{i=1}^{m}s_{i}! \prod_{j=1}^{m} t_{j}!}}.
\end{equation}
Here $\vert S \rangle = \vert s_{1}, \ldots, s_{m} \rangle$ is the input state configuration and $\vert T \rangle = \vert t_{1}, \ldots, t_{m} \rangle$ is the output state, where $\{ s_{i} \}$ and $\{ t_{i} \}$ are the respective lists of mode occupation numbers. Equation \eqref{eq:trans_ampl_linear} depends on the permanent of $U_{S,T}$, which is the $n \times n$ submatrix of $U$ obtained by choosing rows and columns according to $\vert S \rangle$ and $\vert T \rangle$ as described in Ref.\ \cite{AA}. The permanent is defined as:
\begin{equation}
\mathrm{per}(A) = \sum_{\sigma \in S_{n}} \prod_{i=1}^{n} a_{i,\sigma(i)},
\end{equation}
where the sum is extended over all permutations $\sigma$ in $S_{n}$. 

Equation \eqref{eq:trans_ampl_linear} corresponds also to writing the output state as:
\begin{equation}
\vert S \rangle \stackrel{U}{\rightarrow} \sum_{T \in \Phi_{m,n}} \gamma^{U}_{S,T} \vert T \rangle,
\end{equation}
where $\Phi_{m,n}$ is the set of tuples describing $n$ photons in $m$ modes, and $\gamma^{U}_{S,T} = \langle S \vert \varphi(U) \vert T \rangle$. Due to the linearity of the evolution, $\varphi(U)$ is an homomorphism \cite{Aaronson11}. Consider now an evolution $U$ that can be divided as the product of two matrices $U = V W$, thus corresponding to a sequence of two linear interferometers. Given the properties of $\varphi(U)$, the final output state can be written in two equivalent ways:
\begin{equation}
\vert S \rangle \stackrel{U}{\rightarrow} \sum_{T \in \Phi_{m,n}} \gamma^{U}_{S,T} \vert T \rangle = \sum_{T \in \Phi_{m,n}} \left( \sum_{R \in \Phi_{m,n}} \gamma^{W}_{S,R} \gamma^{V}_{R,T}  \right) \vert T \rangle,
\end{equation}
which corresponds to the following identity for matrix permanents when $U = V W$:
\begin{equation}
\label{eq:identity}
\mathrm{per}(U_{S,T}) = \sum_{R \in \Phi_{m,n}} \frac{\mathrm{per}(W_{S,R}) \mathrm{per}(V_{R,T})}{\prod_{i=1}^{m} r_{i}!}.
\end{equation}
This expression is the starting point of the output state expansion in the presence of non-linearities. 

Let us now consider the scenario where a non-linear layer $N$ is inserted between two linear evolutions $V$ and $W$. The non-linear evolution $N$ transforms a state $\vert R \rangle = \vert r_{1}, \ldots r_{m} \rangle$ as:
\begin{equation}
\vert R \rangle \stackrel{N}{\rightarrow}\sum_{R \in \Phi_{m,n}} \mathcal{N}_{r_{1} \ldots r_{m}}^{q_{1} \ldots q_{m}} \vert Q \rangle, 
\end{equation}
where $\vert Q \rangle = \vert q_{1}, \ldots q_{m} \rangle$. The function $\mathcal{N}_{r_{1} \ldots r_{m}}^{q_{1} \ldots q_{m}}$ represents the transition amplitude $\mathcal{A}_{N}(\vert R \rangle \rightarrow \vert Q \rangle)$ due to the non-linear evolution. It must in general satisfy appropriate constraints to represent a physical evolution. 

We now write the overall evolution of input state $\vert S \rangle = \vert s_{1}, \ldots s_{m} \rangle$ according to the full evolution $W \rightarrow N \rightarrow V$. After the first (linear) transformation, the state evolves to:
\begin{equation}
\vert S \rangle \stackrel{W}{\rightarrow} \sum_{R \in \Phi_{m,n}} \gamma^{W}_{S,R} \vert R \rangle.
\end{equation}
Then, after the non-linear term $N$, the state can be written as:
\begin{equation}
\vert S \rangle \stackrel{W,N}{\rightarrow} \sum_{R \in \Phi_{m,n}} \gamma^{W}_{S,R} \sum_{Q \in \Phi_{m,n}} \mathcal{N}_{r_{1} \ldots r_{m}}^{q_{1} \ldots q_{m}} \vert Q \rangle.
\end{equation}
Finally, after the third step corresponding to the (linear) evolution $V$ we obtain the following expression:
\begin{equation}
\vert S \rangle \stackrel{W,N,V}{\rightarrow} \sum_{R \in \Phi_{m,n}} \gamma^{W}_{S,R} \sum_{Q \in \Phi_{m,n}} \mathcal{N}_{r_{1} \ldots r_{m}}^{q_{1} \ldots q_{m}} \sum_{T \in \Phi_{m,n}} \gamma^{V}_{Q,T} \vert T \rangle.
\end{equation}
This equation can be rearranged as:
\begin{equation}
\vert S \rangle \stackrel{W,N,V}{\rightarrow} \sum_{T \in \Phi_{m,n}} \Bigg( \sum_{R \in \Phi_{m,n}}\sum_{Q \in \Phi_{m,n}} \gamma^{W}_{S,R} \mathcal{N}_{r_{1} \ldots r_{m}}^{q_{1} \ldots q_{m}} \gamma^{V}_{Q,T} \Bigg) \vert T \rangle.
\end{equation}
Hence, the transition amplitude from input state $\vert S \rangle$ to $\vert T \rangle$ reads:
\begin{equation}
\label{eq:gen_nonlin}
\mathcal{A}_{W,N,V}\big( \vert S \rangle \rightarrow \vert T \rangle \big) = \sum_{R \in \Phi_{m,n}}\sum_{Q \in \Phi_{m,n}} \frac{\mathrm{per}(W_{S,R}) \mathcal{N}_{r_{1} \ldots r_{m}}^{q_{1} \ldots q_{m}} \mathrm{per}(V_{Q,T})}{\sqrt{\prod_{i=1}^{m}s_{i}! \prod_{j=1}^{m}r_{j}! \prod_{k=1}^{m}q_{k}! \prod_{l=1}^{m}t_{l}!}},
\end{equation}
corresponding to Eq.\ (2) of main text.

\section{Cumulative Boson Sampling distributions} 

\begin{figure}[ht!]
\centering
\includegraphics[width=0.99\textwidth]{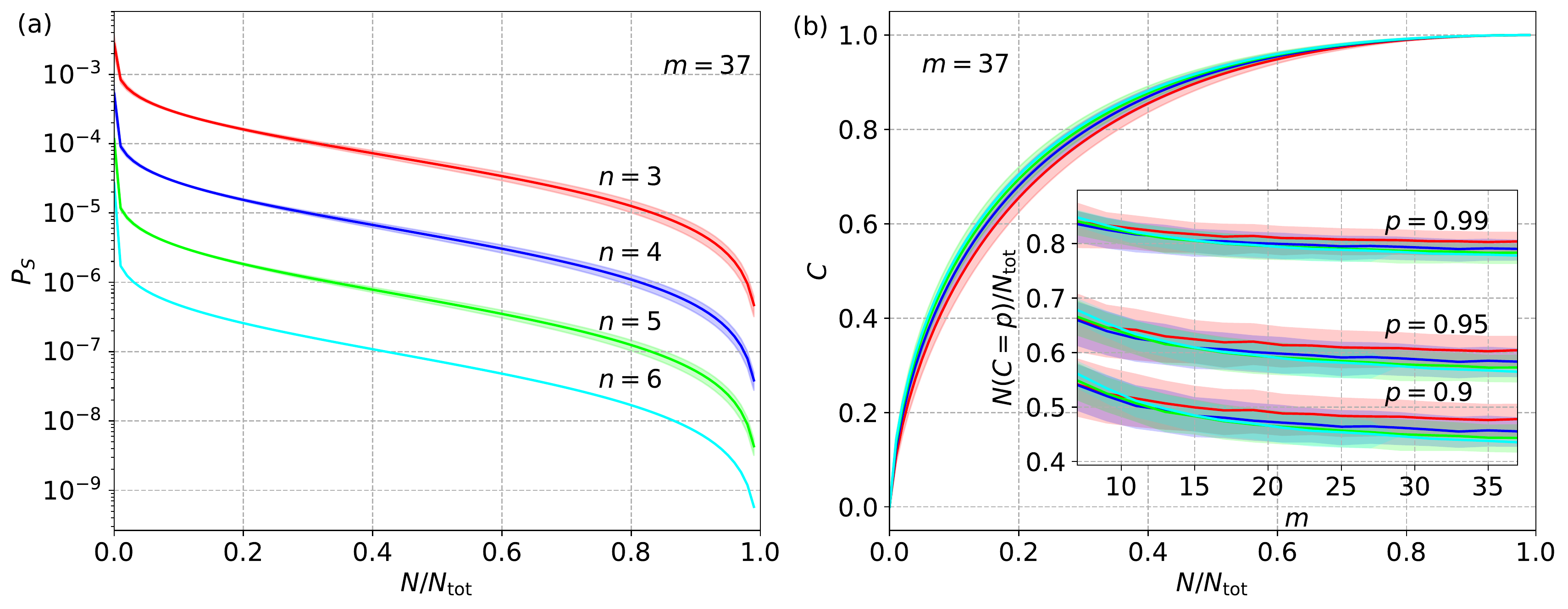}
\caption{Behavior of probability distributions as functions of the fraction $N/N_{\mathrm{tot}}$ of the output configurations. (a) Sorted distributions $P_{S}$ in decreasing probability order for $n=3,4,5,6$ and $m=37$. (b) Cumulative probability $C$ of the sorted probabilities $P_{S}$. Inset: effective fraction $N(C = p)/N_{\mathrm{tot}}$ of the number of configuration to obtain a value of the sorted cumulative $C$ equal to a threshold $p$. Simulations are performed for different values of $n$ and $m$. Curves are shown for thresholds $p=0.9$, $p=0.95$ and $p=0.99$. Solid lines: average values over $N_{\mathrm{unit}}=1000$ Haar-random unitaries. Shaded regions: 1$\sigma$ interval obtained from the numerical simulations. Red: $n=3$, blue: $n=4$, green: $n=5$, cyan: $n=6$.}
\label{fig:cumulative}
\end{figure}

As a first step to analyze the case where a non-linear evolution is introduced in the linear-optical system, we numerically investigate whether a small fraction of matrix permanents can provide a significant contribution to the full transition amplitude. If that were true, a good approximation to the exact amplitude might be achievable by including only a few terms in the sum of Eq.\ (\ref{eq:gen_nonlin}). To that end we perform a numerical simulation by (i) sampling $N_{\mathrm{unit}} = 1000$ linear transformations $U$ according to the Haar measure; (ii) evaluating the full probability distribution $P$ for each $U$; (iii)  sorting each probability distribution in decreasing order to obtain $P_{S}$ as a function of the fraction $N/N_{\mathrm{tot}}$ of output configurations. For each $P_{S}$ we evaluate its cumulative probability $C$ by summing over a given fraction of the combinations, obtained as $C(X) = \sum_{x \leq X} P_{S}(x)$. This cumulative probability provides information on the effective fraction of permanents that contribute to the overall probability mass up to a chosen threshold $p$. The results of a numerical simulation for $n=3,4,5,6$ and $m \in [5, 37]$ are shown in Fig.\ \ref{fig:cumulative}. 

We observe that the sorted probability distributions $P_{S}$ present similar trends, and that the cumulative probabilities $C$ for fixed $m$ are almost independent on the number of photons $n$. An analysis of the fraction $N(C = p)/N_{\mathrm{tot}}$ of configurations that contribute to the full distribution up to a threshold $p$ is then shown in the inset of Fig.\ \ref{fig:cumulative}. We find that fractions $\sim (0.5, 0.6, 0.8)$ of the output combinations are necessary to evaluate $(90\%, 95\%, 99\%)$ of the full probability mass, respectively, and that these values are almost independent of $n$ and $m$. This feature is related to the anti-concentration conjecture for complex Gaussian matrices \cite{AA}, which is relevant for the complexity of computing permanents of random matrices.

\section{non-linear Boson Sampling with a single-mode non-linear phase}

Here we perform the explicit calculation for the scenario in which a non-linear phase shift is introduced in mode $x$, corresponding to Eq.\ (4) in main text. The evolution operator for this transformation can be written as $\hat{U}_{\mathrm{nlp}} = \exp(-\imath \hat{n}^2_{x} \phi)$. Evaluating its action on a generic $m$-mode Fock state $\vert R \rangle$ reads:
\begin{equation}
\vert R \rangle \stackrel{\exp(-\imath \hat{n}^2_{x} \phi)}{\longrightarrow} e^{-\imath r_{x}^{2} \phi} \vert R \rangle,
\end{equation}
thus corresponding to a function $\mathcal{N}$ of the form:
\begin{equation}
\mathcal{N}_{r_{1} \ldots r_{m}}^{q_{1} \ldots q_{m}} = \exp(-\imath r_{x}^{2} \phi) \prod_{i=1}^{m} \delta_{r_{i},q_{i}}.
\end{equation}
Substituting into the general expression (\ref{eq:gen_nonlin}) we obtain:
\begin{equation}
\label{eq:nonlinear_singlephase}
\mathcal{A}^{\mathrm{nlp}}_{W,N,V} \big( \vert S \rangle \rightarrow \vert T \rangle \big) = \sum_{R \in \Phi_{m,n}} \exp(-\imath r_{x}^{2} \phi) \frac{\mathrm{per}(W_{S,R}) \mathrm{per}(V_{R,T})}{\sqrt{\prod_{i=1}^{m}s_{i}! \big(\prod_{j=1}^{m}r_{j}!\big)^{2} \prod_{l=1}^{m}t_{l}!}}.
\end{equation}
This expression can be written in a different form. We first observe that
$\exp{(-\imath r_x^2 \phi)} = \exp(-\imath r_x \phi)$ for $r_x \in \{0,1\}$. Equation \eqref{eq:nonlinear_singlephase} can then be rearranged as:
\begin{equation}
\begin{aligned}
\mathcal{A}^{\mathrm{nlp}}_{W,N,V} \big( \vert S \rangle \rightarrow \vert T \rangle \big) &= \sum_{r_{x}=0,1} \frac{\exp(-\imath r_{x}^2 \phi) \mathrm{per}(W_{S,R}) \mathrm{per}(V_{R,T})}{\sqrt{\prod_{i=1}^{m}s_{i}! \big(\prod_{j=1}^{m}r_{j}!\big)^{2} \prod_{l=1}^{m}t_{l}!}} +
\sum_{r_{x}>1} \frac{\exp(-\imath r_{x}^{2} \phi) \mathrm{per}(W_{S,R}) \mathrm{per}(V_{R,T})}{\sqrt{\prod_{i=1}^{m}s_{i}! \big(\prod_{j=1}^{m}r_{j}!\big)^{2} \prod_{l=1}^{m}t_{l}!}} = \\ 
&= \sum_{r_{x}=0,1} \frac{\exp(-\imath r_{x} \phi) \mathrm{per}(W_{S,R}) \mathrm{per}(V_{R,T})}{\sqrt{\prod_{i=1}^{m}s_{i}! \big(\prod_{j=1}^{m}r_{j}!\big)^{2} \prod_{l=1}^{m}t_{l}!}} +
\sum_{r_{x}>1} \frac{\exp(-\imath r_{x}^{2} \phi) \mathrm{per}(W_{S,R}) \mathrm{per}(V_{R,T})}{\sqrt{\prod_{i=1}^{m}s_{i}! \big(\prod_{j=1}^{m}r_{j}!\big)^{2} \prod_{l=1}^{m}t_{l}!}},
\end{aligned}
\end{equation}
where the sums are extended over $R \in \Phi_{m,n}$ for those terms with $r_{x} = 0,1$ and $r_{x} > 1$ respectively. By summing and subtracting the following term:
\begin{equation}
\sum_{r_{x}>1} \frac{\exp(-\imath r_{x} \phi) \mathrm{per}(W_{S,R}) \mathrm{per}(V_{R,T})}{\sqrt{\prod_{i=1}^{m}s_{i}! \big(\prod_{j=1}^{m}r_{j}!\big)^{2} \prod_{l=1}^{m}t_{l}!}},
\end{equation}
the expression for the transition amplitude can be rearranged as:
\begin{equation}
\label{eq:nonlin_phase_tmp}
\mathcal{A}^{\mathrm{nlp}}_{W,N,V} \big( \vert S \rangle \rightarrow \vert T \rangle \big) = \sum_{R \in \Phi_{m,n}} \frac{\exp(-\imath r_{x} \phi) \mathrm{per}(W_{S,R}) \mathrm{per}(V_{R,T})}{\sqrt{\prod_{i=1}^{m}s_{i}! \big(\prod_{j=1}^{m}r_{j}!\big)^{2} \prod_{l=1}^{m}t_{l}!}} + \sum_{r_{x}>1} \frac{\mathrm{per}(W_{S,R}) [\exp(-\imath r_{x}^{2} \phi) - \exp(-\imath r_{x} \phi)] \mathrm{per}(V_{R,T})}{\sqrt{\prod_{i=1}^{m}s_{i}! \big(\prod_{j=1}^{m}r_{j}!\big)^{2} \prod_{l=1}^{m}t_{l}!}}.
\end{equation}

Let us now define $F$ as the unitary matrix associated to a linear transformation applying a phase shift on mode $x$:  $F = \exp(-\imath \hat{n}_{x} \phi)$. We observe that the transition amplitude between an input state $\vert R \rangle$ and an output state $\vert P \rangle$ which evolve according to $F$ is written as:
\begin{equation}
\mathcal{A}_{F}\big(\vert R \rangle \rightarrow \vert P \rangle \big) = \frac{\mathrm{per}(F_{R,P})}{\sqrt{\prod_{i=1}^{m}r_{i}! \prod_{j=1}^{m}p_{j}!}} = \exp{(- \imath r_x \phi)} \prod_y \delta_{r_y,p_y}.
\end{equation}
This allows us to write the following chain of equalities for the first term of Eq.\ (\ref{eq:nonlin_phase_tmp})
\begin{equation}
\begin{aligned}
& \sum_{R \in \Phi_{m,n}} \frac{\exp(-\imath r_{x} \phi) \mathrm{per}(W_{S,R}) \mathrm{per}(V_{R,T})}{\sqrt{\prod_{i=1}^{m}s_{i}! \big(\prod_{j=1}^{m}r_{j}!\big)^{2} \prod_{l=1}^{m}t_{l}!}} = \sum_{R \in \Phi_{m,n}} \frac{\mathrm{per}(W_{S,R}) \exp(-\imath r_{x} \phi) \sqrt{\big(\prod_{j=1}^{m}r_{j}!\big)^{2}} \mathrm{per}(V_{R,T})}{\sqrt{\prod_{i=1}^{m}s_{i}! \big(\prod_{j=1}^{m}r_{j}!\big)^{4} \prod_{l=1}^{m}t_{l}!}} &= \\
&= \sum_{R,P \in \Phi_{m,n}} \frac{\mathrm{per}(W_{S,R}) \exp(-\imath r_{x} \phi) \sqrt{\prod_{j=1}^{m}r_{j}! \prod_{k=1}^{m}p_{k}!} \big(\prod_y \delta_{r_y,p_y} \big)\mathrm{per}(V_{P,T})}{\sqrt{\prod_{i=1}^{m}s_{i}! \big(\prod_{j=1}^{m}r_{j}!\big)^{2} \big(\prod_{k=1}^{m}p_{k}!\big)^{2} \prod_{l=1}^{m}t_{l}!}} = \\
&=  \sum_{R,P \in \Phi_{m,n}} \frac{\mathrm{per}(W_{S,R}) \mathrm{per}(F_{R,P}) \mathrm{per}(V_{P,T})}{\sqrt{\prod_{i=1}^{m}s_{i}! \big(\prod_{j=1}^{m}r_{j}!\big)^{2} \big(\prod_{k=1}^{m}p_{k}!\big)^{2} \prod_{l=1}^{m}t_{l}!}} = \frac{\mathrm{per}(\bar{U}_{S,T})}{\sqrt{\prod_i s_i! \prod_j t_j!}},
\end{aligned}
\end{equation}
where $\bar{U} = W F V$, and Eq.\ \eqref{eq:identity} has been used to perform the last simplification. By replacing this into Eq.\ \eqref{eq:nonlin_phase_tmp} we obtain Eq.\ (4) of the main text:
\begin{equation}
\label{eq:nonlinear_phase_simplified}
\mathcal{A}^{\mathrm{nlp}}_{W,N,V} \big( \vert S \rangle \rightarrow \vert T \rangle \big) = 
\frac{\mathrm{per}(\bar{U}_{S,T})}{\sqrt{\prod_{i=1}^{m}s_{i}! \prod_{l=1}^{m}t_{l}!}} + \sum_{r_{x}>1} \frac{\mathrm{per}(W_{S,R}) [\exp(-\imath r_{x}^{2} \phi) - \exp(-\imath r_{x} \phi)] \mathrm{per}(V_{R,T})}{\sqrt{\prod_{i=1}^{m}s_{i}! \big(\prod_{j=1}^{m}r_{j}!\big)^{2} \prod_{l=1}^{m}t_{l}!}}.
\end{equation}

\section{Evolution with a non-linear single-mode phase term}

Here we present numerical simulations that quantify the perturbation, in the output distribution, induced by the introduction of a single-mode non-linear phase term $N$. We verify numerically the difference between two output distributions: the desired one (in the presence of $N$) and one obtained by replacing the non-linear phase shift with a linear one defined by the same phase $\phi$ (i.e., the overall transformation $\overline{U} = W F V$). This is suggested by the form of Eq.\ \eqref{eq:nonlinear_phase_simplified}, which shows that a single-mode non-linear phase introduce a departure from a linear phase for those paths where multiple photons propagate in mode $x$. Finally, we verify numerically that replacing the non-linear phase shift with a linear one represents the correct benchmark for the analysis presented in this Section. We find that, in most cases, the closest linear evolution to the non-linear one $W, N, V$ is that given by $\overline{U}$.

\subsection{Evolution in the finite setting} 

We now investigate the effect of introducing a non-linear phase term in a single mode. We begin by considering a scenario with $n=3$ photons in $m=9,16$ modes according to the evolution corresponding to Eq.\ (\ref{eq:nonlinear_singlephase}). More specifically, we insert the non-linear phase in modes $x=5$ and $x=9$, respectively for $m=9$ and $m=16$, after a linear transformation $W$ and before the second linear evolution $V$. 
\begin{figure}[ht!]
\centering
\includegraphics[width=0.99\textwidth]{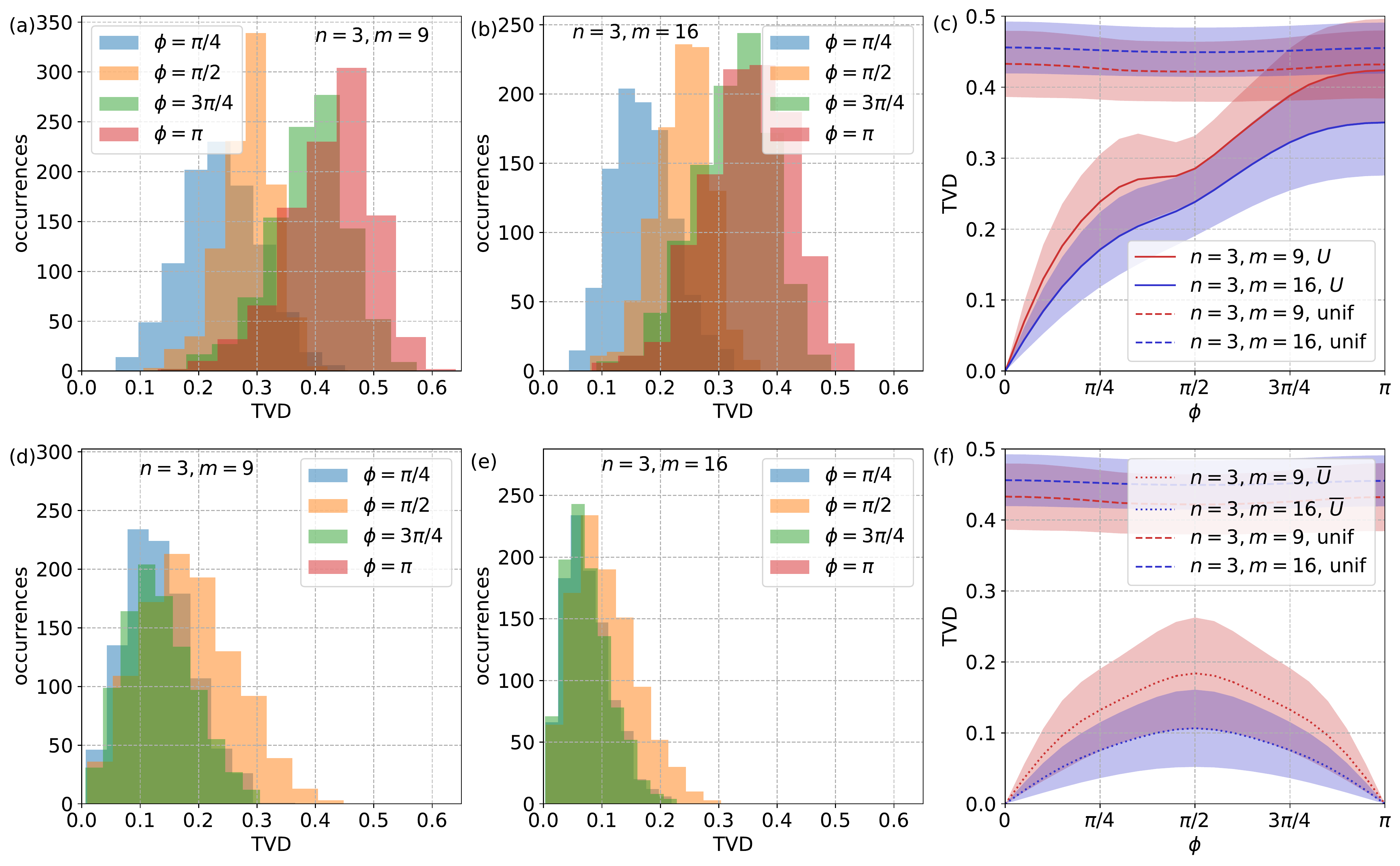}
\caption{(a-c) Analysis of $\mathrm{TVD}$ between $p_{i}$ (evolution $U = WV$) and $q_{i}$ (evolution $W,N,V$) for $N_{\mathrm{unit}} = 1000$ unitaries. (a-b) Histograms for different values of $\phi$. (c) Mean (solid curves) and 1-$\sigma$ intervals (shaded regions) as a function of non-linear phase $\phi$. (d-f) Analysis of $\mathrm{TVD}$ between $q'_{i}$ (evolution $\overline{U} = W F V$) and $q_{i}$ (evolution $W,N,V$) for $N_{\mathrm{unit}} = 1000$ unitaries. (d-e) Histograms for different values of $\phi$. (f) Mean (solid curves) and 1-$\sigma$ intervals (shaded regions) as a function of non-linear phase $\phi$. (c,f) Dashed curves: $\mathrm{TVD}$ for evolution $W,N,V$ with respect to the uniform distribution.}
\label{fig:TVD}
\end{figure}

As a first step, we study the total variation distance $\mathrm{TVD} = 1/2 \sum_{i} \vert p_{i} - q_{i} \vert$ between the output distribution $\{ p_{i} \}$, corresponding to $U = VW$, and the output distribution $\{ q_{i} \}$, which includes the non-linear phase between $W$ and $V$. The results of a numerical simulation with $N_{\mathrm{unit}} = 1000$ sets of unitary transformations $\{W, V\}$ are shown in Fig.\ \ref{fig:TVD} (a-c). We observe that, as expected, the $\mathrm{TVD}$ increases for larger non-linear phases $\phi$. For $m=9$ and $\phi = \pi$, the $\mathrm{TVD}$ reaches a value close to the one relative to a uniform distribution, thus showing that the insertion of a single non-linear phase introduces a significant change in the output distribution. 

Having seen that a non-linear term introduces a significant change in the output distribution, as expected, we then analyze whether the evolution with the non-linear phase can be approximated by a linear transformation that is \emph{different} from $U$. The form of Eq.\ \eqref{eq:nonlinear_phase_simplified} suggests an \emph{ansatz} for a linear approximation for the non-linear dynamics, namely, replacing the non-linear term $N$ with a linear unitary $F=\exp(-i \hat{n}_{x} \phi)$, leading to an overall interferometer $\overline{U} = W F V$. In Fig.\ \ref{fig:TVD} (d-f) we report the $\mathrm{TVD} = 1/2 \sum_{i} \vert q'_{i} - q_{i} \vert$ between the output distribution $\{ q'_{i} \}$, with a linear phase term (i.e.\ evolution $\overline{U}$), and the output distribution $\{ q_{i} \}$, in the presence of the non-linear phase (evolution $W, N, V$). For phase values close to $\phi = 0$ and $\phi = \pi$, the two scenarios present a small $\mathrm{TVD}$, meaning that the non-linear evolution can be approximated by a simple linear phase. Indeed, for $\phi = 0$ there is no phase term in both cases, whereas for $\phi = \pi$ the terms $\exp(-\imath n \phi)$ and $\exp(-\imath n^2 \phi)$ provide the same phase factor for all values of $n$. Conversely, the maximum difference is obtained for $\phi = \pi/2$.

\begin{figure}[ht!]
\centering
\includegraphics[width=0.99\textwidth]{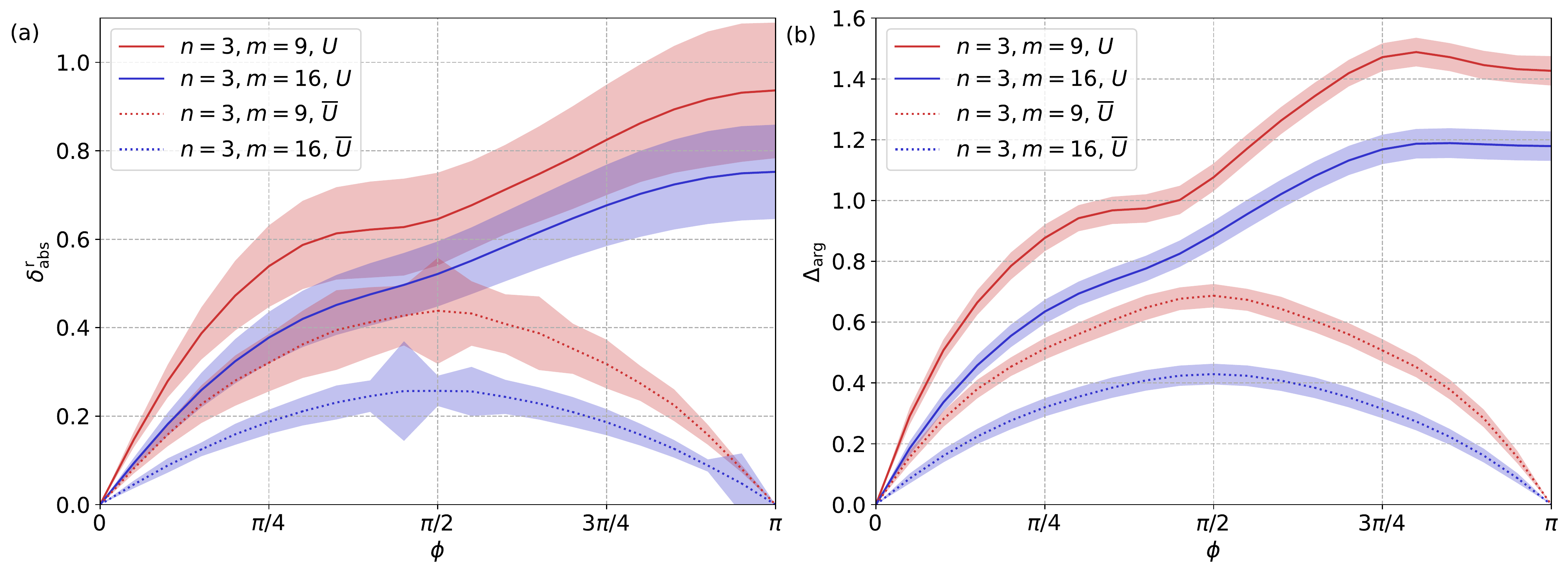}
\caption{Average trends of (a) $\delta_{\mathrm{abs}}^{\mathrm{r}}$ and (b) $\Delta_{\mathrm{arg}}$ as functions of the non-linear phase $\phi$, obtained by calculating all amplitudes for $N_{\mathrm{unit}} = 1000$ unitaries. Solid lines: comparison between evolution $W,N,V$ and evolution $U = WV$. Dashed lines: comparison between evolution $W,N,V$ and the one obtained with evolution $\overline{U} = W F V$. Shaded regions: 1$\sigma$ intervals obtained from the numerical simulation.}
\label{fig:ampl}
\end{figure}

\begin{figure}[ht!]
\centering
\includegraphics[width=0.99\textwidth]{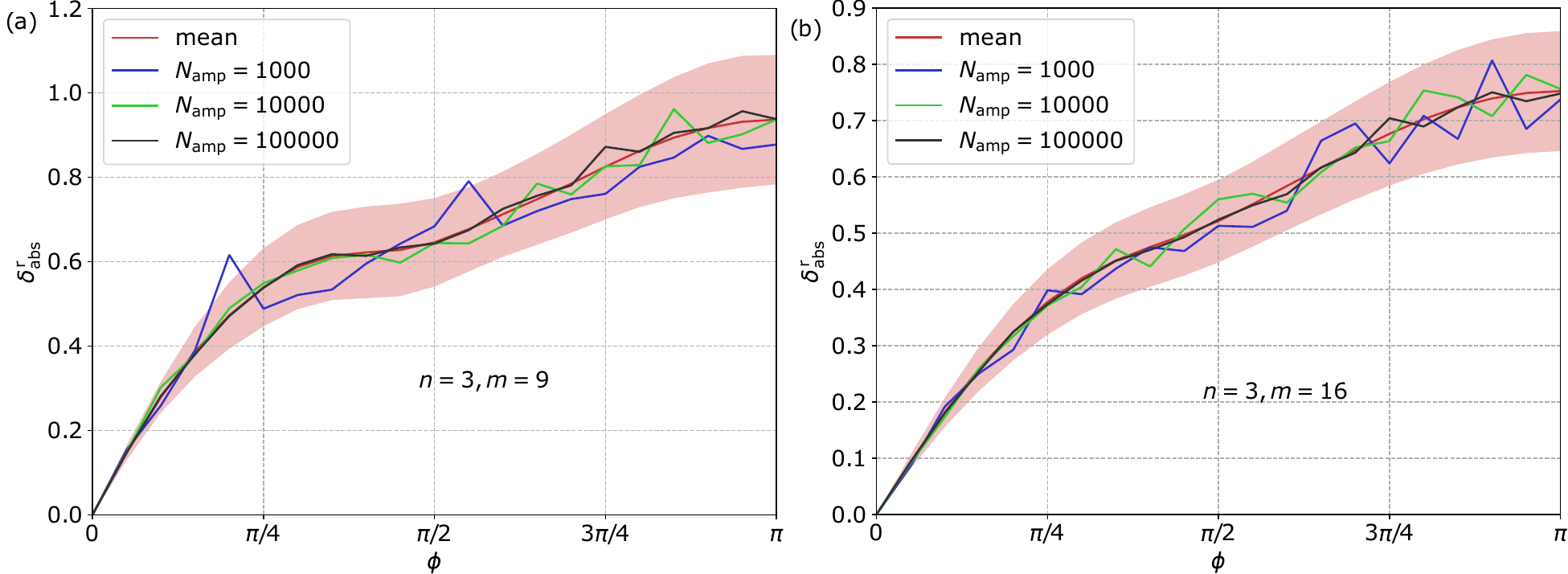}
\caption{Convergence of $\delta_{\mathrm{abs}}^{\mathrm{r}}$ for (a) $n=3, m=9$ and (b) $n=3, m=16$. In each plot, the average behavior (red curves) obtained for all amplitudes corresponding to $N_{\mathrm{unit}} = 1000$ unitaries are compared with the trend where only a single amplitude is calculated for $N_{\mathrm{amp}}$ different unitaries. Shaded regions: 1$\sigma$ intervals for the simulation with $N_{\mathrm{unit}}$ unitaries.}
\label{fig:conv}
\end{figure}

\begin{figure}[ht!]
\centering
\includegraphics[width=0.99\textwidth]{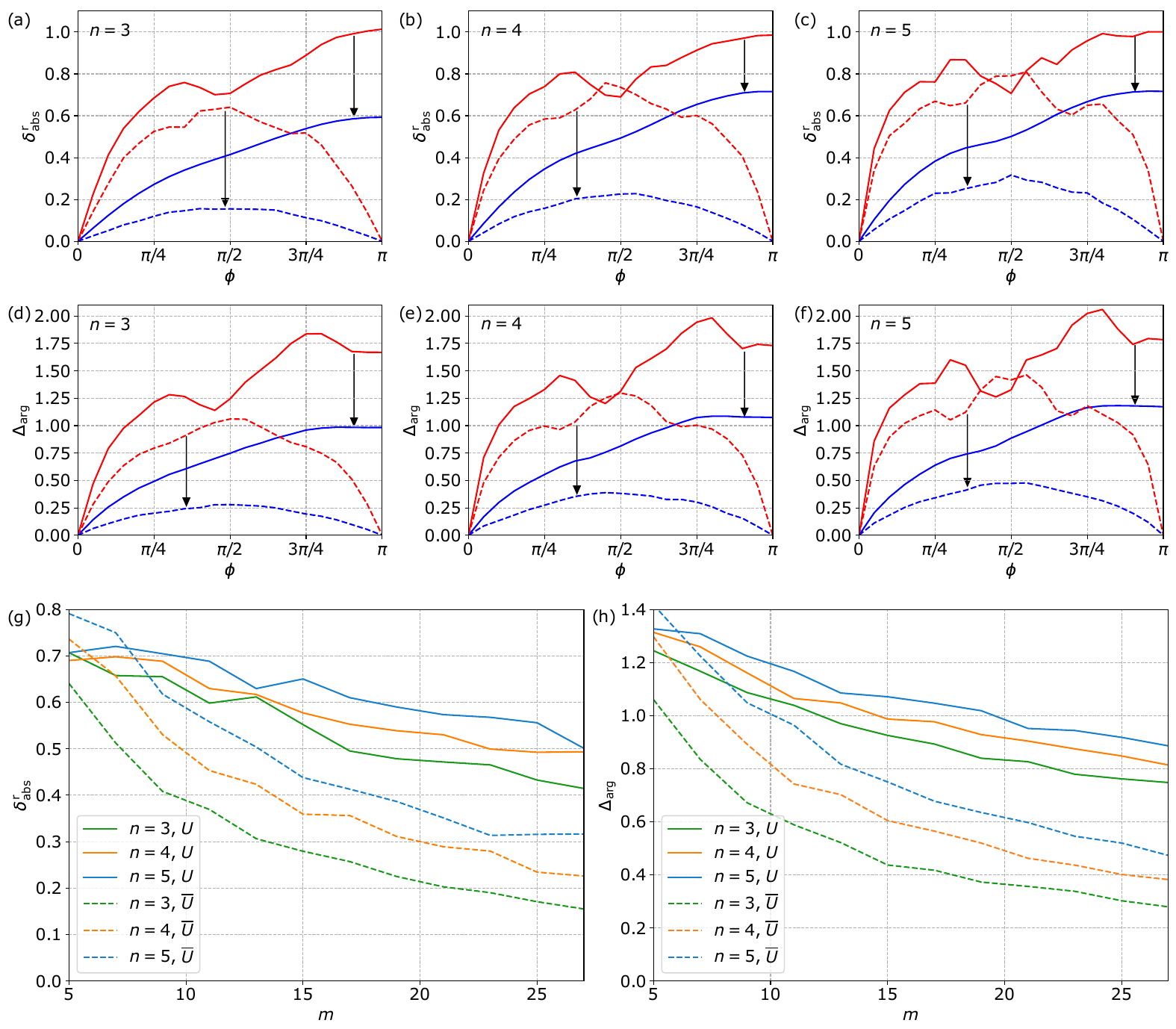}
\caption{Analysis of $\delta_{\mathrm{abs}}^{\mathrm{r}}$ and $\Delta_{\mathrm{arg}}$ for different values of $n,m$, obtained by evaluating a single amplitude for $N_{\mathrm{unit}} = 10^4$ unitaries. (a-f) Simulations as a function of the non-linear phase $\phi$. Red: $m=5$. Blue: $m=27$. The arrow shows the trend for increasing values of $m$. (g-h) Analysis for $\phi=\pi/2$ as a function of $m$, for $n=3$ (green) and $n=4$ (orange) and $n=5$ (cyan). In all plots, solid lines are the comparison between evolution $W,N,V$ and evolution $U = WV$, while dashed lines are the comparison between evolution $W,N,V$ and the one obtained with evolution $\overline{U} = W F V$.}
\label{fig:trend}
\end{figure}

As a following step, we verify numerically that similar trends are obtained by directly looking at single amplitudes. In Fig.\ \ref{fig:ampl} we observe the same trend, as displayed by the $\mathrm{TVD}$, but now for the differences in the moduli and in the argument of the amplitudes. More specifically, we consider the behavior of relative differences $\delta_{\mathrm{abs}}^{\mathrm{r}} = \vert (\vert \mathcal{A}^{\mathrm{nlp}}_{W,N,V} \vert - \vert \mathcal{A}_{W,V} \vert)\vert/\vert\mathcal{A}_{W,V} \vert$ and the absolute difference $\Delta_{\mathrm{arg}} = \vert \arg(\mathcal{A}^{\mathrm{nlp}}_{W,N,V}) - \arg(\mathcal{A}_{W,V}) \vert$. Furthermore, numerical simulations show that the average trend for these quantities can be calculated by sampling $N_{\mathrm{amp}} = 10^{4}$ unitaries, and by calculating a single amplitude for each sampled unitary (see Fig.\ \ref{fig:conv}). This allows us to obtain an approximated estimate of $\delta_{\mathrm{abs}}^{\mathrm{r}}$ and $\Delta_{\mathrm{arg}}$ without having to calculate the full distributions.
Following these results, we studied how the perturbation introduced by the non-linear phase scales with the number of photons $n$ and the number of modes $m$. The results are shown in Fig.\ \ref{fig:trend} for $n=3,4,5$, and different values of $m$. We observe that the change induced by the non-linear phase relative to the linear transformation $\overline{U} = W F V$ decreases with the number of modes $m$. This trend is shown by dashed lines of Fig.\ \ref{fig:trend} (g-h). This can be explained by the second term of Eq.\ \eqref{eq:nonlinear_phase_simplified}, which shows that the departure from the linear phase evolution  depends on the weight of the bunching terms at the non-linearity. This is discussed in the main text, and will be also addressed in Sec. \ref{sec:simulation}.

\subsection{Approximating non-linear evolution with linear transformations} \label{sec:approx}

In the previous section, we employed the linear transformation $\overline{U}$ as a benchmark to assess the disturbance induced by the non-linear phase shift. In particular, we reported numerical evidence that the insertion of $\exp(-\imath \hat{n}^{2}_{x} \phi)$ within linear evolutions $W$ and $V$ introduces a non-negligible departure from the linear behaviour. This choice of benchmark is suggested by the expression of the transition amplitude reported in Eq.\ (4) of the main text and Eq.\ \eqref{eq:nonlinear_phase_simplified} above. There, we observe that the overall evolution can be written as a combination of a linear term, identified by unitary $\overline{U}$, that includes a linear phase shift, and a second term which is present only when multiple photons propagate through the non-linearity.

\begin{figure}[ht!]
\centering
\includegraphics[width=0.99\textwidth]{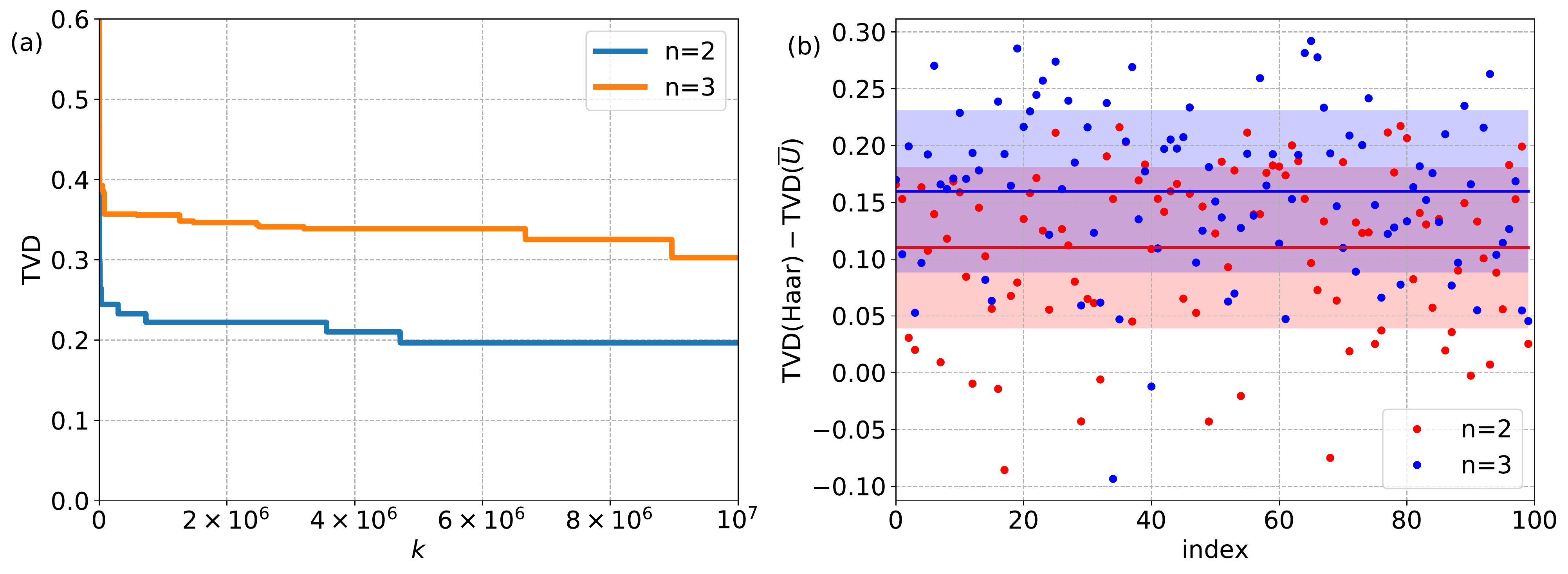}
\caption{Approximation via Haar random matrices. (a) Evolution of the best TVD between the non-linear case and the sampled unitaries aiming at the best effective unitary matrix approximating the full evolution (where $k$ is the number of sampled unitaries). Blue: sampling for $n=2$ and $m=9$ for a single choice of $W$, $N$, $V$. Orange: similar analysis for $n=3$ and $m=9$. (b) Scatterplot of the difference $\mathrm{TVD}(\mathrm{Haar}) - \mathrm{TVD}(\overline{U})$ for $100$ sampled evolutions $W$ and $V$, with $\phi = \pi/2$. Here, $\mathrm{TVD}(\mathrm{Haar})$ is the total variation distance of the non-linear evolution with respect to the best Haar random matrix found after $N_{\mathrm{unit}}$ iterations, and $\mathrm{TVD}(\overline{U})$ is the total variation distance with respect to evolution $\overline{U} = W F V$.}
\label{fig:optimization}
\end{figure}

Now we report on additional numerical analysis to support the choice of benchmark evolution $\overline{U}$. More specifically, we verified whether a different effective linear transformation $H$ can have a smaller distance with respect to the full evolution $W, N, V$. We consider the case of $n=2,3$ photons and $m=9$ modes, focusing on the $\phi = \pi/2$ case. We then sample two Haar random unitary matrices $W$ and $V$, and search numerically for a linear evolution $H$ whose output distribution has a smaller TVD relative to the output considering the non-linear evolution $W, N, V$. This search is performed by randomly sampling from the Haar measure.

In Fig.\ \ref{fig:optimization} (a) we show the progress of the best TVD as a function of the number of sampled unitaries $k$ up to $N_{\mathrm{unit}} = 10^7$ sampled Haar-random matrices. The result suggests that the TVD approximately saturates after $N_{\mathrm{unit}} = 10^6$ sampled matrices. 
We then repeated this numerical search for $100$ different pairs ${W,V}$ [see Fig.\ \ref{fig:optimization} (b)]. We observe that, for most tested cases, the linear transformation $\overline{U}$ is closer to the non-linear evolution $W, N, V$ than the best Haar random matrix found via the brute force numerical search.

These numerical results then give supporting evidence that $\overline{U}$, obtained by replacing the non-linear phase shift with a linear one, can represent a suitable benchmark for the chosen non-linearity as per Eq.\ \eqref{eq:nonlinear_phase_simplified}.

\section{Simulation of non-linear Boson Sampling with linear optics and ancillary photons}
\label{sec:simulation}

In this Section we give more details on the algorithm for simulating non-linear Boson Sampling using a linear-optical gadget acting on  auxiliary modes and photons. 

We begin by arguing that, at least in principle, there should always exist some linear-optical gadget that simulates a given non-linear transformation. There are likely to be  decompositions that are much more efficient than the one we describe here in terms of required ancillary photons, success probability of the gadget, computational complexity of constructing the gadget and so on. We leave this optimization as an interesting avenue for future research, focusing on small-scale numerical investigations.
We then provide some numerical insight on the role of the number of auxiliary photons, $k$, by looking at the contribution of bunching terms in Haar-random photon number distributions. Finally we provide the explicit construction of the simulation algorithm, and discuss the calculation of the effective linear-optical matrices required for the simulation. Finally we provide some numerical analysis to complement the discussion reported in the main text, and  leverage the linear-optical simulation into a classical simulation algorithm.

\subsection{Universality of non-linear gadgets} 

We begin by quoting a result due to Oszmaniec and Zimboras \cite{Oszmaniec18} regarding the ability of some non-linear gates to extend linear optics to full universality within the state space of $n$ photons in $m$ modes.

In particular, let $\mathcal{H}_{m,n}$ be the restriction of Fock space to $n$ photons in $m$ modes, and let LO$_b$ be set of all transformations generated by linear optics acting on $\mathcal{H}_{m,n}$. Let $V$ be some unitary gate that is not in LO$_b$ (hence, a non-linear gate), and let $\langle \textrm{LO}_b, V \rangle$ be the group of dynamics generated by sequences of elements from $\textrm{LO}_b \cup \{V\}$.

Theorem 1 in \cite{Oszmaniec18} guarantees that $\langle \textrm{LO}_b, V \rangle$ is the full unitary group acting on $\mathcal{H}_{m,n}$ if either:
\begin{itemize}
    \item[(i)] $m > 2$, or
    \item[(ii)] $m = 2$ and $\left[V \otimes V, L\right]\neq0$.
\end{itemize}

Here, $L = \left \lvert \Psi\right \rangle \left \langle \Psi \right \lvert$ is a project acting on two copies of $\mathcal{H}_{m,n}$ defined by
\begin{equation*}
    \left \lvert \Psi \right \rangle = \sum_{l=0}^{n} (-1)^l \left \lvert D_l \right \rangle \left \lvert D_{n-l} \right \rangle,
\end{equation*}
and where $\left \lvert D_{l} \right \rangle$ is the two-mode state with $l$ particles in the first mode and $n-l$ particles in the second. In our description of the result of \cite{Oszmaniec18} we have also skipped one set of conditions which applies to fringe cases and do not concern us here.

This result allows us to check which conditions a particular (unitary) non-linear gate must satisfy in order to be able to approximately decompose any other non-linear gate. In other words, suppose we have some gate $V$ that satisfies the above criteria. If our non-linear Boson Sampling setup includes some non-linearity $N$, we know that it is possible, in principle, to approximate $N$ by a sequence of linear-optical elements interspersed with some copies of $V$.

By the result of \cite{Oszmaniec18}, \emph{any} non-linear unitary gate $V$ suffices for the above purposes if $m>2$, which means that we can always replace $N$ by a 3-mode circuit consisting of beam splitters and applications of $V$. However, $N$ is a single-mode non-linearity. Therefore it is likely that applying the above result for $m=3$ is less efficient (in the sense of number of applications of $V$ required to obtain an approximation of $N$ to desired accuracy) than applying it for $m=2$ whenever that option is available. Thus, we can test some reasonable non-linear gates to check whether they satisfy condition (ii) when $m=2$. The authors of \cite{Oszmaniec18} proved that for a two-mode cross-Kerr gate, given by $\exp{(-i \hat{n}_1 \hat{n}_2 \phi)}$, this is true for some values of $\phi$ such as $\pi/3$. It is easy to check that the following gates also satisfy condition (ii):
\begin{itemize}
    \item[(i)] The gate which we focus on in this paper, namely $\hat{U}_{\mathrm{nlp}} = \exp(-\imath \hat{n}^2_{x} \phi)$, for $\phi = \pi/3$. If the number of photons $n$ is odd, then this gate is also universal for $\phi=\pi/2$.
    \item[(ii)] The generalized non-linear sign shift gate $\textrm{NS}_k$ \cite{ScheelA05}, which leaves all states with up to $k-1$ photons invariant and applies $\left \lvert k \right \rangle \rightarrow -\left \lvert k \right \rangle$.
\end{itemize}

So far we considered the ability of some non-linear gates to simulate others, but our main goal is to map this into a post-selected linear-optical gadget. Two issues arise if we try to directly apply the result of \cite{Oszmaniec18} to this question. The first is that the result is not constructive. It provides a test to decide whether some non-linear gate is universal when supplemented with linear optics, but does not output a sequence that approximates a specific desired gate. This can be solved by invoking the standard Solovay-Kitaev theorem \cite{NielsenBook} (with the caveat that, besides $V$, we also need to be able to apply its inverse, though that is trivial for the gates listed above).

The second issue is that, given a maximum number of photons present in the input state, $n$, we need there to always exist \emph{some} unitary non-linear gate that can be implemented as a post-selected linear-optical gadget. To the best of our knowledge, this remains an open question. In subsequent sections we show that there exists a gadget for $\hat{U}_{\mathrm{nlp}}$  when $\phi = \pi/2$ and for up to 4 photons, while previous work \cite{ScheelA05} has given evidence that the gate $\textrm{NS}_k$ can be implemented for arbitrary $k$ with success probability that decreases as $1/k^2$.

We can combine the observations above into a procedure to obtain a post-selected gadget for any non-linearity $N$. We begin by using the Solovay-Kitaev algorithm to provide a sequence that approximates $N$ as a sequence of beam-splitters and some standard non-linear gate such as $\hat{U}_{\mathrm{nlp}}$ or $\textrm{NS}_k$. Then, we replace every copy of that gate by a postselected gadget. We do not pursue this approach as a practical means to perform this task, only as an in-principle argument that it should always be possible. The circuit generated by this reasoning is likely to lead to a very inefficient gadget for a typical $N$, and subsequently a very inefficient classical algorithm. However, we conjecture that is unnecessary, and that some gadget that uses only at most $n$ additional photons should exist for any $n$-photon single-mode non-linearity. As evidence, throughout this work we show that, for many different gates, a direct gadget exists that does not require a Solovay-Kitaev type decomposition.

\subsection{Haar-random distribution with photon-number threshold} 

As a first step, we need to evaluate to evaluate the contribution of bunching terms in the photon number distribution after the first Haar-random interferometer. This is important because the measurement-induced scheme of Ref.\ \cite{Scheel03} allows us to simulate single-mode non-linearities acting on up to $n$ input photons by using a linear-optical gadget and $n$ auxiliary photons. Therefore, if the rate bunching is not too large after the first interferometer, a smaller number of auxiliary photons might suffice.

\begin{figure}[ht!]
\centering
\includegraphics[width = 0.99 \textwidth]{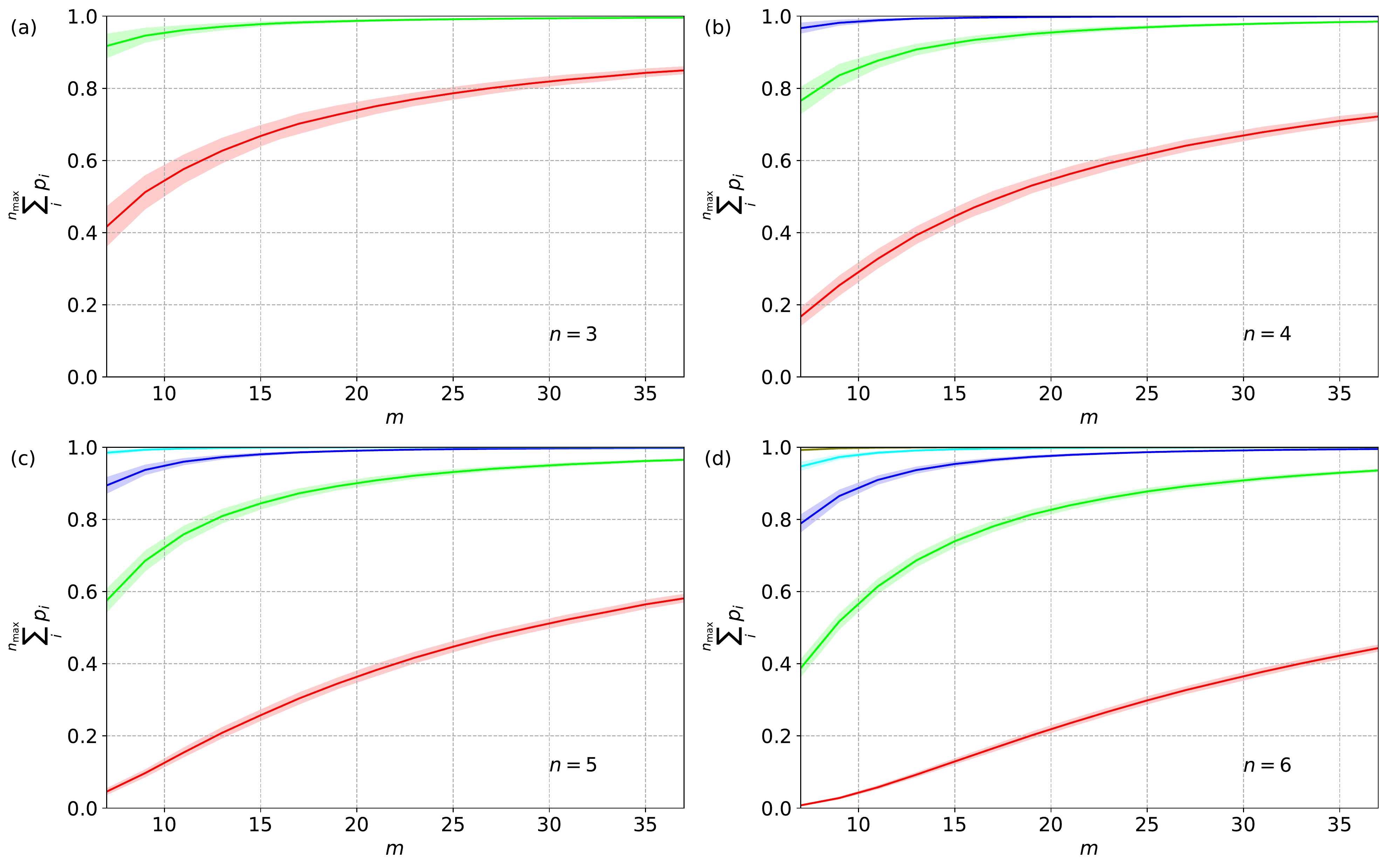}
\caption{Numerical simulation of the average probability $\sum_{i}^{n_{\mathrm{max}}} p_{i}$ of measuring an output event with at most $n_{\mathrm{max}}$ photons for each output mode at the output of a Haar-random interferometer. (a) $n=3$, (b) $n=4$, (c) $n=5$, (d) $n=6$. Red: $n_{\mathrm{max}} = 1$. Green: $n_{\mathrm{max}} = 2$. Blue: $n_{\mathrm{max}} = 3$. Cyan: $n_{\mathrm{max}} = 4$. Orange: $n_{\mathrm{max}} = 5$. Solid lines: average values over $N=1000$ Haar-random unitaries. Shaded regions: 1$\sigma$ interval obtained from the numerical simulation.}
\label{fig:truncation}
\end{figure}

In the asymptotic limit, the Bosonic Birthday Paradox holds, stating that bunching contributions are negligible for $m \gg n^2$. To verify this limit for finite sizes, we performed numerical simulations that investigate the contribution of bunching terms at the output of Haar random matrices for $n = 3,4,5,6$ and $7 \leq m \leq 37$. For each system size ($n$, $m$) we numerically evaluated the overall average probability to obtain an output state with contributions having at most $n_{\mathrm{max}}$ photons for each output mode (i.e.\ excluding  configurations where at least one output mode has $n_{\mathrm{out}} > n_{\mathrm{max}}$ photons). The results are shown in Fig.\ \ref{fig:truncation}. 

We observe that, for $m=n^2$, there is a significant portion ($\sim 5 \%$) of the distribution probability where bunching terms with $3$ photons on the same mode are occur, while higher order terms are almost negligible. This is a relevant aspect that must be taken into account for the simulation algorithm described below.

\subsection{Simulating non-linearities with auxiliary photons and extended linear evolution} 

Simulation of non-linear Boson Sampling with single-mode non-linearities can be performed according to the approach reported in the main text. In particular, the scheme of Fig. 2 employs a set of ancillary photons and modes, and an auxiliary transformation $U_{\mathrm{eff}}$. Such transformation depends on the actual non-linear transformation to be simulated (more details on the calculation of $U_{\mathrm{eff}}$ can be found in Sec. \ref{sec:gadget}). The non-linear evolution is successfully simulated conditioned to the detection of state $\vert 1 \rangle$ on each of the output auxiliary modes.

The scheme of Fig. 2 thus defines both a linear-optical simulation algorithm, which can be implemented by building the corresponding experimental apparatus and thus running the device, and a classical simulation algorithm. In the second case, sampling can be performed via the Clifford and Clifford algorithm \cite{Clifford18}, with the additional ingredient of discarding those events which do not correspond to a correct post-selected configuration.

We can now define below the classical simulation algorithm for non-linear Boson Sampling.

\textbf{Algorithm 1.} The following classical algorithm provides an approximate simulation of non-linear Boson Sampling with single-mode non-linearities:
\begin{enumerate}
    \item[(i)] Compute the overall $m+k$-mode linear transformation resulting from the sequential applications of $W$, $U_{\mathrm{eff}}$ and $V$.
    \item[(ii)] Sample a single event from the overall transformation with an input state having single photons in modes $(1, \ldots, n)$ and $(m+1, \ldots, m+k)$, by using the exact  algorithm due to Clifford and Clifford \cite{Clifford18}.
    \item[(iii)] If one photon per mode is obtained for modes $(m+1, \ldots, m+k)$, retain the event as valid. Otherwise, discard the event and repeat point (ii).
    \item[(iv)] Iterate the procedure until the number of required samples $N_{\mathrm{sample}}$ is obtained.
\end{enumerate}

\subsection{Calculation of the effective unitary $U_{\mathrm{eff}}$}

\label{sec:gadget}

A central step in the  linear-optics based simulation algorithm os the calculation of the effective unitary $U_{\mathrm{eff}}$. This transformation depends on the employed non-linear evolution $N$, and a recipe for obtaining it has been reported in Ref.\ \cite{Scheel03}.

Consider an input state composed by an arbitrary superposition of up to $k$-photon Fock terms:
\begin{equation}
\vert \psi_{\mathrm{in}} \rangle = c_{0} \vert 0 \rangle + c_{1} \vert 1 \rangle + c_{2} \vert 2 \rangle + \ldots + c_{k} \vert k \rangle.
\end{equation}
We focus on the specific non-linear evolution $N$ considered here, corresponding to a non-linear phase shift acting as $\exp(- \imath \hat{n}^2 \phi)$. The desired output state after transformation $N$ is given by:
\begin{equation}
\label{eq:psi_out}
\vert \psi_{\mathrm{out}} \rangle = c_{0} \vert 0 \rangle + c_{1} e^{- \imath \phi} \vert 1 \rangle + c_{2} e^{- \imath 4 \phi} \vert 2 \rangle + \ldots + c_{k} e^{- \imath k^{2} \phi} \vert k \rangle.
\end{equation}

\begin{figure}[ht!]
\centering
\includegraphics[width=0.99\textwidth]{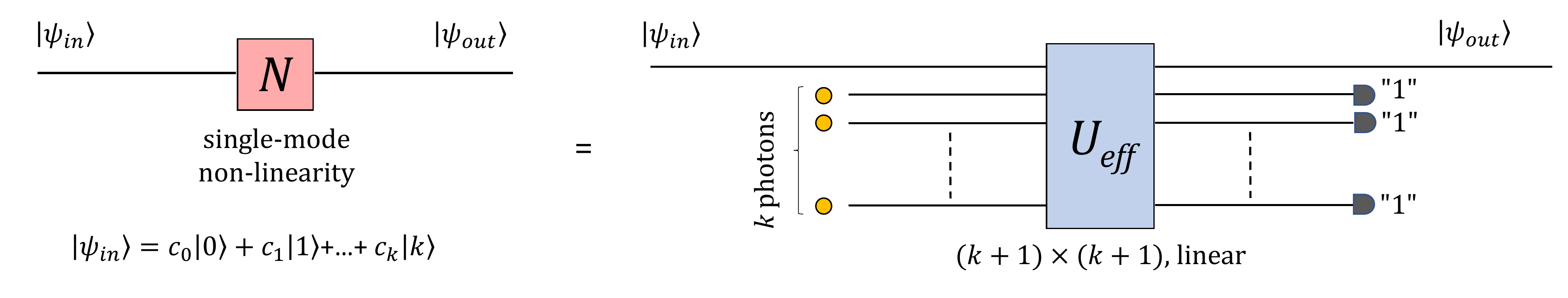}
\caption{Scheme of the linear-optics gadget for simulation of single-mode non-linearities with an input of up to $k$ photons.}
\label{fig:gadget}
\end{figure}

The main idea is to consider the gadget shown in Fig.\ \ref{fig:gadget}.
It is composed by a $(k+1) \times (k+1)$ unitary transformation $U_{\mathrm{eff}}$. The input state $\vert \psi_{\mathrm{in}} \rangle$ is injected into  mode $1$, while a single photon is injected in each auxiliary mode (from $2$ to $k+1$). Conditioned on the detection a single photon in each output mode from $2$ to $k+1$, the output state on mode $1$ can be written as:
\begin{equation}
\label{eq:psi_prime_0}
\vert \psi^{\prime} \rangle = c_{0} \mathrm{per}(U_{\mathrm{eff}}^{0,1,\ldots,1}) \vert 0 \rangle + c_{1} \mathrm{per}(U_{\mathrm{eff}}^{1,1,\ldots,1}) \vert 1 \rangle + c_{2} \mathrm{per}(U_{\mathrm{eff}}^{2,1,\ldots,1}) \vert 2 \rangle + \ldots + c_{k} \mathrm{per}(U_{\mathrm{eff}}^{k,1,\ldots,1}) \vert k \rangle.
\end{equation}
Here, $U_{\mathrm{eff}}^{l,1,\ldots,1}$ is the $(l+k) \times (l+k)$ matrix obtained by repeating $l$ times the first row and the first column of $U_{\mathrm{eff}}$, and by repeating a single time each of the remaining rows and columns of $U_{\mathrm{eff}}$. Assuming that $\mathrm{per}(U_{\mathrm{eff}}^{0,1,\ldots,1})$ is different from zero, Eq.\ \eqref{eq:psi_prime_0} can be rearranged as:
\begin{equation}
\label{eq:psi_prime_rearr}
\vert \psi^{\prime} \rangle = \mathrm{per}(U_{\mathrm{eff}}^{0,1,\ldots,1}) \left[ c_{0}  \vert 0 \rangle + c_{1} \frac{\mathrm{per}(U_{\mathrm{eff}}^{1,1,\ldots,1})}{\mathrm{per}(U_{\mathrm{eff}}^{0,1,\ldots,1})} \vert 1 \rangle + c_{2} \frac{\mathrm{per}(U_{\mathrm{eff}}^{2,1,\ldots,1})}{\mathrm{per}(U_{\mathrm{eff}}^{0,1,\ldots,1})} \vert 2 \rangle + \ldots + c_{k} \frac{\mathrm{per}(U_{\mathrm{eff}}^{k,1,\ldots,1})}{\mathrm{per}(U_{\mathrm{eff}}^{0,1,\ldots,1})} \vert k \rangle \right].
\end{equation}
In this expression, the common term $\mathrm{per}(U_{\mathrm{eff}}^{0,1,\ldots,1})$ is directly related to the success probability of the process, which is obtained as $\mathrm{Pr}_{\mathrm{succ}} = \vert \mathrm{per}(U_{\mathrm{eff}}^{0,1,\ldots,1}) \vert^{2}$. The conditions that $U_{\mathrm{eff}}$ must satisfy in order to correspond to the desired phase shift are obtained by imposing that, up to the global term, coefficients for state $\vert \psi_{\mathrm{out}} \rangle$ in Eq.\ \eqref{eq:psi_out} are equal to those for state $\vert \psi^{\prime} \rangle$ in Eq.\ \eqref{eq:psi_prime_rearr}. This procedure leads to the following set of equations:
\begin{equation}
\label{eq:conditions_Ueff}
\begin{cases}
\mathrm{per}(U_{\mathrm{eff}}^{1,1,\ldots,1}) &= \mathrm{per}(U_{\mathrm{eff}}^{0,1,\ldots,1}) e^{-\imath \phi}\\
\mathrm{per}(U_{\mathrm{eff}}^{2,1,\ldots,1}) &= \mathrm{per}(U_{\mathrm{eff}}^{0,1,\ldots,1}) e^{-\imath 4 \phi}\\
&\ldots \\
\mathrm{per}(U_{\mathrm{eff}}^{k,1,\ldots,1}) &= \mathrm{per}(U_{\mathrm{eff}}^{0,1,\ldots,1}) e^{-\imath k^2 \phi}
\end{cases}
\end{equation}

Finding $U_{\mathrm{eff}}$ such that all conditions \eqref{eq:conditions_Ueff} are satisfied is not a trivial task. Indeed, evaluation of these conditions involves matrix permanents of size up to $(2k) \times (2k)$. Here, to determine the matrices $U_{\mathrm{eff}}$ to simulate the non-linear phase shift evolution up to $k=4$, we  employed a numerical minimization approach. More specifically, we minimized the following quantity:
\begin{equation}
\label{eq:minimiz}
\mathcal{D} = \sum_{l=1}^{k} \left\{ \left[\mathrm{Re}\left(\mathrm{per}(U_{\mathrm{eff}}^{l,1,\ldots,1})\right) - \mathrm{Re}\left(\mathrm{per}(U_{\mathrm{eff}}^{0,1,\ldots,1}) e^{-\imath l^{2} \phi}\right)\right]^{2} + \left[\mathrm{Im}\left(\mathrm{per}(U_{\mathrm{eff}}^{l,1,\ldots,1})\right) - \mathrm{Im}\left(\mathrm{per}(U_{\mathrm{eff}}^{0,1,\ldots,1}) e^{-\imath l^{2} \phi}\right)\right]^{2} \right\}   .
\end{equation}
Parametrization of $U_{\mathrm{eff}}$ is obtained by using the Reck decomposition \cite{Reck}, which is employed as a mathematical tool to define the set of parameters (beam-splitter transmittivities and phase shifts) over which the minimization is performed.

Performing an unconstrained minimization, we found that the algorithm can collapse into a minimum (a value of $U_{\mathrm{eff}}$) which provides a very small success probability $\mathrm{Pr}_{\mathrm{succ}}$. To avoid this issue, we  searched for $U_{\mathrm{eff}}$ by performing a constrained numerical optimization, imposing that $\mathrm{Pr}_{\mathrm{succ}} \geq p_{\mathrm{th}}$. Here, $p_{\mathrm{th}}$ is a threshold which has to be manually inserted before the minimization process. By trial and error, we found that the optimization process is sensitive to the value of $p_{\mathrm{th}}$, which thus should be carefully chosen.

By using this numerical approach, we calculated the effective unitaries $U_{\mathrm{eff}}$ for the linear-optics simulation gadget, for $k=2,3,4$ and $\phi = \pi/2$. For $k=2$ we found:
\begin{equation}
\label{eq:Ueff_2}
U_{\mathrm{eff}}^{(2)} = \begin{pmatrix}
0. -0.4574 \imath & -0.8426 + 0.0223 \imath & 0.2822 - 0.0261 \imath\\
-0.0969 - 0.0943 \imath & -0.1689 + 0.1830 \imath & -0.6028 + 0.7458 \imath\\
0.6775 + 0.5599 \imath & -0.2940 + 0.3756 \imath & 0. + 0. \imath
\end{pmatrix}
\end{equation}
which leads to a success probability of $\mathrm{Pr}_{\mathrm{succ}} \simeq 0.209$. For $k=3$ we found:
\begin{equation}
U_{\mathrm{eff}}^{(3)} = \begin{pmatrix} 
0.0032 + 0.2218 \imath & 0.0889 - 0.8075 \imath & 0.1756 - 0.0772 \imath & 0.1455 - 0.4826 \imath \\
0.6671 + 0.1569 \imath & 0.1942 - 0.2732 \imath & 0.1871 - 0.0443 \imath & -0.1623 + 0.5956 \imath\\
0.0606 - 0.1733 \imath & 0.2204 - 0.2742 \imath & -0.8767 + 0.1685 \imath & -0.2133 - 0.0099 \imath\\
0.2418 - 0.6237 \imath & 0.3134 + 0.0717 \imath & 0.3240 + 0.1560 \imath & -0.4179 - 0.3802 \imath
\end{pmatrix}
\end{equation}
which leads to a success probability of $\mathrm{Pr}_{\mathrm{succ}} \simeq 0.04$. Finally, for $k=4$ we found that:
\begin{equation}
U_{\mathrm{eff}}^{(4)} = \begin{pmatrix}
-0.0006 - 0.1994 \imath & -0.5735 + 0.0763 \imath & 0.0071 - 0.0505 \imath & 0.2902 - 0.3501 \imath & 0.1843 - 0.6181 \imath \\
0.3200 + 0.2740 \imath & 0.3072 + 0.5019 \imath & -0.0749 - 0.0098 \imath & -0.0761 + 0.3436 \imath & 0.4135 - 0.4190 \imath \\
0.4328 - 0.1960 \imath & -0.0933 + 0.4354 \imath & -0.1877 + 0.3742 \imath & 0.4265 - 0.1473 \imath & -0.0404 + 0.4421 \imath \\
0.4356 + 0.6058 \imath & 0.0671 - 0.2111 \imath & 0.2851 + 0.0872 \imath & - 0.0559 - 0.5462 \imath & -0.0572 - 0.0219 \imath \\
0.0123 + 0.0017 \imath & 0.2591 + 0.0667 \imath & -0.3623 - 0.7721 \imath & 0.2273 - 0.3355 \imath & 0.1105 + 0.1560 \imath
\end{pmatrix}
\end{equation}
which leads to a success probability of $\mathrm{Pr}_{\mathrm{succ}} = \simeq 0.008$.

We first observe that, in all cases, all matrices present a finite success probability. Furthermore, we observe a decreasing trend in $\mathrm{Pr}_{\mathrm{succ}}$ with respect to the photon number threshold $k$. We also find that, due to the presence of larger matrix permanents and to the increasing number of parameters, the computation becomes progressively more expensive for increasing $k$. 

As a further analysis, we have then focused on the $k=2$ scenario, in particular with reference to the results of Ref.\ \cite{Eisert2005}. In that paper, upper bounds on the success probabilities for measurement-induced non-linear gates have been derived. For instance, in the case of a non-linear phase gate of the form: 
\begin{equation}
\label{eq:nonlinear_gate}
c_{0} \vert 0 \rangle + c_{1} \vert 1 \rangle + c_{2} \vert 2 \rangle \rightarrow c_{0} \vert 0 \rangle + c_{1} \vert 1 \rangle + c_{2} e^{\imath \varphi} \vert 2 \rangle,
\end{equation}
the success probability is bounded by: 
\begin{equation}
\label{eq:bound_0}
\mathrm{Pr}_{\mathrm{succ}} \leq [3 - \cos(\pi - \varphi)]^2/16.
\end{equation} 
This bound is derived by assuming unlimited resources for the measurement-induced scheme, that is, unbounded number of ancillary modes and photons.

This result can be applied directly to the non-linear phase shift $\mathrm{exp}(-\imath \hat{n}^2 \phi)$. More specifically, the output state of up to $2$ photon terms after this non-linear evolution is 
\begin{equation}
\label{eq:nonlinear_phase_twophot}
\exp(- \imath \hat{n}^2 \phi) \left(c_{0} \vert 0 \rangle + c_{1} \vert 1 \rangle + c_{2} \vert 2 \rangle \right) = c_{0} \vert 0 \rangle + c_{1} e^{-\imath \phi} \vert 1 \rangle + c_{2} e^{- \imath 4 \phi} \vert 2 \rangle  
\end{equation}
The state of Eq.\ \eqref{eq:nonlinear_phase_twophot} can be mapped to Eq.\ \eqref{eq:nonlinear_gate} by means of a linear phase shift operator as:
\begin{equation}
c_{0} \vert 0 \rangle + c_{1} e^{-\imath \phi} \vert 1 \rangle + c_{2} e^{- \imath 4 \phi} \vert 2 \rangle = \exp(-\imath \hat{n} \phi) \left( c_{0} \vert 0 \rangle + c_{1} \vert 1 \rangle + c_{2} e^{- \imath 2 \phi} \vert 2 \rangle \right) 
\end{equation}
Thus, the bound of Eq.\ \eqref{eq:bound_0} applies to the non-linear phase shift considered here up to $2$ input photons, by identifying $\varphi = -2 \phi$:
\begin{equation}
\label{eq:bound_1}
\mathrm{Pr}_{\mathrm{succ}} \leq \frac{[3 - \cos(\pi + 2 \phi)]^2}{16}    
\end{equation}

As a first step, we focus on the $\phi = \pi/2$ scenario. In this case, the bound of Eq.\ \eqref{eq:bound_1} gives $\mathrm{Pr}_{\mathrm{succ}} \leq 0.25$. The effective unitary $U_{\mathrm{eff}}^{(2)}$ of Eq.\ \eqref{eq:Ueff_2}, found with the previous method, corresponds to $\mathrm{Pr}_{\mathrm{succ}} \simeq 0.209$, and thus is close to the bound of Ref.\ \cite{Eisert2005}.

\begin{figure}[ht!]
\centering
\includegraphics[width=0.6\textwidth]{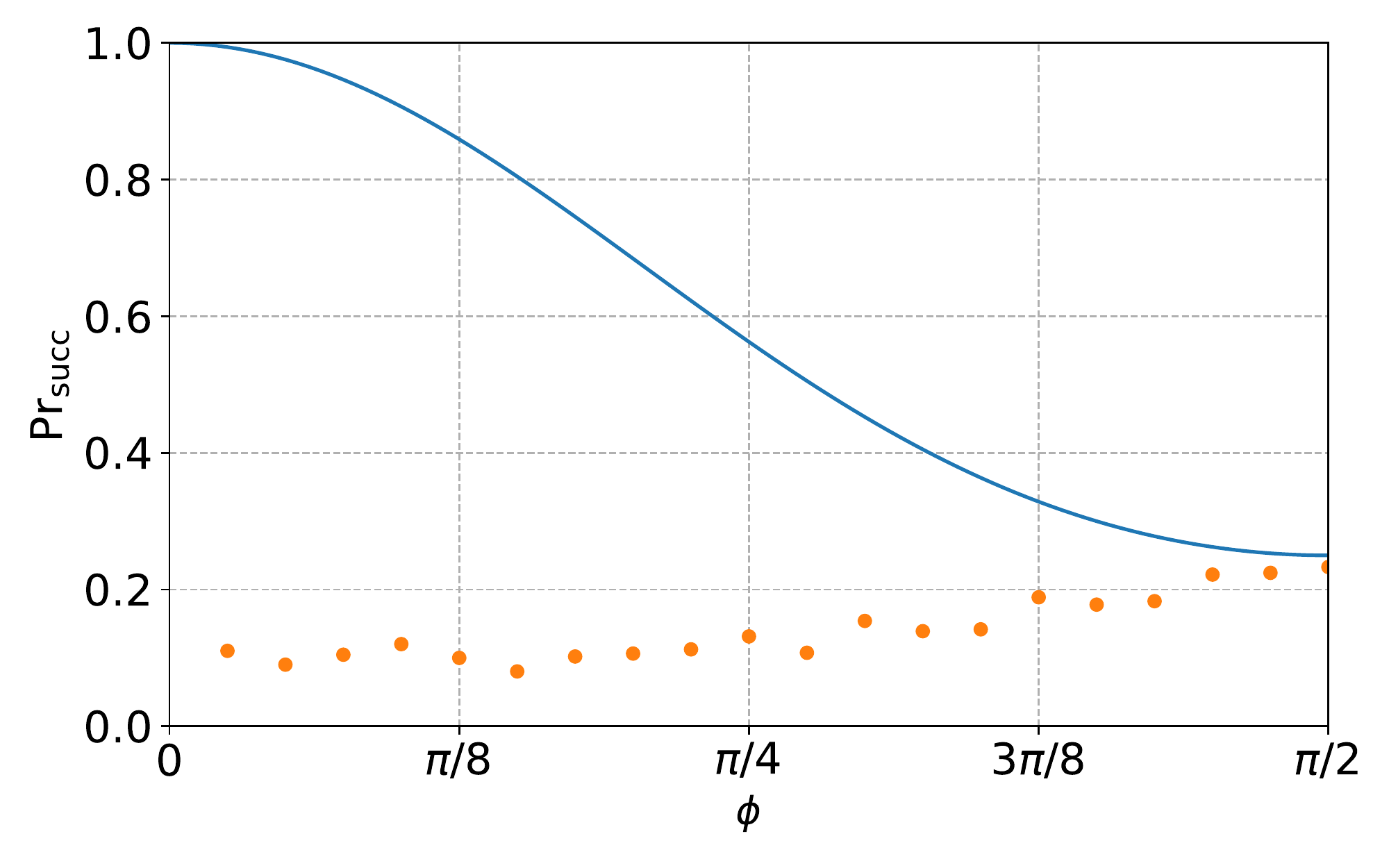}
\caption{Success probability for the effective unitaries, found with the approach discussed in the text, for $k=2$ and different values of $\phi$. Blue solid line: bound for $\mathrm{Pr}_{\mathrm{succ}}$ obtained from Eq.\ \eqref{eq:bound_1} as per Ref.\ \cite{Eisert2005}. Orange points: $\mathrm{Pr}_{\mathrm{succ}}$ found for the best unitaries obtained via the constrained numerical minimization of Eq.\ \eqref{eq:minimiz}.}
\label{fig:psucc}
\end{figure}

We then investigated whether it is possible to use the same approach to obtain effective unitaries with finite success probability for all values of $\phi \in [0, \pi/2]$. The results are shown in Fig.\ \ref{fig:psucc}, and compared with the bound of Eq.\ \eqref{eq:bound_1}. We observe that, in fact, for all tested values of $\phi$, $\mathrm{Pr}_{\mathrm{succ}} \gtrsim 0.1$. Thus, it is possible to implement the simulation algorithm with non-negligible success probability. On the other hand, for small $\phi$ we also observe larger deviations from the bound of Eq.\ \eqref{eq:bound_1}. One reason could be the fact that this bound allows for measurement-induced schemes with unbounded resources, whereas the gadget we consider here uses a specific number of photons and modes (see Fig.\ \ref{fig:gadget}). It is also relevant that at this stage we are only considering the maximum success probability for an \emph{exact} implementation of the non-linear phase. Thus, even though there is a trivial implementation with $\mathrm{Pr}_{\mathrm{succ}}=1$ for the specific value $\phi=0$ (i.e.\ the identity matrix), as soon as $\phi$ deviates from $0$ it might become necessary to have more modes and photons achieve success probability greater than $\approx 0.1$.

However, in the regime where $\phi$ is small, an exact implementation might not be the best approach. As discussed in Sec.\ \ref{sec:approx}, in that regime it might be better to incur a small error in TVD and use a linear phase as approximation to the non-linear one (which would also lead to a smaller runtime in the simulation, since it would not require simulation via extra photons). We leave the investigation of the tradeoff between these figures of merit---runtime and approximation error---as direction for future work.

\subsection{Correlations between bunching and TVD}

\begin{figure}[ht!]
\centering
\includegraphics[width = 0.99 \textwidth]{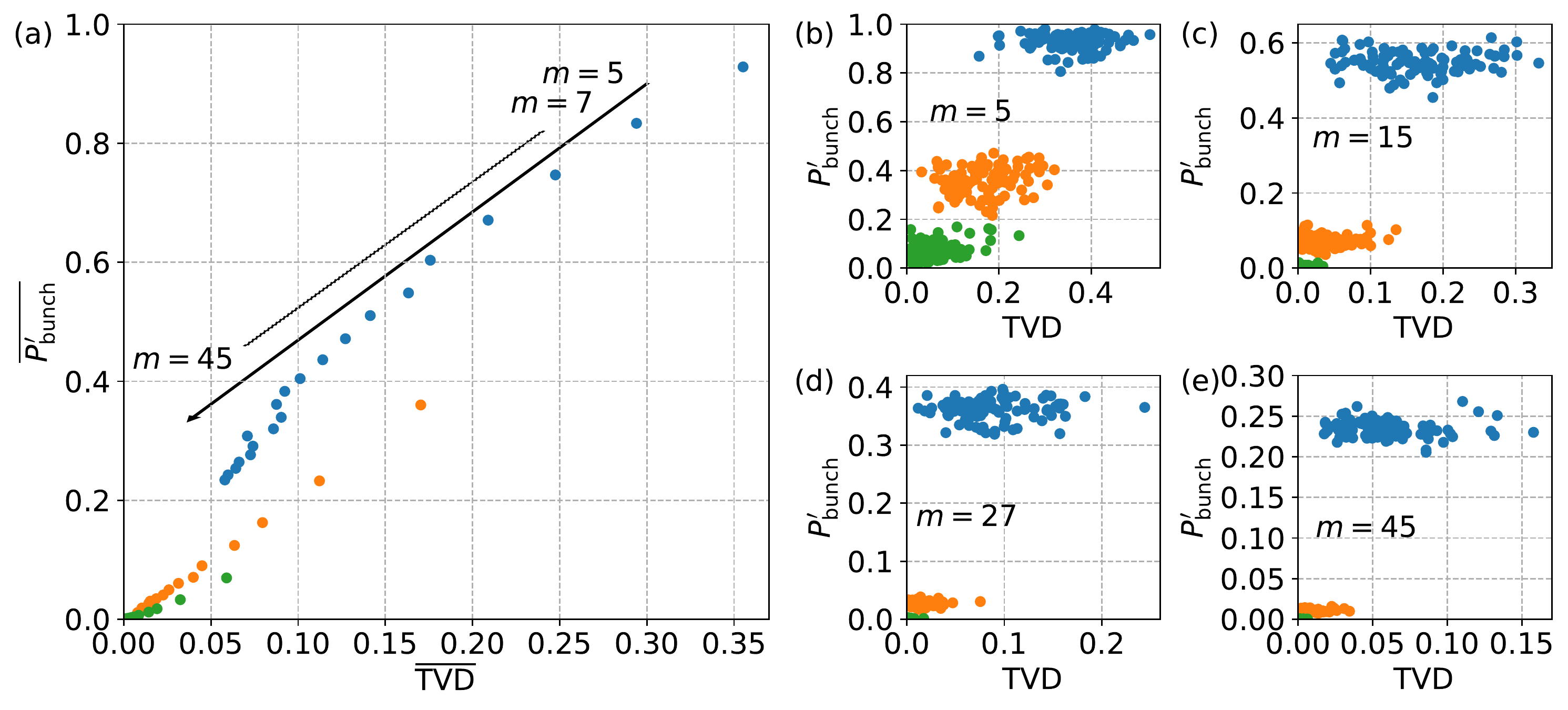}
\caption{Analysis of total variation distance (TVD) and bunching probability, for $n=4$ photons and different number of modes. The TVD is calculated between the exact probability distribution, whose single transition amplitudes are obtained from Eq.\ \eqref{eq:nonlinear_phase_simplified}, and the post-selected distribution corresponding to the linear optics measurement-based simulation algorithm for $k<n$. In these plots, the exact distribution is calculated as the output one from the simulation scheme with $n$ ancillary photons and modes. At variance with Fig.\ 3 of the main text, the bunching probability $P^{\prime}_{\mathrm{bunch}}$ corresponds to the overall probability obtained by summing up all terms having more than $k$ photons in the full distribution. (a) Parametric plot showing the relative trend between the $\overline{\mathrm{TVD}}$ and the bunching probability $\overline{P^{\prime}_{\mathrm{bunch}}}$ at the non-linear site, averaged over $100$ different evolutions $W$ and $V$ for fixed $\phi = \pi/2$.  Each point corresponds to a different value of $m$, in the range $[5, 7, \ldots, 45]$. (b)-(e) Correlation between TVD and $P^{\prime}_{\mathrm{bunch}}$ for fixed number of modes $m$. Here, each point corresponds to different evolutions $W$ and $V$ for fixed $\phi = \pi/2$.
In all plots: blue points corresponds to $k=1$, orange points to $k=2$, and green points to $k=3$.}
\label{fig:bunching_TVD_SI}
\end{figure}

In the main text we have shown that the bunching probability at the non-linear site is related to the TVD between the exact distribution and the one obtained with the simulation scheme based on linear optics and auxiliary photons and modes. In Fig.\ \ref{fig:bunching_TVD_SI} we report a similar analysis to the one shown in Fig.\ 3 of the main text. However, here we replace the bunching probability at the non-linear site $P_{\mathrm{bunch}}$ with the overall bunching probability $P^{\prime}_{\mathrm{bunch}}$ calculated for the full distribution.

In this scenario, we still observe an overall trend between the average $\overline{\mathrm{TVD}}$ and the average bunching probability $\overline{P^{\prime}_{\mathrm{bunch}}}$ for varying number of modes [see Fig.\ \ref{fig:bunching_TVD_SI} (a)]. However, when looking at the distribution for fixed number of modes $m$, we observe no correlations  between $\mathrm{TVD}$ and the bunching probability (over all modes). This highlights the facy that, as shown in the main text and in Eq.\ \eqref{eq:nonlinear_phase_simplified}, the TVD is related to the bunching probability restricted to the mode with the non-linearity.

\begin{figure}[ht!]
\centering
\includegraphics[width = 0.99 \textwidth]{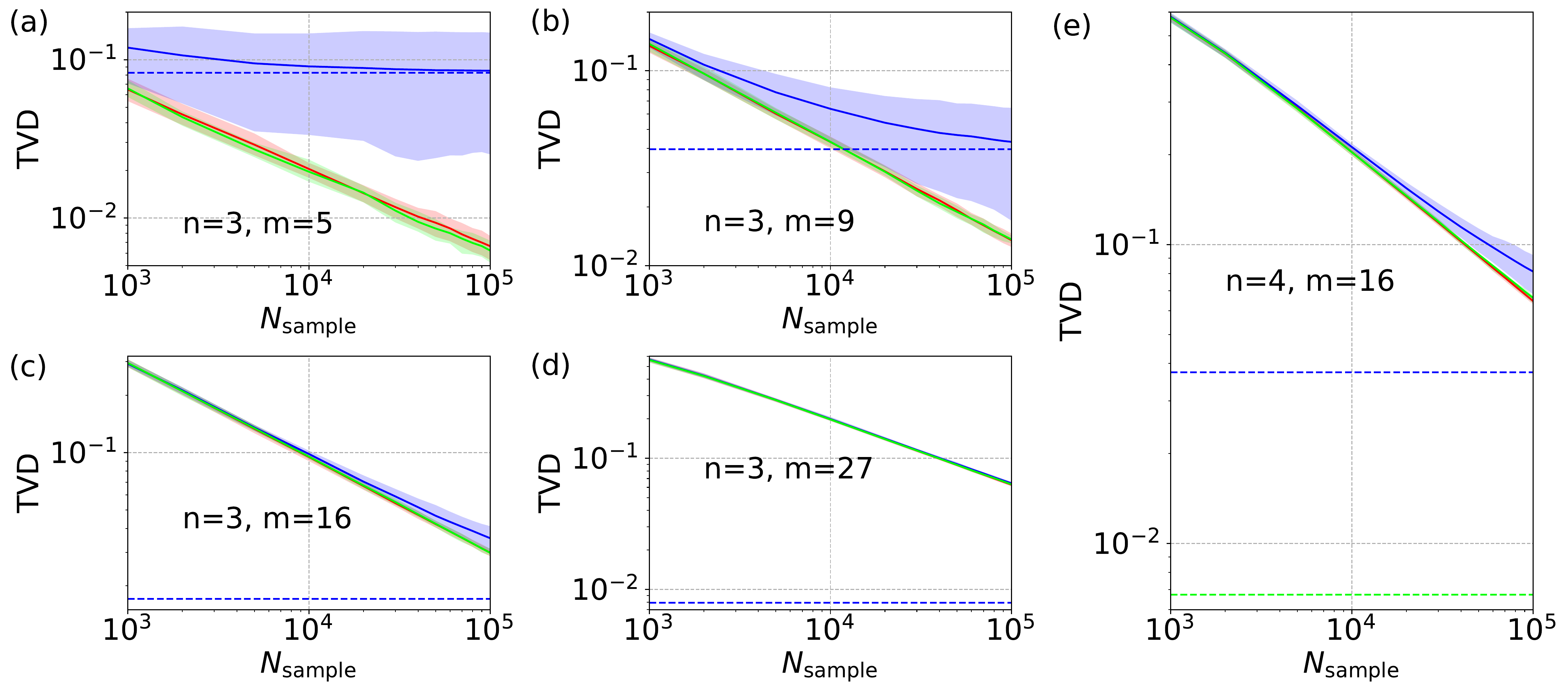}
\caption{Classical simulation of non-linear Boson Sampling with a single non-linear phase $\phi = \pi/2$ via algorithm based on linear optics and ancillary photons. Quality of the simulation is measured via the total variation distance (TVD), defined as $\mathrm{TVD} = 1/2 \sum_i \vert p_i - q_i \vert$ between the output distribution $p_i$ calculated from Eq.\ (\ref{eq:nonlinear_singlephase}) with $\phi = \pi/2$ and a sample $q_i$ drawn by a chosen sampling algorithm. (a-d) $n=3$ with $m=5,9,16,27$ and (e) $n=4, m=16$. Red: $\mathrm{TVD}$ obtained from exact simulation. Blue: $\mathrm{TVD}$ obtained from the linear optics based algorithm with $k=2$. Green: $\mathrm{TVD}$ obtained from the linear optics based algorithm with $k=3$. Horizontal dashed lines correspond to the TVD in the asymptotic limit of very large sample size.}
\label{fig:convergence_sample_size}
\end{figure}

\subsection{Classical simulation algorithm implementation with finite sample size}

In this section we report on numerical simulations performed to test the implementation of Algorithm 1 as a classical simulation for non-linear Boson Sampling. The results of these numerical simulations are shown in Fig.\ \ref{fig:convergence_sample_size} for different values of $n$ and $m$. In all cases, we considered a single non-linear phase positioned in the central mode between linear transformations $W$ and $V$ according to the scheme described in the main text. We then verified the convergence, in total variation distance, of a sample composed of $N_{\mathrm{sample}}$ events with respect to the exact output distribution.

More specifically, for each tested value of $n$ and $m$ we have drawn samples by using different approaches. These include brute-force sampling from the exact distribution obtained with the Feynman integral approach, and sampling with the Algorithm 1 for $k=n$ and $k<n$. From the plots of Fig.\ \ref{fig:convergence_sample_size}, we first observe that sampling from the exact distribution and from the linear-optics simulation algorithm with $k=n$ provides the same trend as a function of $N_{\mathrm{sample}}$. This provides an additional numerical evidence that using Algorithm 1 with $k=n$ leads to an exact simulation algorithm for non-linear Boson Sampling. Conversely, when $k<n$, the simulation saturates to an asymptotic value provided by the post-selected distribution at the output of the linear-optics gadget.